\pdfoutput=1

\documentclass[11pt,twoside,a4paper,cmspaper,final,collab]{cms-tdr}
\def\svnVersion{277505}\def\cmsCernNo{2013-037}\def\cmsCernDate{\today}\def\cmsMessage{Published in the Journal of Instrumentation as \href{http://dx.doi.org/10.1088/1748-0221/10/02/P02006}{\doi{10.1088/1748-0221/10/02/P02006.}}}

\begin{document}\cmsNoteHeader{JME-13-003}

\hyphenation{had-ron-i-za-tion}
\hyphenation{cal-or-i-me-ter}
\hyphenation{de-vices}
\RCS$Revision: 272672 $
\RCS$HeadURL: svn+ssh://svn.cern.ch/reps/tdr2/papers/JME-13-003/trunk/JME-13-003.tex $
\RCS$Id: JME-13-003.tex 272672 2014-12-24 21:55:30Z alverson $

\newcommand{\fulllumi}{19.7\xspace}
\newcommand{\fulllumierr}{0.5\xspace}
\newcommand{\W}{\PW\xspace}
\newcommand{\SigmaZ}{\ensuremath{\sigma_0}}
\newcommand{\SigmaS}{\ensuremath{\sigma_{\mathrm{s}}}}
\newcommand{\XX}{\ensuremath{X}}
\newcommand{\Wenu}{\ensuremath{\PW\to \Pe\nu}}
\newcommand{\sig}{\ensuremath{\mathcal{S}}}
\newcommand{\pchisq}{\ensuremath{\mathcal{P}_2(\sig)}}
\newcommand{\vet}{\ensuremath{\vec\varepsilon}}
\newcommand{\like}{\ensuremath{\mathcal{L}}}
\newcommand{\chisq}{\ensuremath{\chi^2}}
\newcommand{\eslash}{\ensuremath{{\hbox{$E$\kern-0.6em\lower-.05ex\hbox{/}\kern0.10em}}}}
\newcommand{\bigeslash}{{\hbox{$E$\kern-0.38em\lower-.05ex\hbox{/}\kern0.10em}}}
\newcommand{\vecmet}{\ensuremath{\vec{\eslash}_\mathrm{T}}\xspace} 
\newcommand  {\met}{\ensuremath{\eslash_\mathrm{T}}\xspace}
\newcommand{\mex}{\ensuremath{\eslash_{x}}\xspace} 
\newcommand{\mey}{\ensuremath{\eslash_{y}}\xspace} 
\newcommand{\pfmet}{PF\,\MET}
\newcommand{\pfvecmet}{PF\,\vecmet}
\newcommand{\pfsumet}{PF\, \ensuremath{\sum E_\mathrm{T}}\xspace}
\newcommand{\Zmm}{\ensuremath{\cPZ \to \Pgmp\Pgmm}\xspace}
\newcommand{\Zee}{\ensuremath{\cPZ \to \Pep\Pem}\xspace}
\newcommand{\Wen}{\ensuremath{\PW \to \Pe\cPgn}\xspace}
\newcommand{\WJ}{\ensuremath{\PW + \text{jets}}\xspace}
\newcommand{\ZJ}{\ensuremath{\cPZ + \text{jets}}\xspace}
\newcommand{\GJ}{\ensuremath{\gamma + \text{jets}}\xspace}
\newcommand{\GG}{\ensuremath{\gamma\gamma}\xspace}
\newcommand{\sumet}{\ensuremath{\sum E_\mathrm{T}}\xspace}
\newcommand{\sumET}{\ensuremath{\sum E_\mathrm{T}}\xspace}
\newcommand{\bigmet}{\ensuremath{\bigeslash_\mathrm{T}}\xspace}
\newcommand{\qt}{\ensuremath{{q}_\mathrm{T}}\xspace}
\newcommand{\vecpt}{\ensuremath{\vec{p}_\mathrm{T}}\xspace}
\newcommand{\vqt}{\ensuremath{\vec{q}_\mathrm{T}}\xspace}
\newcommand{\vut}{\ensuremath{\vec{u}_\mathrm{T}}\xspace}
\newcommand{\vpt}{\ensuremath{\vec{p}_\mathrm{T}}\xspace}
\newcommand{\hatqt}{\ensuremath{{\hat q}_\mathrm{T}}\xspace}
\newcommand{\upar}{\ensuremath{u_\Vert}\xspace}
\newcommand{\upara}{\ensuremath{u_\Vert}\xspace}
\newcommand{\uperp}{\ensuremath{u_\perp}\xspace}
\newcommand{\redupara}{\ensuremath{u_\Vert + \qt}\xspace}
\newcommand{\reso}[1]{\ensuremath{ \sigma(#1) }\xspace}
\newcommand{\resp}{\ensuremath{- \langle \upar \rangle / \qt}\xspace}
\newcommand{\DeltaR}{\ensuremath{\Delta R}\xspace}
\newcommand{\metsig}{\ensuremath{\mathcal{S}}\xspace}

\cmsNoteHeader{JME-13-003}
\title{Performance of the CMS missing transverse momentum reconstruction in pp data at $\sqrt{s}=8$\TeV}

\date{\today}

\abstract{
The performance of missing transverse energy reconstruction algorithms is presented
using $\sqrt{s}=8$\TeV proton-proton (pp) data collected with the CMS detector.
Events with anomalous missing transverse energy are studied, and
the performance of algorithms used to identify and remove these events is presented.
The scale and resolution for missing transverse energy,
including the effects of multiple pp interactions (pileup), are measured using
events with an identified Z boson or isolated photon, and are found to
be well described by the simulation.
Novel missing transverse energy reconstruction algorithms
developed specifically to mitigate the effects of large numbers of pileup interactions
on the missing transverse energy resolution are presented.
These algorithms significantly reduce the dependence of the missing transverse energy resolution on pileup interactions.
Finally, an algorithm that provides an estimate of the significance of the missing transverse energy is presented, which is used to estimate
 the compatibility of the reconstructed missing transverse energy with a zero nominal value.
}

\hypersetup{%
pdfauthor={CMS Collaboration},%
pdftitle={Performance of the CMS missing transverse momentum reconstruction in pp data at sqrt(s) = 8 TeV pp data},%
pdfsubject={CMS},%
pdfkeywords={CMS, Missing ET, performance, pileup}}

\maketitle

\section{Introduction}
\label{sec:intro}

The CMS detector~\cite{Chatrchyan:2008zzk} can detect almost all stable or long-lived particles produced in the proton-proton (pp) collisions
provided by the LHC at CERN.  Notable exceptions are
neutrinos and hypothetical neutral weakly interacting particles.
Although these particles do not leave a
signal in the detector, their presence can be inferred from the momentum
imbalance in the plane perpendicular to the beam direction, a quantity known as
missing transverse momentum and denoted by \vecmet.  Its magnitude is
denoted by \met and will be referred to as missing transverse energy.

The \vecmet\ plays a critical role in many physics
analyses at the LHC.  It is a key variable in many searches for physics
beyond the standard model, such as supersymmetry and extra dimensions, as
well as for collider-based dark matter searches.  It also played an important
role in studies contributing to the discovery of the Higgs boson,
in particular in channels with the $\PW\PW$,
$\Z\Z \rightarrow \ell\ell\nu\nu$, where $\ell$ is $\Pe$ or $\mu$, and
$H\to\tau\tau$ final states~\cite{Higgs2013Paper}.  In addition, the precise
measurement of \vecmet\ is critical for measurements of standard
model physics involving \PW~bosons and top quarks.

The \vecmet\ reconstruction is sensitive to detector malfunctions and to
various reconstruction effects that result in the mismeasurement of
particles or their misidentification.  Precise calibration of all
reconstructed physics objects ($\Pe$, $\mu$, $\tau$, $\gamma$, jets,
etc) is crucial for the \vecmet\ performance. The \vecmet\ is
particularly sensitive to additional pp interactions in the same,
earlier, and later bunch crossings (pileup interactions).  It is
therefore essential to study \vecmet\ reconstruction in detail with
data. This paper describes the \vecmet\ reconstruction algorithms and
associated corrections, together with performance studies
conducted in 8\TeV pp data.
The average number of interactions per bunch crossing in this dataset is
approximately 21.  Previous studies of the missing transverse energy
reconstruction in 7\TeV data were presented in Ref.~\cite{METJINST}.

This paper is organized as follows.  A brief description of the CMS
detector is presented in Section~\ref{sec:detector}.  In
Section~\ref{sec:samples}, the data and Monte Carlo (MC) simulation samples
used for the present study, together with the event selection criteria, are
described.  In Section~\ref{sec:metreco}, the different algorithms for
reconstructing \vecmet\ are presented.  In Section~\ref{sec:tails},
sources of anomalous \vecmet\ measurements from known detector artifacts
and methods for identifying them are described.  In
Section~\ref{sec:resolution}, the \vecmet\ scale and resolution are
reported based on the measurements made with event samples containing
isolated photon or \Z\
boson candidates.  Studies presented in Section~\ref{sec:resolution}
include a detailed evaluation of \vecmet resolution degradation caused by
pileup interactions.
Section~\ref{sec:nopumet} reports the performance of novel \vecmet\
reconstruction algorithms developed to cope with large numbers of pileup
interactions.  The algorithm that provides an estimate of the \met significance is described and its performance
presented in Section~\ref{sec:significance}.
Conclusions are given in Section~\ref{sec:conclusions}.

\section{The CMS detector}
\label{sec:detector}

The central feature of the CMS apparatus is a
superconducting solenoid, of 6\unit{m} internal diameter, providing a field of 3.8\unit{T}.
Within the field volume are the silicon pixel and strip tracker, the crystal
electromagnetic calorimeter (ECAL), and the brass/scintillator hadron
calorimeter (HCAL). Muons are measured in gas-ionization detectors
embedded in the steel flux-return yoke.

The ECAL consists of 75\,848 lead tungstate crystals, which provide coverage
in pseudorapidity $\abs{\eta}< 1.479 $ in a barrel region
and $1.479 <\abs{\eta} < 3.0$ in two endcap regions. A preshower
detector consisting of two planes of silicon sensors interleaved with
a total of $3X_0$ of lead is located in front of the endcap.
The ECAL has an energy resolution of better than 0.5\%
for unconverted photons with transverse energy $E_{\rm T}>
100\GeV$.

The HCAL comprises the following subdetectors: a barrel detector  covering
$\abs{\eta}<1.3$, two endcap detectors  covering $1.3<\abs{\eta}<3.0$, and two
forward detectors covering $2.8<\abs{\eta}<5.0$.
The HCAL, when combined with the ECAL, measures hadrons with a resolution
$\Delta E/E \approx 100\%\sqrt{\smash[b]{E\,[\GeVns{}]}} \oplus 5\%$.
In the region $\abs{\eta}< 1.74$, the HCAL cells have
widths of 0.087 in pseudorapidity and 0.087\,rad in
azimuth. In the $(\eta,\phi)$ plane,
and for $\abs{\eta}< 1.48$, the HCAL cells map onto
$5{\times}5$ ECAL crystal arrays to form calorimeter towers
projecting radially outwards from close to the nominal
interaction point. In addition to the barrel and endcap
detectors, CMS has extensive forward calorimetry.

The muons are measured in the pseudorapidity window $\abs{\eta}< 2.4$,
with detection planes made of three technologies:
drift tubes, cathode strip chambers, and resistive-plate chambers.
A global fit of the measurements from the muon system and the central tracker
results in a \pt resolution between 1 and 5\%, for \pt values up to 1\TeV.

The inner tracker measures charged particles within the $\abs{\eta} < 2.5$
pseudorapidity range. It consists of 1440 silicon pixel and 15\,148 silicon
strip detector modules.
The tracker provides an impact parameter resolution of about 15\mum and a
\pt resolution of about 2.0\% for 100\GeV particles.

The first level of the CMS trigger system, composed of custom hardware processors, uses
information from the calorimeters and muon detectors to select, in less than 3.2\mus, the
most interesting events. The high-level trigger processor farm further
decreases the event rate from around 100\unit{kHz} to $\sim$400\unit{Hz}, before data storage.

A more detailed description of the CMS apparatus can be found in Ref.~\cite{Chatrchyan:2008zzk}.

\section{Data samples, particle reconstruction, and event selection}
\label{sec:samples}

Data samples used for the studies presented in this paper were collected
from February through December 2012 in pp collisions at a
centre-of-mass energy $\sqrt{s}=8$\TeV, and correspond to an integrated
luminosity of $\fulllumi{}\pm\fulllumierr{}$\fbinv \cite{CMS-PAS-LUM-13-001}.  For all studies,
we require at least one well-identified event vertex whose $z$ position
is less than 24\unit{cm} away from the nominal centre of the detector,
whose transverse distance from the $z$-axis is less than 2\unit{cm}, and which is reconstructed with at least four tracks.  The
vertex with the largest value of $\sum \pt^2$ taken over all
associated tracks is considered to be the primary vertex
that corresponds to the origin of the hard-scattering process.

The CMS experiment uses global event reconstruction, also called particle-flow (PF)
event reconstruction~\cite{CMS-PAS-PFT-09-001,CMS-PAS-PFT-10-001}, which consists of
reconstructing and identifying each particle with an optimized
combination of all subdetector information. In this process, the
identification of the particle type (photon, electron, muon, charged hadron, or neutral hadron) plays an important role in the determination
of the particle direction and energy. Photons, such as those from \Pgpz\
decays or from electron bremsstrahlung, are identified as ECAL energy
clusters not matched to the extrapolation of any charged-particle
trajectory to the ECAL.
Electrons are identified as primary charged-particle tracks reconstructed by a Gaussian-sum
filter (GSF) algorithm \cite{0954-3899-31-9-N01} and matched to ECAL energy clusters;
the matching allows for associated bremsstrahlung photons.

Muons, such as those from \cPqb-hadron semileptonic decays, are identified as
tracks in the central tracker consistent with either a track or several hits
in the muon system, associated with minimum ionizing particle depositions in the calorimeters.
Muon reconstruction and identification are described in detail in Ref.~\cite{MUO-10-004}.
 Charged hadrons are defined to be charged-particle tracks identified
 neither as electrons nor muons. Finally,
neutral hadrons are identified as HCAL energy clusters not matched to any
charged-hadron trajectory, or as ECAL and HCAL energy excesses with
respect to the expected charged-hadron energy deposit.

The energy of photons is directly obtained from the ECAL measurement and
corrected for zero-suppression effects~\cite{ZeroSuppression}.
The energy of electrons is
determined from a combination of the track momentum at the main
interaction vertex, the corresponding ECAL cluster energy, and the
energy sum of all associated bremsstrahlung photons. The
energy of muons is obtained from the corresponding track momentum. The
energy of charged hadrons is determined from a combination of the track
momentum and the corresponding ECAL and HCAL energies, corrected for
zero-suppression effects, and calibrated for the nonlinear response of
the calorimeters. Finally the energy of neutral hadrons is obtained from
the associated calibrated ECAL and HCAL energy deposits.

For each event, hadronic jets are clustered from these reconstructed
particles with the infrared and collinear-safe anti-\kt
algorithm~\cite{Cacciari:2008gp, Cacciari:2011ma}, with a distance
parameter $R=0.5$.  The jet momentum is determined as the vectorial sum
of all particle momenta in the jet, and is found in simulated samples to
be below 2 to 5\% of the true momentum over the entire \pt range of
interest and over the detector acceptance.  The jet energy corrections
are derived from simulation and are confirmed by in-situ measurements
exploiting the energy balance of dijet and photon+jet
events~\cite{CMS-PAS-JME-10-010}. Jet energy resolution (JER) after PF
reconstruction is typically about 25\%, 10\%, and 5\% at $E=10$, 100,
and 1000\GeV, respectively; this may be compared to approximately 40\%,
12\%, and 5\% obtained when the calorimeters alone are used for jet
clustering without PF reconstruction.

The data are compared to simulated events
generated either with \PYTHIA v6.4.24 Monte Carlo \cite{pythia} for the QCD and \GG{} processes,
or with \MADGRAPH v5.1.3.30~\cite{MadGraph5,MadGraph} interfaced with \PYTHIA v6.4.24 for top
(\ttbar and single-top), \ZJ, \WJ, \GJ, and diboson (VV) processes. The \PYTHIA v6.4.24 program has been
set up with a parameter set description for the underlying event referred to as tune Z2*~\cite{pythia-z2,CMS-FWD-11-003}. The generated events are passed through the CMS detector simulation, which
is based on \GEANTfour~\cite{geant4}.  The detector geometry description
includes realistic subsystem conditions, such as the simulation of
non-functioning channels.

\begin{figure}[htb]
  \centering
  \includegraphics[width=0.45\textwidth]{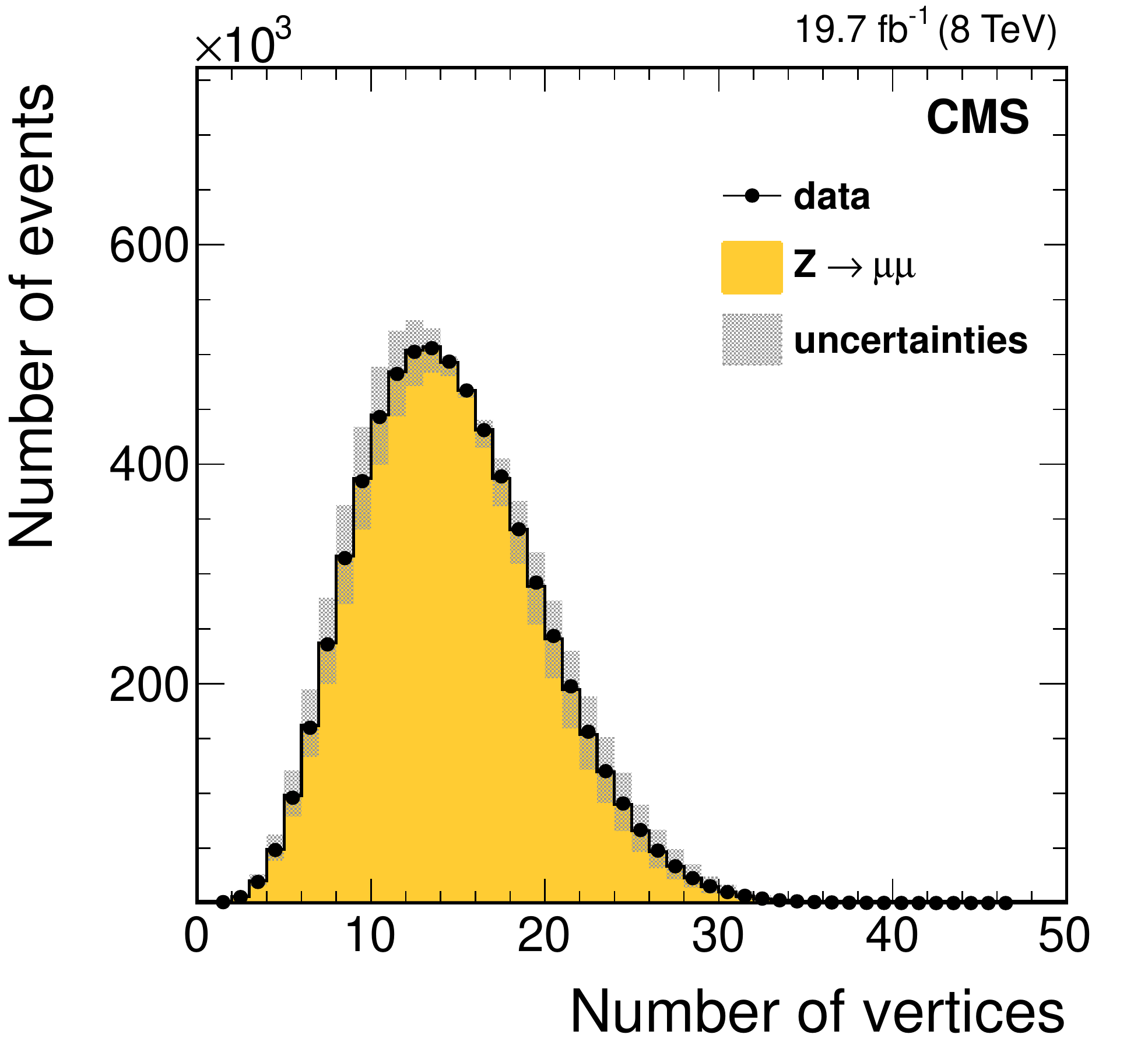}
  \caption{Multiplicity of reconstructed vertices for \Zee\ candidate events.
    The grey error band displays the systematic uncertainty of the simulation, and
    is dominated by the uncertainty in the total inelastic pp scattering cross
    section measurement~\cite{Antchev:1495764,CMS-PAS-LUM-13-001}.}
  \label{fig:nPU}
\end{figure}

The simulated events are weighted such that the distribution of the
simulated pileup interaction multiplicity matches the expected distribution,
as based on measurements of the instantaneous luminosities in data.
This is demonstrated in Fig.~\ref{fig:nPU}, which shows agreement
in the reconstructed vertex multiplicity ($N_\text{vtx}$) distribution between data
and simulated samples.  The total uncertainty in the
$N_\text{vtx}$ distribution is dominated by the uncertainty in the
total inelastic pp scattering cross section
measurement~\cite{Antchev:1495764,CMS-PAS-LUM-12-001}, which affects
the pileup profile in the simulated sample. The other uncertainty source
is in the luminosity measurement, which constitutes $\sim$30\%
of the total uncertainty.

\subsection{The dijet event selection}
\label{subsec:Dijetsample}

The dijet data sample is used in the studies of anomalous high-\met events are presented in Section~\ref{sec:tails}
and in the \met significance studies in Section~\ref{sec:significance}.
It was collected with a single-jet trigger that requires at least one jet in the event with $\pt>320\GeV$.
Dijet events are selected offline by requiring a leading jet with $\pt>400\GeV$
and at least one other jet with $\pt>200\GeV$.

\subsection{The \texorpdfstring{$\Z\to\ell^+\ell^-$}{Zll} event selection}
\label{subsec:Zsample}

The $\Z\to\ell^+\ell^-$ events, where $\ell$ is either a muon or an electron, are used
in the \met scale, resolution, and significance studies presented in Sections~\ref{sec:resolution}, \ref{sec:nopumet}, and
\ref{sec:significance}.

In order to discriminate  between prompt leptons
and leptons that are produced inside a jet through the decay of a hadron, we
define an isolation variable $R_\text{Iso}$ as the ratio of
\pt of particles near the lepton to the \pt of the lepton itself,
\begin{equation}
  R_\text{Iso}(\pt^\ell)
  \equiv \frac{1}{\pt^\ell}\Bigg[ \sum_{~\text{HS}\pm} \!\!\pt
    + \max \!\Big(0, ~\sum_\text{neu} \pt
      + \!\sum_\text{pho} \pt
      - \tfrac{1}{2}\!\!\sum_{~\text{PU}\pm} \!\!\pt \Big)
  \Bigg],
  \label{eq:LepIso}
\end{equation}

The scalar \pt\ sums $\sum_{\mathrm{HS}\pm}\pt$, $\sum_\text{neu}\pt$, and
$\sum_\text{pho}\pt$ are taken over particles from the primary hard-scatter
(HS) vertex, neutral hadrons, and photons, respectively; all particles entering the sums
must lie within a distance $\Delta R \equiv \sqrt{\smash[b]{(\Delta\phi)^2+ (\Delta\eta)^2}}<0.3$
of the lepton candidate. Well-isolated leptons, unlikely to have originated
from semi-leptonic decay within a jet, are characterized by low values of $R_\text{Iso}$.
The final negative sum over charged hadrons from pileup (PU) vertices
 compensates the additional energy produced by photons and
 neutral hadrons stemming from pileup interactions. The relative balance between charged particles
 and neutral particles produced by pileup interactions is taken into account using a factor 0.5 in the final sum.

The $\Zmm$ events were collected using a trigger that requires the
presence of two muons passing \pt thresholds of 17 and 8\GeV, respectively.
The muon candidates must be reconstructed in the
tracker and in the muon chambers, must satisfy
$\pt> 20$\GeV and lie in the
pseudorapidity range $\abs{\eta}<2.1$. In order to veto candidates from non-prompt processes,
muons must further satisfy $R_\text{Iso}(\pt^\mu) < 0.1$.

The $\Zee$ candidate events were collected using a double-electron
trigger with \pt thresholds of 17 and 8\GeV.  The events are required to
have two electron candidates within the ECAL
fiducial volume defined by $\abs{\eta} < 1.44$ and $1.56 < \abs{\eta} < 2.5$. To
reject jets or photons misidentified as electrons, requirements are
applied on the shower shape and the matching of the energy cluster with the associated GSF track, in
both $\phi$ and $\eta$. In addition,
electrons must satisfy $R_\text{Iso}(\pt^{\Pe})<0.1$ and
$\pt>20\GeV$.

\begin{figure}[htb]
  \centering
  \includegraphics[width=0.42\textwidth]{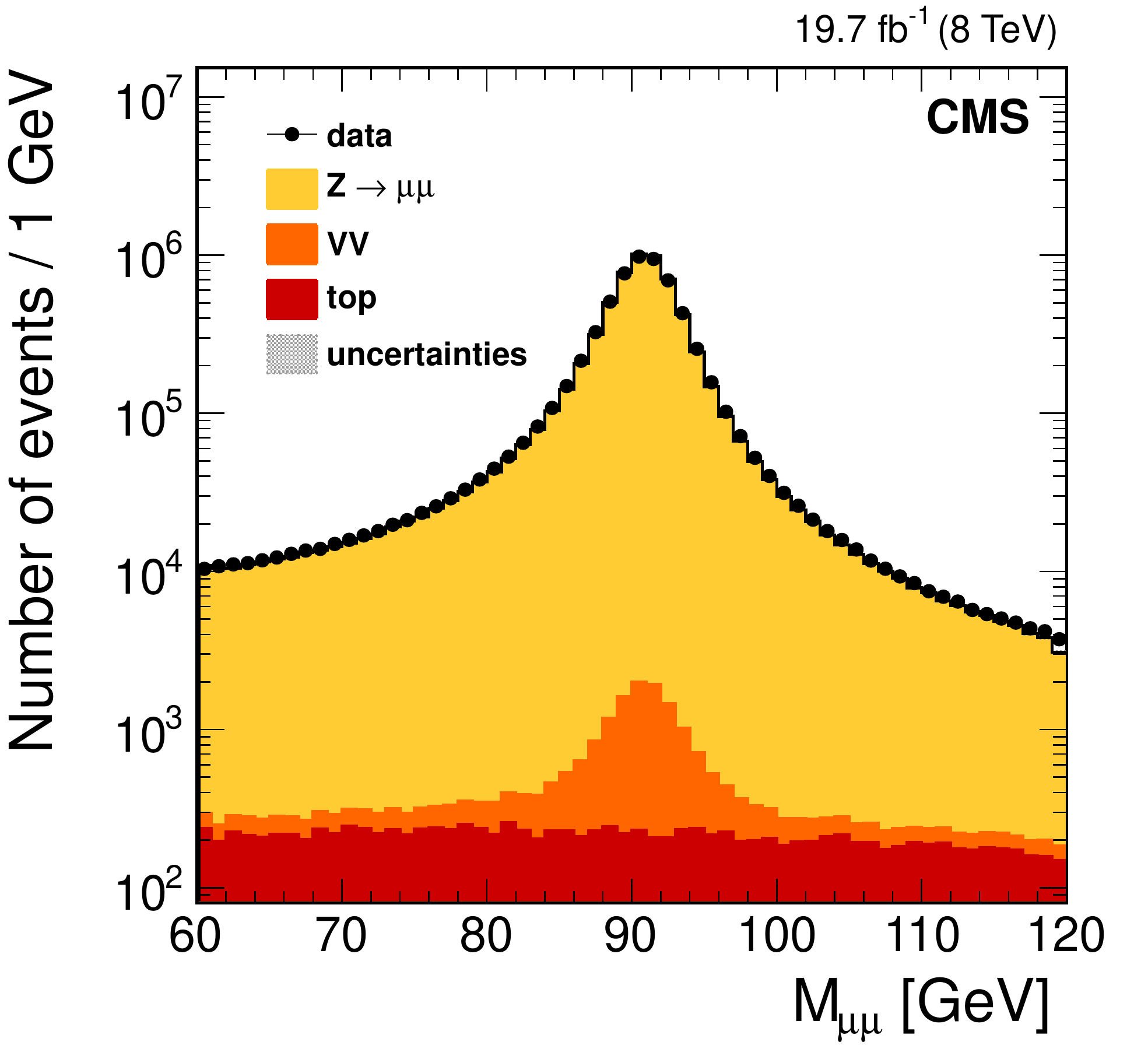}
  \includegraphics[width=0.42\textwidth]{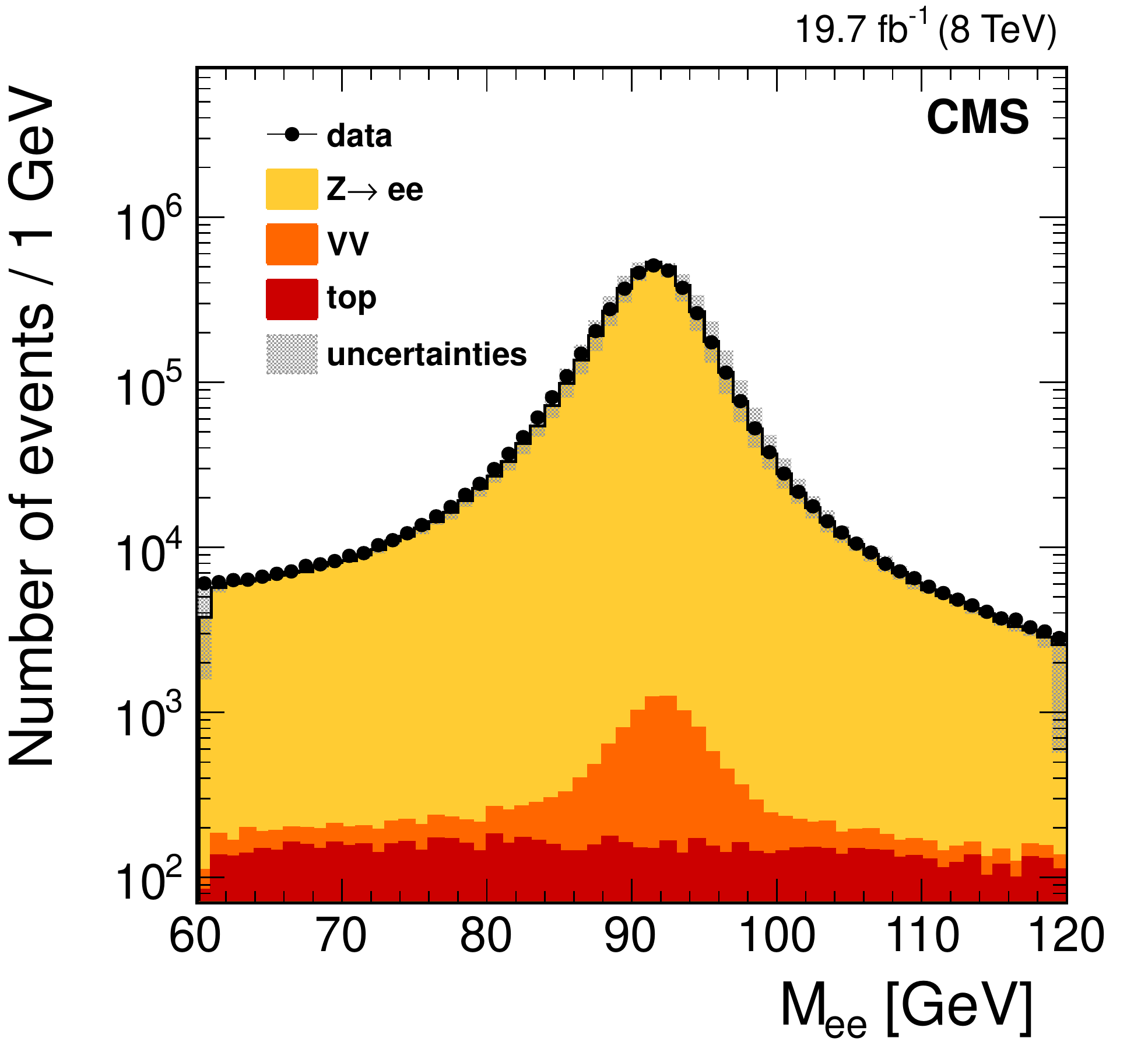}
  \caption{Dilepton invariant mass distributions in events passing the
    $\Zmm$ (left) and $\Zee$ (right) selections.  The VV contribution
    corresponds to processes with two electroweak bosons produced in the
    final state. The top contribution corresponds to the top pair and
    single top production processes. The grey error band displays the
    systematic uncertainty of the simulation, due to the muon (left), or electron (right) energy scale.
    As the invariant mass selection is performed before the computation of the systematic uncertainty 
    on the energy scale, a large event migration is observed for \Zee events. }
  \label{fig:ZbosonMass}
\end{figure}

\begin{figure}[htb]
  \centering
  \includegraphics[width=0.42\textwidth]{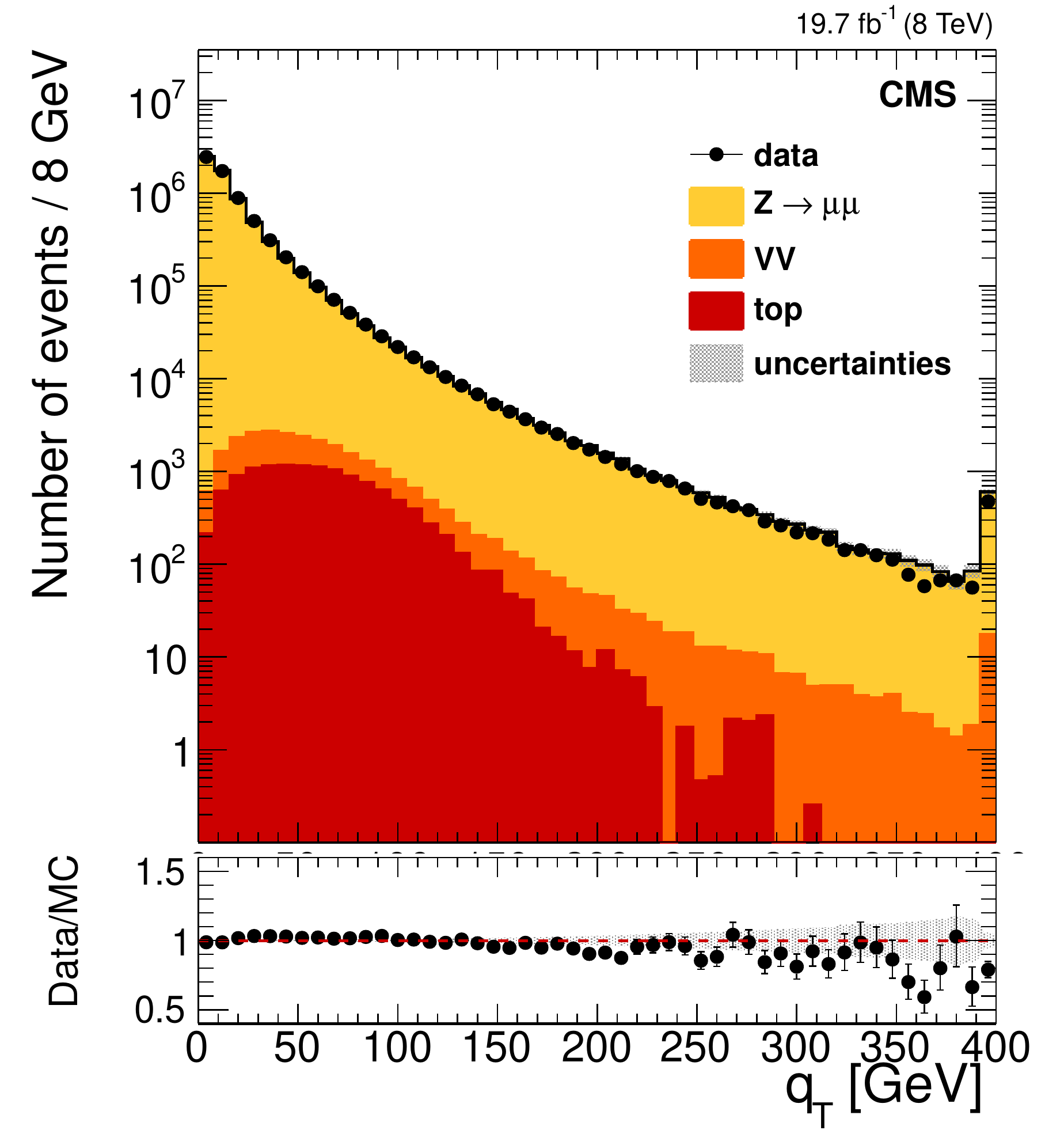}
  \includegraphics[width=0.42\textwidth]{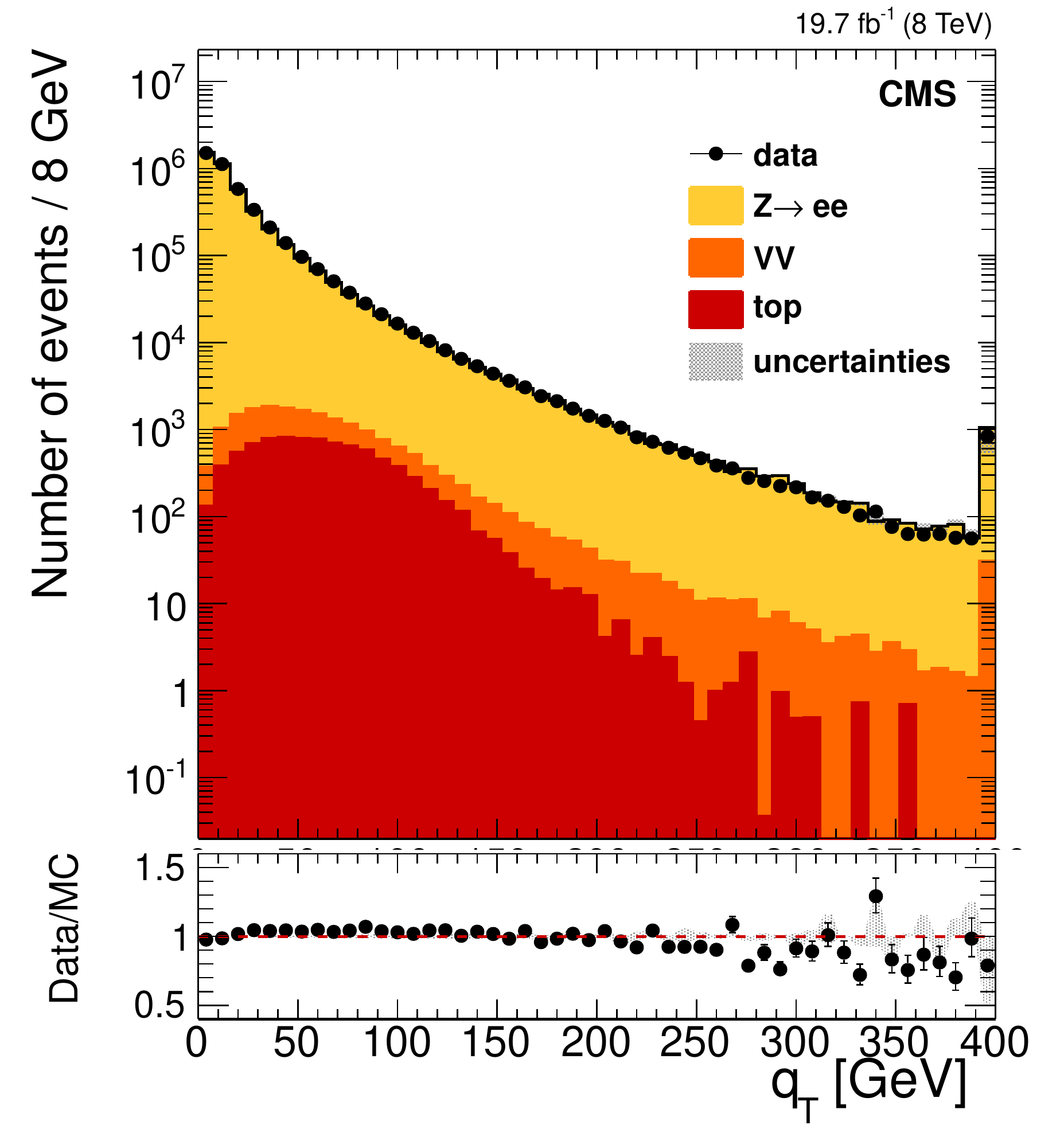}\\
  \includegraphics[width=0.42\textwidth]{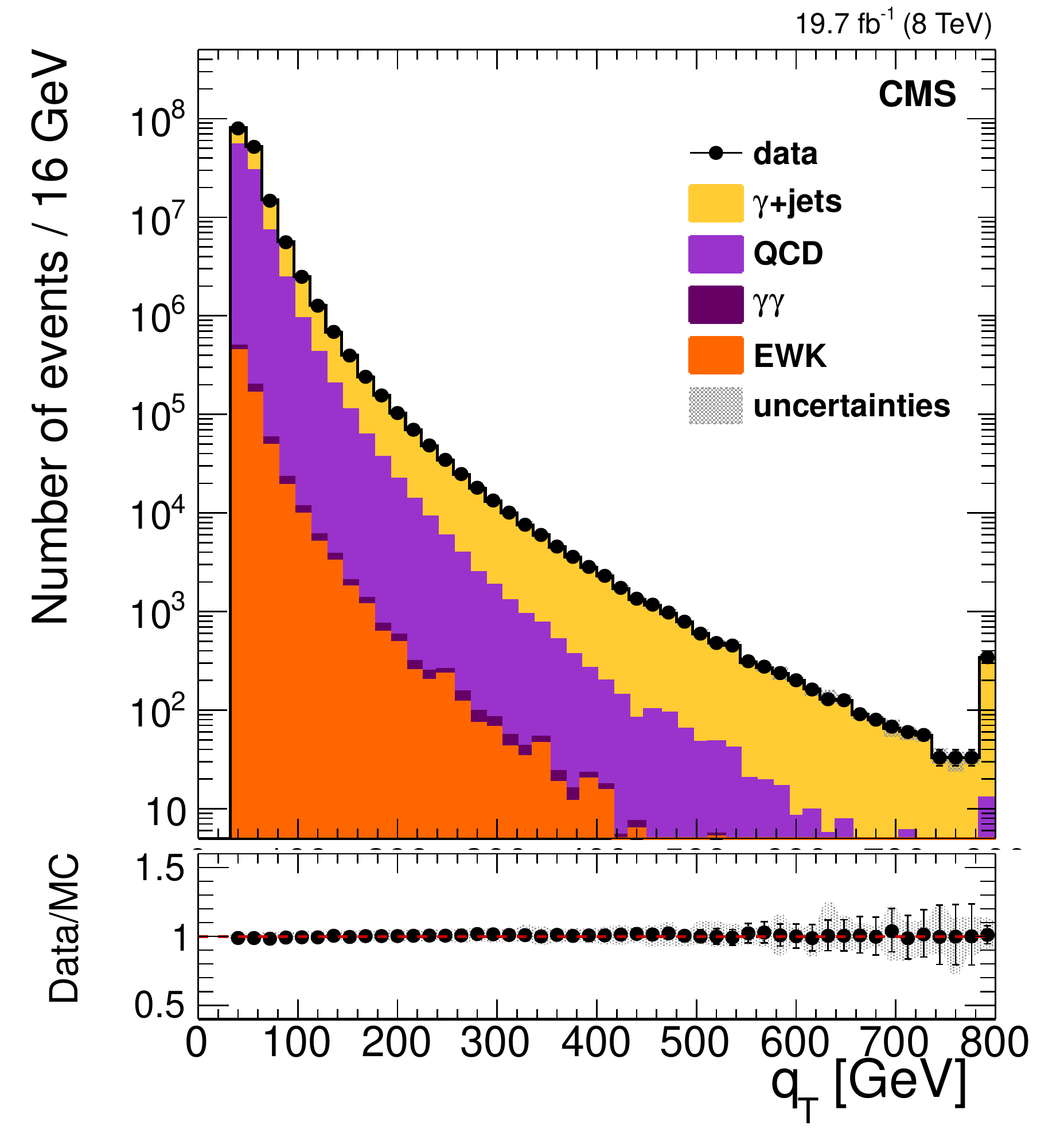}
  \caption{Distributions of $\Z/\gamma$ transverse momentum \qt in $\Zmm$ (left), $\Zee$ (right), and direct-photon events (bottom).
    The points in the lower panel of each plot show the data/MC ratio, including the
    statistical uncertainties of both data and simulation;
    the grey error band displays the systematic uncertainty of the simulation. The last bin contains overflow content.
    The VV contribution corresponds to processes with two electroweak
    bosons produced in the final state. The top contribution corresponds
    to the top pair and single top production processes. The EWK contribution corresponds to the $\Z\gamma$ and $\W\gamma$
    production processes as well as $\W\rightarrow\text{e}\nu$ events.} 
  \label{fig:ZbosonQT}
\end{figure}

Events with an invariant mass of the dimuon or dielectron system outside
of the \cPZ-boson mass window $60\GeV < {M}_{\ell\ell} < 120\GeV$ are
rejected. The \ttbar and single-top (top) processes as well as dibosons
(VV) processes are the dominant backgrounds in both the \Zee and \Zmm
samples. The spectra for the invariant mass and transverse momentum, $\vqt$, of magnitude \qt, of the $\Z\to\ell^+\ell^-$
candidate are presented in Figs.~\ref{fig:ZbosonMass} and \ref{fig:ZbosonQT}, respectively.
The data distributions are well modeled by the simulation.

\subsection{\texorpdfstring{$\Wen$}{W e nu} and \texorpdfstring{\ttbar}{ttbar} event selection}
\label{subsec:Wsample}

The \Wen\ and semi-leptonic \ttbar\ events are used in the \met significance studies presented in Section~\ref{sec:significance}.
The \Wen\ candidate events are collected with a single-electron trigger that requires the presence of an electron object
with $\pt>27\GeV$. Offline, we require the presence of an electron candidate passing the medium working point of a set of quality
requirements and also satisfying $\pt>30\GeV$ and $\abs{\eta}<2.5$. This working point is identical to the one used for the selection 
of \Zee events. We reject events with two or more
electrons if at least one of the additional electrons satisfies $\pt>20\GeV$, $\abs{\eta} < 2.5$, and passes the loose working
point of a set of quality requirements (the same set just mentioned above). The medium and loose working points for
the electron quality requirements have been defined so that they select electrons
with an efficiency of 80\% and 95\%, respectively~\cite{DP-2013-003}.

In the semi-leptonic \ttbar\ channel, we select single-muon and single-electron events. Each event is required to
pass either an e+jet or a $\mu$+jet trigger.  Offline we require at least 2 b-tagged
jets with $\pt > 45\GeV$, at least 3 jets with $\pt > 45\GeV$, and at
least 4 jets with $\pt > 20\GeV$.  Jet energies are fully corrected and required to
satisfy the jet identification criteria ~\cite{SUS-12-011}  described in Section~\ref{sec:tails}. For b-tagging, we
use the combined secondary vertex tagger with the tight working
point \cite{BTV-12-001}.  Exactly one identified and isolated lepton is required.

\subsection{The direct-photon event selection}
\label{subsec:photonsample}

A direct-photon sample corresponding to final states containing at least one photon and
at least one jet
is used for the measurements of \met scale and
resolution presented in Sections~\ref{sec:resolution} and
\ref{sec:nopumet}. Photon events were collected with a set of triggers based on the
measured \pt of the hardest reconstructed photon candidate in the event. The \pt
thresholds of the triggers were 30, 50, 75, 90, 135, and 150\GeV. The rates
of the first five triggers were randomly reduced (prescaled) because
of the limited data acquisition bandwidth. The approximate effective
values of the prescaling factors were 5000, 900, 150, 71, and 1.33
respectively. Events are selected offline by requiring the
highest \pt reconstructed photon candidate to pass the selection criteria
described below.

Photon candidates are selected from clusters of energy in the ECAL
within the pseudorapidity coverage $\abs{\eta}<1.44$. Various identification
criteria, such as the consistency between the cluster width and a typical
photon electromagnetic shower shape, are applied in order to correctly identify
photons with high efficiency and to suppress the misidentification of
electrons, jets, or spurious ECAL signals as photons \cite{EGM-10-005,EGM-11-001}.
An isolation requirement ensures that hadronic jets misidentified as photons are rejected efficiently: activity from charged hadrons, neutral hadrons, and other photons in the event is determined
by calculating the scalar sum of their transverse momenta in a cone of $\DeltaR < 0.3$ around the photon trajectory. Separate requirements on these
isolation sums suppress photon candidates inside jets and jets misidentified as photons:
$\sum \pt < 2.6\GeV$,  $\sum \pt < 3.5+0.04 \qt\GeV$ and $\sum \pt < 1.3 + 0.005 \qt\GeV$ for charged hadrons, neutral hadrons and photons, respectively.
Finally, to prevent the misidentification of electrons as photons, the photon
candidate must not match any track with hits in the pixel detector that is associated with the primary vertex
and reconstructed in the pixel detector. Events satisfying these criteria
form our signal sample.

The background processes that are considered for the direct photon sample are QCD multijet events,
diphoton production, production of single $\W$ bosons, and single photons
produced in association with the $\W$ or $\Z$ boson, referred as the electroweak (EWK) contribution. Although the majority
of QCD multijet events fail the photon selection, they
constitute a dominant background due to the large production
cross section and occasional misidentification of jets with large
electromagnetic fraction as photons. Jets that pass the photon
selection are typically enriched in $\pi^0\to\gamma\gamma$ and contain
little hadronic activity; therefore, the detector response to these jets
is similar to that of single photons. To have a robust description of the
QCD background, its expected contribution is estimated from data.

We utilize the following procedure to estimate the expected contribution
of QCD multijet background processes for a given kinematic variable. We begin
with a sample of data events where the highest \pt photon
candidate failed the charged-hadron isolation requirement but passed all
other requirements; we denote this sample of events as the charged hadron
isolation sideband.
For each kinematic variable studied, we take the distribution of this variable
from data in the charged hadron isolation sideband and remove non-QCD background
processes by subtracting their simulated distributions.
The remaining distribution forms our initial estimate for the shape
of the kinematic variable's distribution in the QCD background in the signal sample.
We set the normalization of this expected QCD multijet background by
scaling the number of events in data from the charged hadron isolation sideband
to match the number of events in data from the main signal sample, after subtracting
the respective expected contributions of other backgrounds.

In order to account for the differences in detector response to
photon candidates between the signal sample and the charged hadron isolation sideband,
we correct these distributions with information from simulated QCD multijet events.
The magnitude of these corrections depends upon the algorithm used for \met
reconstruction; for \pfvecmet (defined in Section~\ref{sec:metreco}),
the magnitude of the correction falls within 6--8\%. For No-PU \pfvecmet and MVA \pfvecmet
(both defined in Section~\ref{sec:nopumet}), the magnitudes of the corrections fall
within 2--4\%.

Figure~\ref{fig:ZbosonQT} shows a comparison between the photon transverse momentum \qt distribution in
data and the expected signal and background contributions. Note that the signal and background
contributions for the prediction have been reweighted in \qt to match the distribution observed in data.

\section{Reconstruction of \texorpdfstring{\bigmet}{MET}}
\label{sec:metreco}

We define $\vecmet\equiv-\sum\vpt$, where the sum is over all
observed final-state particles; by momentum
conservation, \vecmet\ is also equal to the total
transverse momentum of all unobserved particles, such as neutrinos
or other weakly interacting objects.
CMS has developed several distinct and complementary algorithms to reconstruct
\vecmet, already presented in Ref.~\cite{METJINST}.  The \vecmet\
reconstructed using a particle-flow technique (PF~\vecmet) is used in the majority of
CMS analyses.
It is defined as the negative vectorial sum over the transverse momenta of all PF
particles. The PF~\sumet is the associated scalar sum of
the transverse momenta of the PF particles.
The less commonly used Calo~\vecmet is calculated
using the energies contained in calorimeter towers and their directions
relative to the centre of the detector.  The sum excludes energy deposits
below noise thresholds but is corrected for the calorimeter deposits of muons, when they are present,
by adding their momentum to the sum~\cite{METPas07}.

In the following sections, we present the
performance of PF~\vecmet and Calo~\vecmet, giving primary attention
to \pfvecmet.  In addition, two advanced \vecmet\
reconstruction algorithms specifically developed to mitigate effects
from large numbers of pileup interactions are discussed in
Section~\ref{sec:nopumet}.

The magnitude of the \vecmet can be underestimated or overestimated for a variety of
reasons, including minimum energy thresholds in the calorimeters, \pt
thresholds and inefficiencies in the tracker, and the nonlinearity of the response of the
calorimeter for hadronic particles due to its non-compensating nature.
This bias is significantly reduced by correcting the \pt of jets to
the particle-level \pt using jet energy
corrections~\cite{JETJINST}:
\begin{equation}
\vecmet^\text{corr}
=\vecmet - \vec{\Delta}_\text{jets}
=\vecmet - \sum_\text{jets} (\vec{p}_{\mathrm{T},\text{jet}}^\text{corr}-\vec{p}_{\mathrm{T},\text{jet}}),
\label{eq:Type1MET}
\end{equation}
where the superscript ``corr'' refers to the corrected values.
The sum extends over all jets with an electromagnetic energy fraction below 0.9
and a corrected $\pt>10\GeV\,(20\GeV)$ for PF~\vecmet~(Calo~\vecmet).

Further corrections improve
the performance of the \vecmet\ reconstruction in events
with
large numbers
of pileup interactions.
The contribution to the genuine \vecmet\ from such interactions is close
to zero, as the probability to produce neutrinos is small in inelastic pp scattering (minimum bias) interactions.
The vectorial \vecpt sum of charged particles
is therefore expected to be well balanced by that of neutral particles.
However, the nonlinearity and
minimum energy thresholds in the calorimeters cause
\vecmet to point on average in the direction
of the vectorial \vecpt sum of neutral particles.

We correct for this effect by using the
vectorial \vecpt sum of charged particles associated with pileup vertices
as an estimator of the induced \vecmet.
The correction is parametrized by
$f(\vec{v}) = c_1  (1.0+\erf(-c_2 {\abs{\vec{v}}^{c_3}}))$
where $\vec{v}= \sum_{\text{charged}}{\vecpt}$
is the vectorial \vecpt sum of charged particles associated with a given pileup vertex.
The coefficients $c_1=-0.71$, $c_2=0.09$, and $c_3=0.62$ are obtained
by fitting the \vecmet component
parallel to the $\vec{v}$ direction as a function of $\abs{\vec{v}}$
in simulated minimum bias events with exactly one generated pp interaction.
When this correction is applied to the data and simulation
samples with pileup interactions,
the factor $f(\vec{v}) \vec{v}$, which gives the expected total \vecmet
for each pileup interaction, is summed over all pileup vertices
and is subtracted from the reconstructed \vecmet:
\begin{equation}
\vecmet^\text{corr}
=\vecmet - \vec{\Delta}_\text{PU}
=\vecmet - \sum_\text{PU} f(\vec{v})\vec{v}.
\label{eq:type0}
\end{equation}

Although particles are on average produced uniformly in $\phi$,
some $\phi$ asymmetry is observed in the \vecpt sums
of calorimeter energy deposits, tracks, and particles
reconstructed by the particle-flow algorithm,
leading to a $\phi$ asymmetry in \vecmet. The $\phi$ asymmetry is present not only in the data but also in
simulated events.
The sources of the asymmetry have been identified as
imperfect detector alignment, inefficiencies,
a residual $\phi$ dependence of the calibration, and a
shift between the centre of the detector and the beamline~\cite{CMS-PAS-TRK-10-003}.

The observed \vecmet\ $\phi$ asymmetry is due to a shift in the \vecmet components
 along the x and y detector axes (denoted by \mex\ and \mey\ respectively),
which increases approximately linearly
with the number of reconstructed vertices.
This correlation is utilized for a correction procedure.
The $\phi$-asymmetry corrections are determined separately
for data and simulated events.
Linear functions are fitted to the correlation of \mex\ and \mey\
to $N_\text{vtx}$, the number of reconstructed vertices:
\begin{equation}\begin{split}
\left<\mex\right> &=  c_{x_o} + c_{x_s}  N_\text{vtx},\\
\left<\mey\right> &=  c_{y_o} + c_{y_s}  N_\text{vtx}.
\label{eq:metPhiAsymmetryFitModel}
\end{split}\end{equation}

The linear dependence of $\langle\mex\rangle$ and $\langle\mey\rangle$
on $N_\text{vtx}$ is used to correct \vecmet\ on an event-by-event basis as:
\begin{equation}\begin{split}
\mex{}^\text{corr} &=  \mex - \left<\mex\right>= \mex -(c_{x_0} + c_{x_s}  N_\text{vtx}),  \\
\mey{}^\text{corr} &=  \mey - \left<\mey\right> = \mey -(c_{y_0} + c_{y_s}  N_\text{vtx}).
\end{split}\end{equation}

The coefficients $c_{x_0}$, $c_{x_s}$, $c_{y_0}$, and $c_{y_s}$ are determined
separately from \Zmm\ candidate events in data and simulation samples.
These coefficients for the PF~\vecmet\ are shown in Table~\ref{tab:METshifts}.

\begin{table}[bthp]
\centering
\topcaption{\label{tab:METshifts} The parameters for the \vecmet\
  $\phi$-asymmetry corrections for PF~\vecmet for data and
  simulation. As the detector alignment and $\phi$-intercalibrations are
  different between data and simulation, the values of the respective parameters
  are expected to be different.}

\begin{tabular}{lrrrr}

& \multicolumn{1}{c}{$c_{x_0}$ (\GeVns{})} & \multicolumn{1}{c}{$c_{x_s}$ (\GeVns{})} & \multicolumn{1}{c}{$c_{y_0}$ (\GeVns{})} & \multicolumn{1}{c}{$c_{y_s}$ (\GeVns{})} \\ \hline
Data       &  $0.05$ & $0.25$ & $-0.15$ & $-0.08$ \\
Simulation &  $0.16$ & $-0.24$ & $0.36$ & $-0.13$ \\
\end{tabular}

\end{table}

In this paper, the correction $\vec{\Delta}_\text{jets}$
defined in Eq.~\eqref{eq:Type1MET}
is applied to both PF and Calo~\vecmet, while
the pileup correction
$\vec{\Delta}_\text{PU}$
defined in Eq.~\eqref{eq:type0}
is applied only to PF~\vecmet, as the information from tracking needed
for determination of $\vec{\Delta}_\text{PU}$ is not used in the Calo~\vecmet calculation.
All the \met distributions are further corrected for the $\phi$ asymmetry.
In simulated events,
jet momenta are smeared in order to account for the jet resolution differences
between data and simulation~\cite{JETJINST}, and the \vecmet\ is recomputed based
on the smeared jet momenta.

\section{Large \texorpdfstring{\bigmet}{MET} due to misreconstruction}
\label{sec:tails}

Spurious detector signals can cause fake \vecmet signatures
that must be identified and suppressed.
In Ref.~\cite{METJINST} we showed the results of studies of anomalous high-\vecmet
events in the data collected during 2010 LHC running,
associated with particles striking sensors in the ECAL barrel detector, as well as those caused by beam-halo
particles and dead cells in the ECAL. Studies of anomalous \vecmet\ events
caused by (1) HCAL hybrid photodiode and readout box
electronics noise and
(2) direct particle interactions with the light guides and photomultiplier tubes of the
forward calorimeter are discussed in Ref.~\cite{Apresyan:1479437}.

In the 2012 data, we have identified several new types of anomalous
events populating the high \vecmet tail. There are a few channels in the
ECAL endcaps that occasionally produce high-amplitude anomalous pulses.
The affected events are identified by the total energy and the number of
low-quality hits within the same super-cluster, and are removed.  A misfire of the HCAL laser
calibration system in the HCAL barrel (HB), endcap (HE), or forward (HF)
regions can produce false signals in almost all channels in a
subdetector.  If this misfire overlaps with a bunch crossing resulting
in a trigger, the event can be contaminated, inducing a large, fake
\vecmet.  The affected events are identified by the hit occupancies in
the channels used for signal and calibration readout and are removed
from the sample.

Another source of fake \vecmet\ comes from the track reconstruction.
The silicon strip tracker can be affected by coherent noise, which can generate
${\sim}10^4$ clusters widely distributed in the silicon detectors. A significant fraction of
these events are vetoed at early stages of the online trigger selection;
however, the veto is not fully efficient and some of these events are read out and reconstructed.
In such events the transverse momentum of misreconstructed spurious tracks can exceed 100\GeV.
These tracks can mimic charged particles, which are then clustered into jets with high \pt
creating large spurious \met.
The affected events can be identified by the number of clusters in the silicon strip
and pixel detectors.

Although the rejection of anomalous high-\met events due to noise in HB and HE was studied in Ref.~\cite{METJINST},
further developments have proven necessary to cope with the evolving LHC running conditions,
including high luminosities and the shortening of the bunch crossing interval
from 100\unit{ns} to 50\unit{ns}.
A noise-rejection algorithm was developed to exploit the differences between noise and signal pulse
shapes.
The CMS hadron calorimeter signals are digitized in time intervals of 25 ns,
and signals in neighboring time intervals are used to define the pulse shape;
measured and expected signal pulse shapes are compared and several
compatibility tests to a signal hypothesis are performed. The energy reconstructed in channels having anomalous signals is removed during event processing, so that the affected channels do not contribute to the reconstructed physics objects.

\begin{figure}[tb!]
  \centering
    \includegraphics[width=0.6\textwidth]{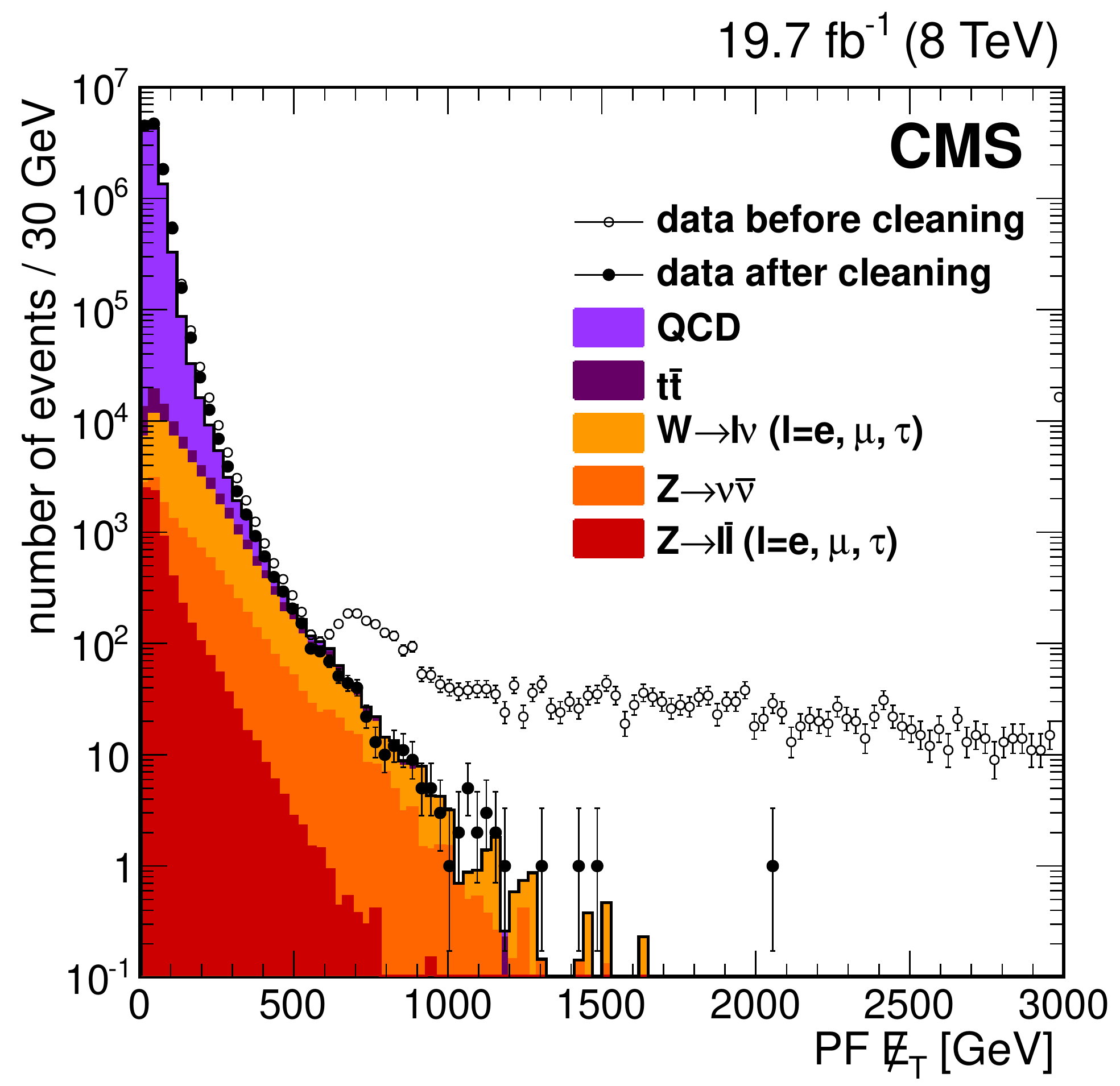}
   \caption{The \pfvecmet distributions for events passing the dijet selection
  without cleaning algorithms applied (open markers),
  with cleaning algorithms applied including the one based on jet identification requirements
  (filled markers), and simulated events (filled histograms).
  }
  \label{fig:PFMETtail}
\end{figure}

Figure~\ref{fig:PFMETtail} shows a comparison of the \pfvecmet
distribution before and after the application of the algorithms to
remove anomalous events in the dijet sample described in
Section~\ref{subsec:Dijetsample}.  The anomalous events with \pfvecmet
around 600\GeV are mainly due to misfires of the HCAL laser calibration
system, and the anomalous events with \pfvecmet above 1.5\TeV are mainly
caused by the electronics noise in HB and HE.  Even after applying all
the anomaly-removal algorithms developed for the 2012 data, we still
find a small residue of anomalous \vecmet events in the tail of the
\pfvecmet distribution.  Imposing jet identification criteria that limit
the maximum neutral hadron energy fraction to 0.9 and the maximum photon
energy fraction to 0.95 guarantees efficient removal of such events.
These requirements are presented in Ref.~\cite{SUS-12-011} and are
frequently used in CMS data analyses.
The event is rejected if any jet fails the jet identification criteria.
The \pfvecmet distribution for events passing all cleaning algorithms and jet identification requirements
shows a substantial reduction of the high \pfvecmet tail,
and agrees well with the simulated distributions for \pfmet above 500\GeV (Fig.~\ref{fig:PFMETtail}).

\section{Missing transverse energy scale and resolution}
\label{sec:resolution}

In this section, we present studies of the performance of \vecmet reconstruction algorithms using
events where an identified Z boson or isolated photon is present.
The bulk of such events contain no genuine \vecmet, and thus a balance exists between the
well-measured vector boson momentum and the
hadronic system, which dominates the \vecmet measurement.
Using the vector boson momentum as a reference, we are able to
measure the scale and resolution of \met in an event sample with a
hadronic system that is kinematically similar to
standard model processes such as \ttbar+jets and \PW+jets,
which are typically important backgrounds in searches where \vecmet
is an essential signature.

Even if no genuine \vecmet is expected in physical processes, many
physics and detector effects can significantly affect the \vecmet
measurement, inducing nonzero \vecmet in these events. The detector noise,
particle misreconstruction, detector energy resolution, and jet energy
corrections are part of the detector sources of \vecmet, while the
pileup, underlying event activity, and fluctuations in jet composition
are physical sources of \vecmet.

\begin{figure}[t!hb]
\centering
\includegraphics[width=0.32\textwidth]{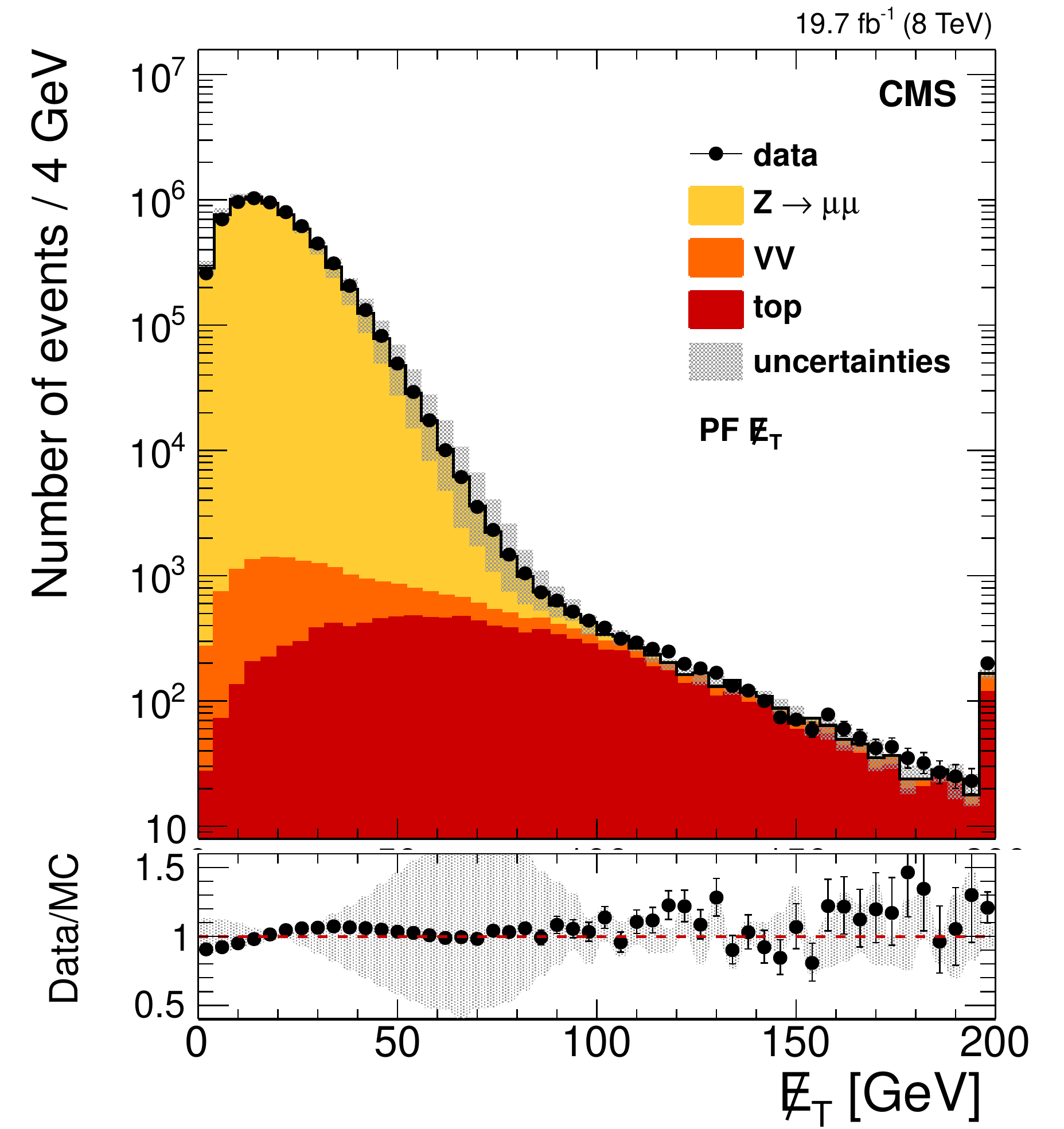}
\includegraphics[width=0.32\textwidth]{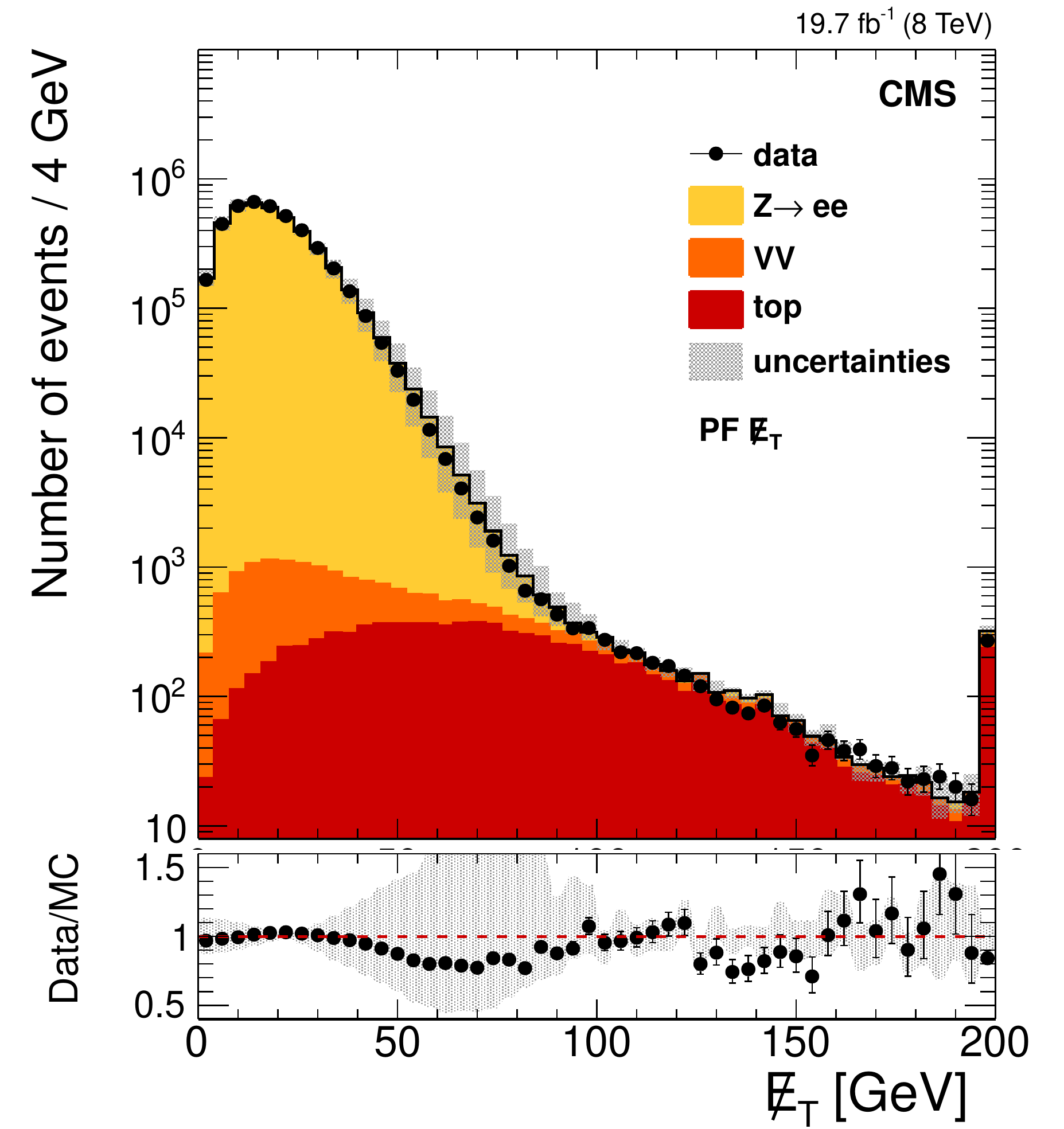}
\includegraphics[width=0.32\textwidth]{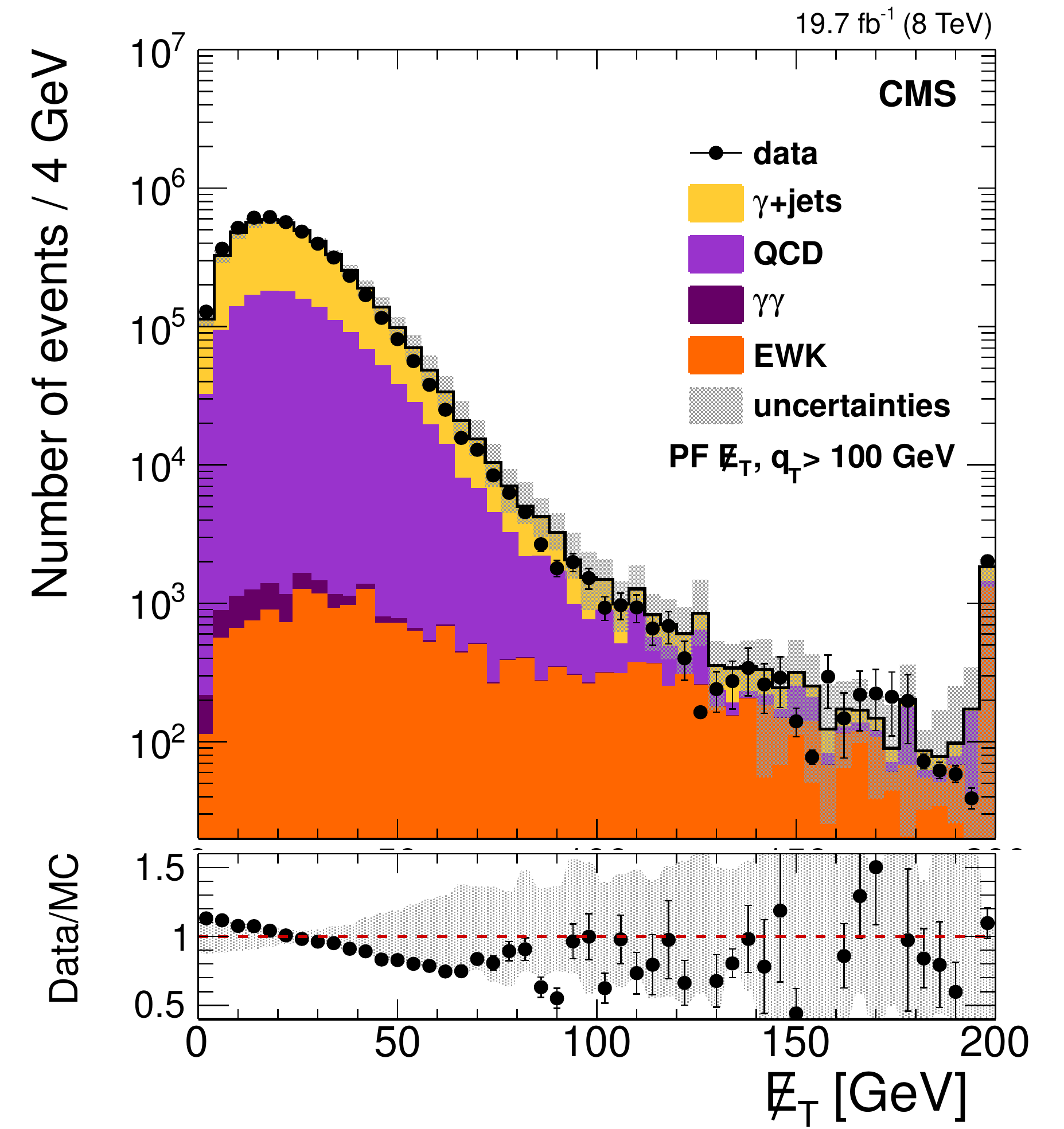}
\caption{The \pfvecmet distribution in $\Zmm$ (left), $\Zee$ (middle), and direct-photon events (right).
The points in the lower panel of each plot show the data/MC ratio, including the
statistical uncertainties of both data and simulation;
the grey error band displays the systematic uncertainty of the simulation. The last bin contains the overflow content.
}
\label{fig:ZbosonMET}
\end{figure}

The \pfvecmet distributions in $\Zmm$, $\Zee$, and direct-photon events are
presented in Fig.~\ref{fig:ZbosonMET}. Note that for the direct-photon distribution we require
$\qt > 100$\GeV in order to avoid biases from the prescales of the lower \pt photon triggers.
 Good agreement between data and simulation is observed in all distributions.
Momenta of leptons from \Z{}-boson decays (direct photons) are
reconstructed with resolutions of $\sigma_{\pt}/\pt \sim 1$--4
(1--3)\%~\cite{MUO-10-004,EGM-10-005}, while jet energies are
reconstructed with resolutions of
$\sigma_E/E\sim10$--15\%~\cite{JME-10-014}. Thus the \vecmet resolution
in \cPZ{} or \GJ events is dominated by the resolution with which the
hadronic activity in the event is reconstructed.

Uncertainty bands for the distributions of $\Zmm$, $\Zee$, and direct-photon
events include uncertainties in the lepton and photon energy scales
(0.2\% for muons, 0.6\% for barrel electrons and photons, and 1.5\%
for endcap electrons), jet energy scale (2--10\%), jet energy
resolution (6--15\%), and the energy scale of low-energy particles,
defined as the unclustered energy (arbitrary 10\%, covering for all differences observed between the data and the simulation).
 In addition, for the direct-photon
events only, we account for the systematic uncertainty in the \met response
correction applied to events used to estimate the QCD multijet
contribution to the direct-photon sample (2--10\%).

The increase in the uncertainty band in Fig.~\ref{fig:ZbosonMET} around 70\GeV stems from
 the large impact of jet energy resolution uncertainties in events with no genuine \pfvecmet: as this region of \pfvecmet is mostly filled with direct-photon or \cPZ{} events with at least one jet, the impact of a modification of the jet energy on the \vecmet reconstruction will be maximized in this area.
For higher values of \pfvecmet, where processes with genuine \vecmet dominate such as the \ttbar process, the relative uncertainty is much smaller.

\begin{figure}[th!b]
\centering
\includegraphics[width=0.4\textwidth]{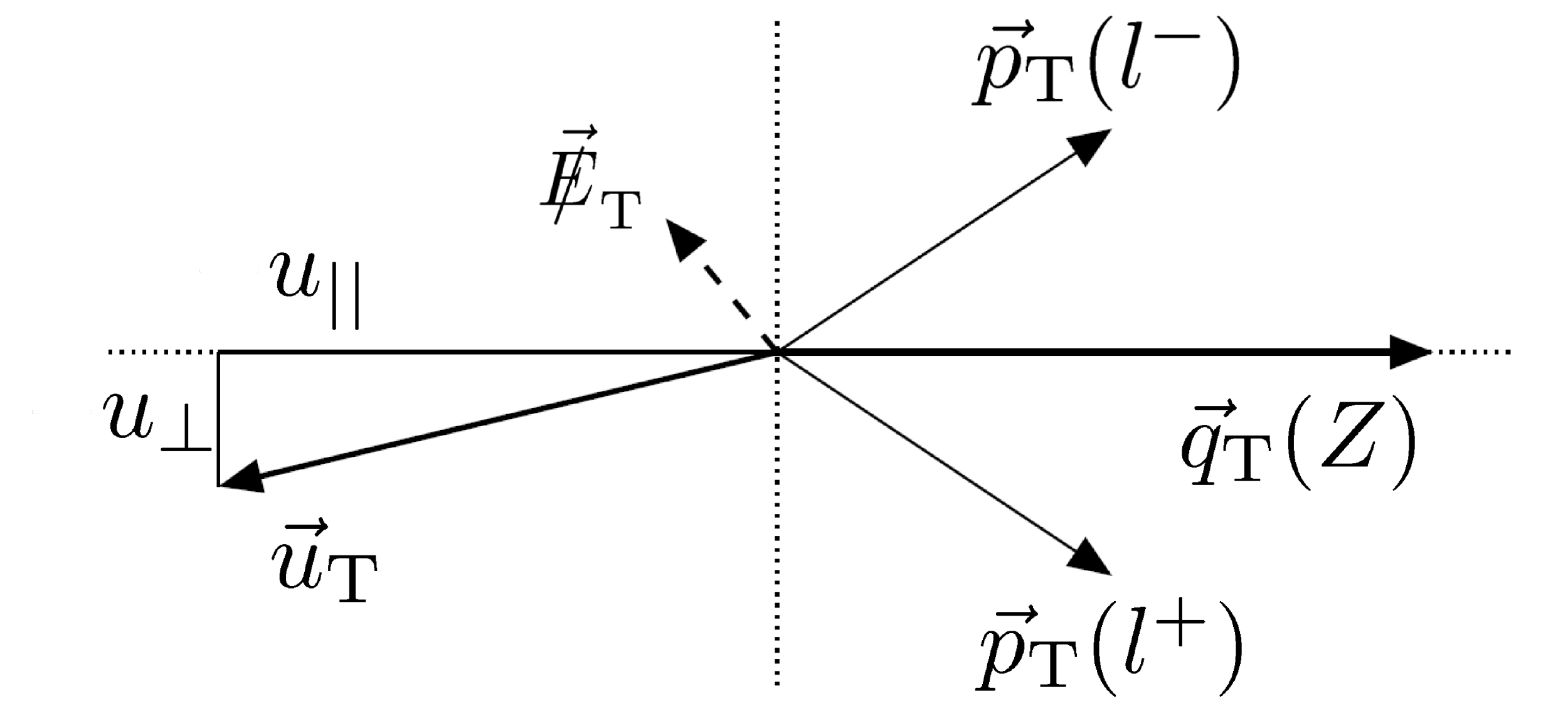}
\includegraphics[width=0.4\textwidth]{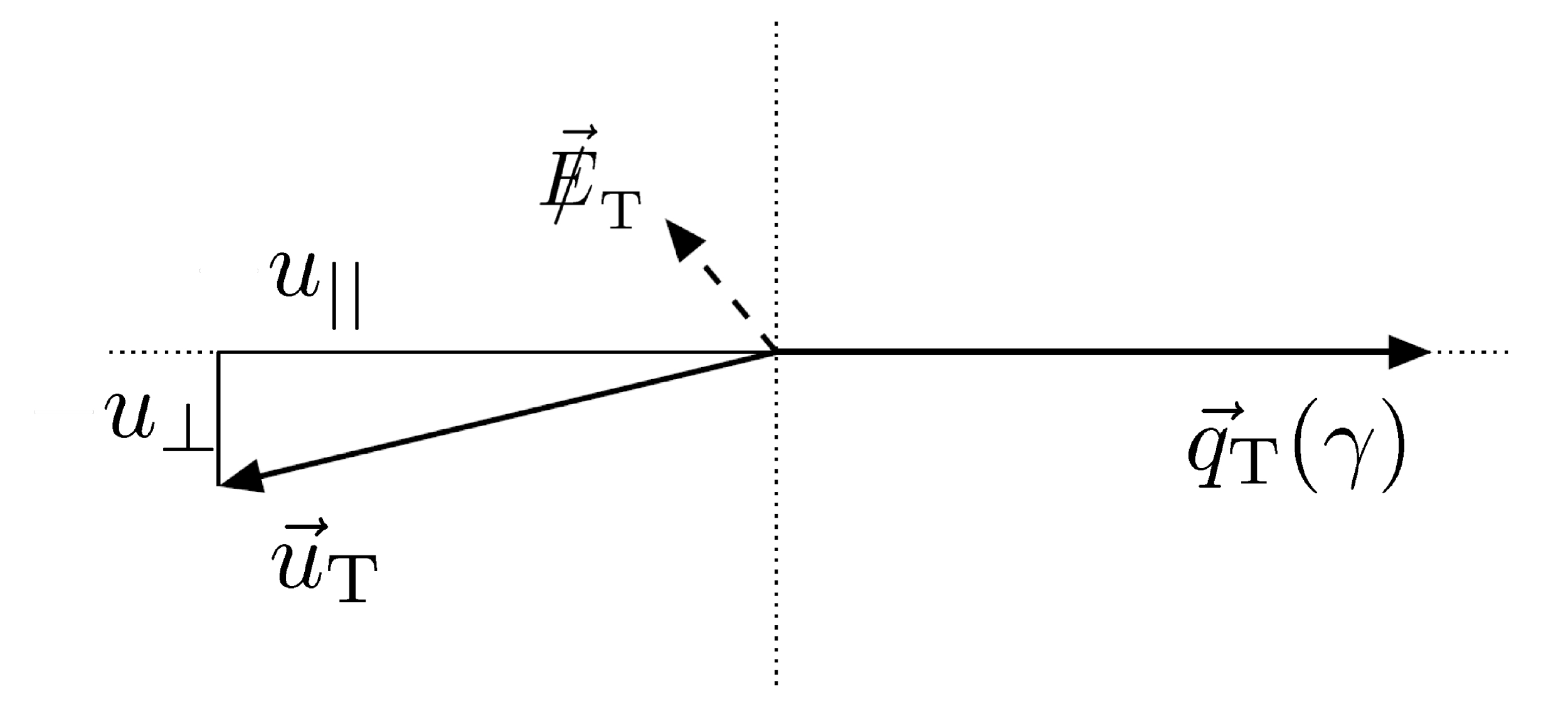}
\caption{Illustration of $Z\to\ell^+\ell^-$ (left) and direct-photon (right) event kinematics in the transverse plane.
The vector \vut denotes the vectorial sum of the transverse momentum of all particles reconstructed in the event
except for the two leptons from the \Z{} decay (left) or the photon (right).}
\label{fig:recoilDef}
\end{figure}

We denote the vector boson
momentum in the transverse plane by $\vqt$, and the hadronic recoil,
defined as the vectorial sum of the transverse momenta of all particles
except the vector boson (or its decay products, in the case of \Z{}
bosons), by \vut.  Momentum conservation in the transverse plane
requires $\vqt+\vut+\vecmet=0$. By definition, the recoil is therefore
the negative sum of the induced \vecmet and \vqt. Figure~\ref{fig:recoilDef}
summarizes these kinematic definitions.

The presence of a well-measured \Z{} boson or direct photon provides both a
momentum scale, $\qt\equiv\abs{\vqt}$, and a unique event axis, along the
unit vector \hatqt.  The hadronic recoil can be projected onto this
axis, yielding two signed components, parallel (\upar) and perpendicular
(\uperp) to the event axis. The direction of \uperp is defined by considering the
coordinate frame based on the \vqt axis. Since $\upar\equiv \vut\cdot\hatqt$, and
because the observed hadronic system is usually in the hemisphere opposite
the boson, \upar\ is typically negative.
The scalar quantity $-\langle \upar \rangle /\qt$ is referred to as the
\vecmet response, and the dependence of $-\langle \upar \rangle /\qt$
versus \qt\ as the response curve.

The \vecmet energy resolution is assessed with a parametrization of the
\redupara{} and \uperp distributions by a Voigtian function, defined by
the convolution of a Breit--Wigner distribution and a Gaussian
distribution,
as it is found to describe the observed \redupara{} and \uperp
distributions very well.  The resolutions of \upar and \uperp, denoted by
$\sigma(\upar)$ and $\sigma(\uperp)$, are given by the full width at half
maximum of the Voigtian profile, divided by $2\sqrt{2\ln2}\simeq 2.35$.

\subsection{Measurement of PF \texorpdfstring{\bigmet}{MET} scale and resolution}

\begin{figure}[htbp!]
\centering
\includegraphics[width=0.32\textwidth]{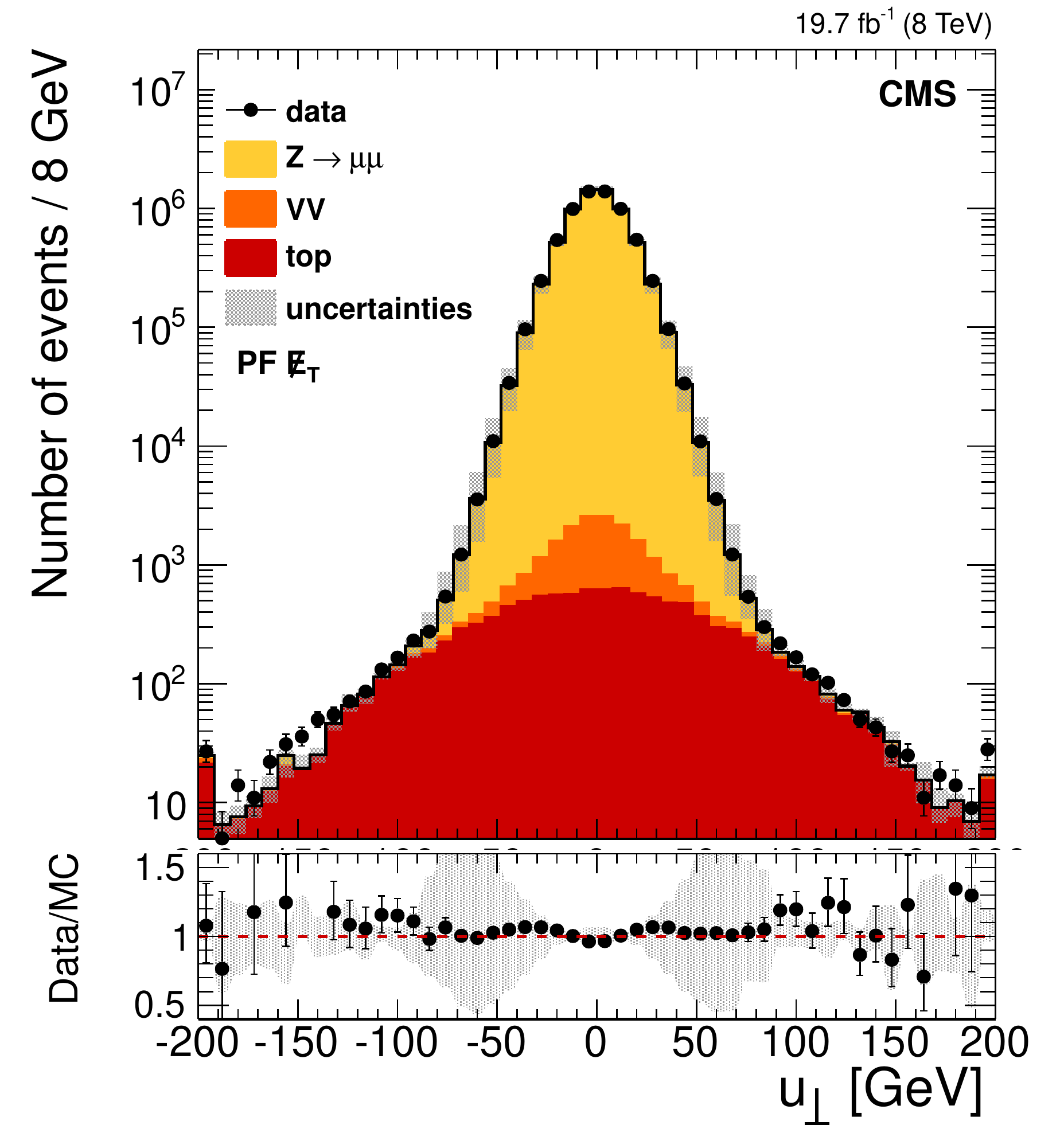}
\includegraphics[width=0.32\textwidth]{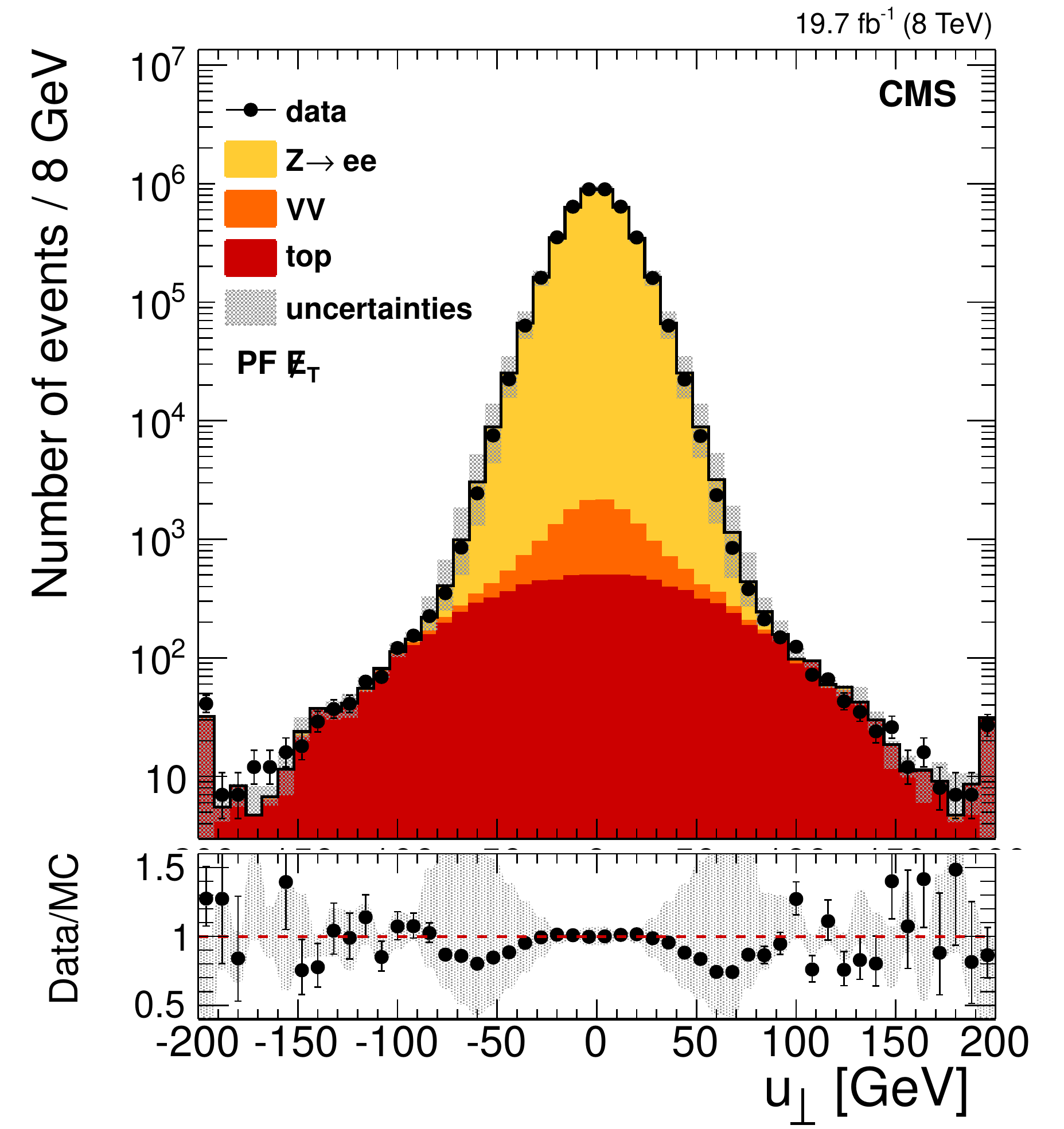}
\includegraphics[width=0.32\textwidth]{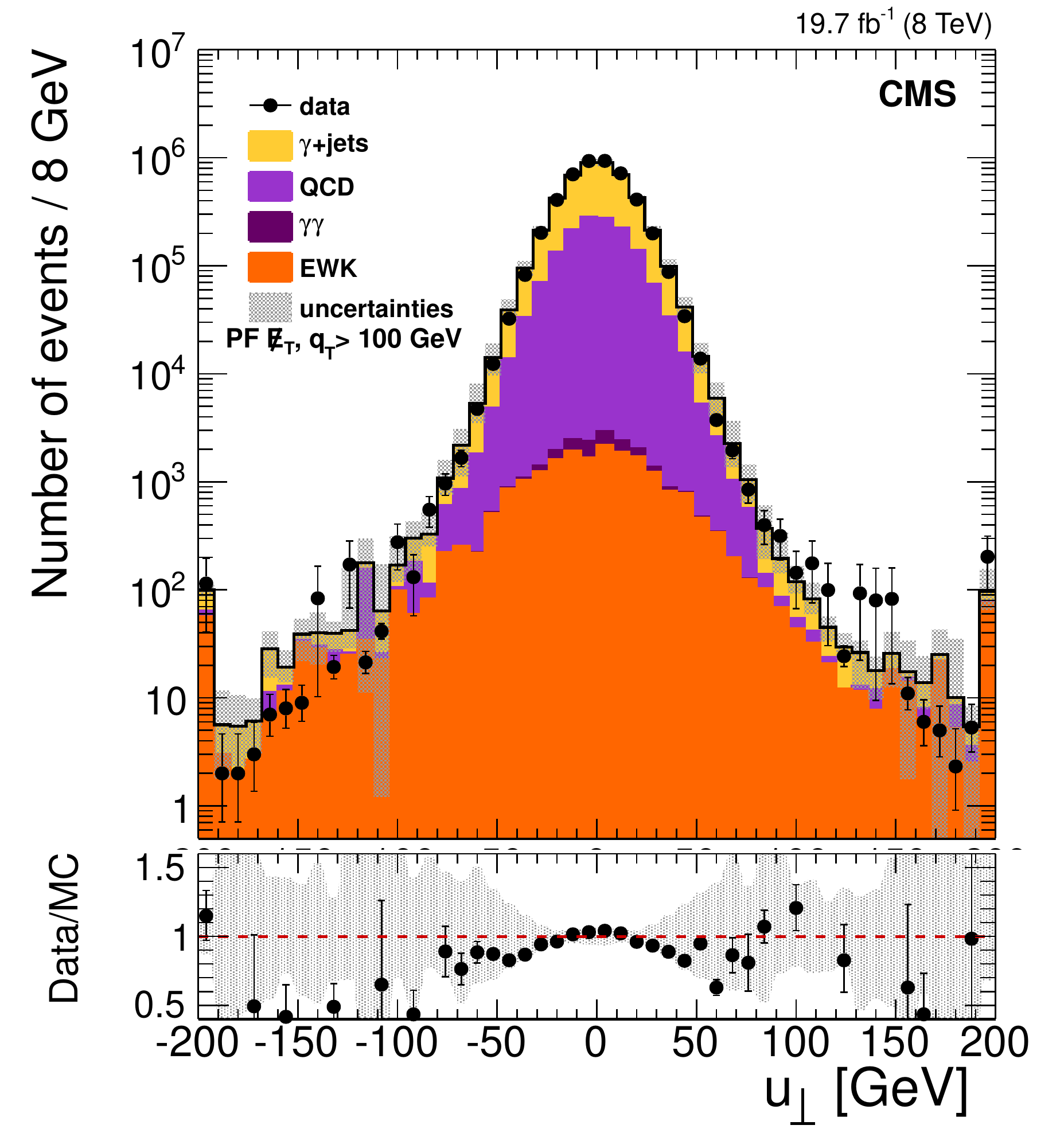} \\
\includegraphics[width=0.32\textwidth]{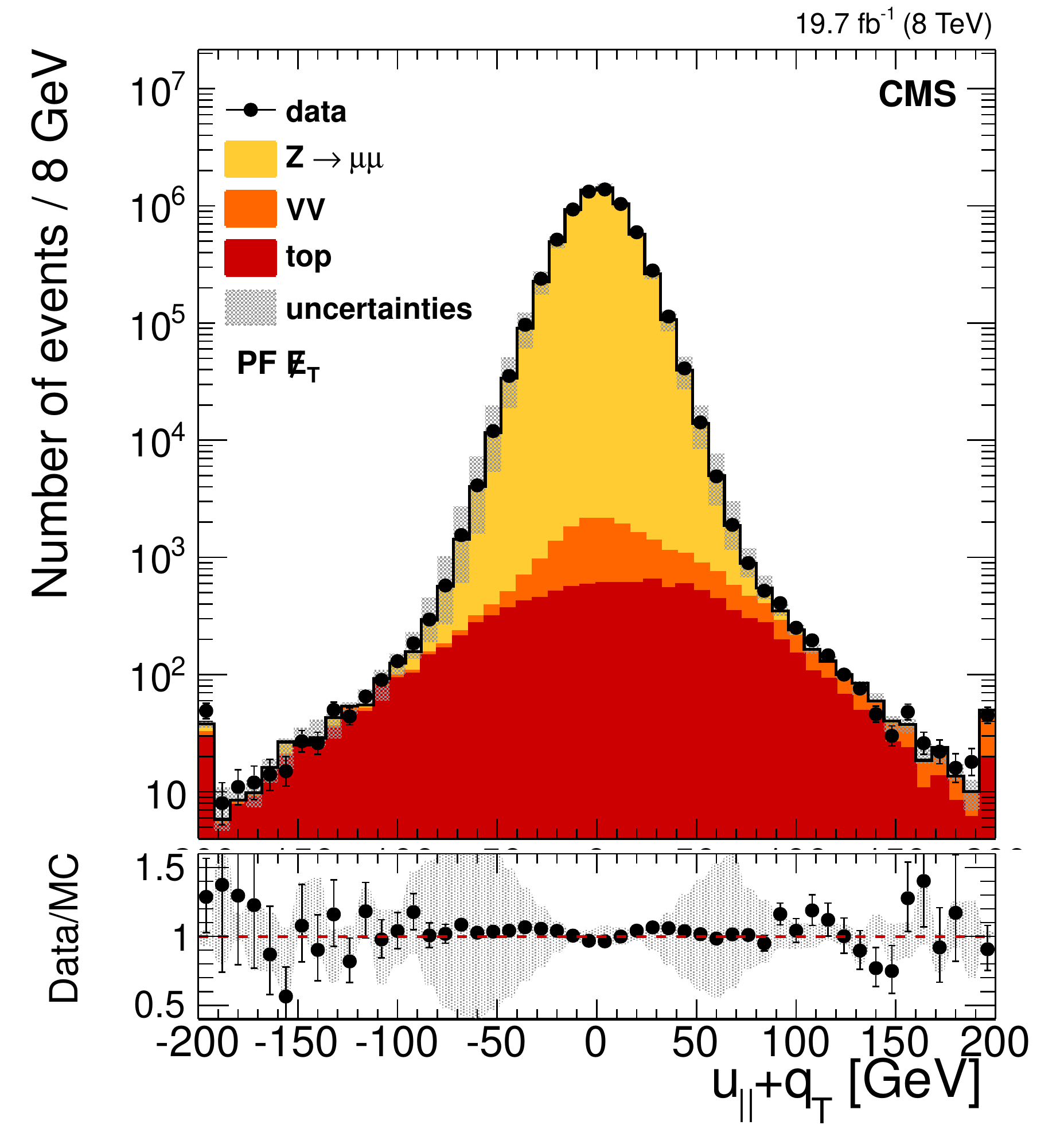}
\includegraphics[width=0.32\textwidth]{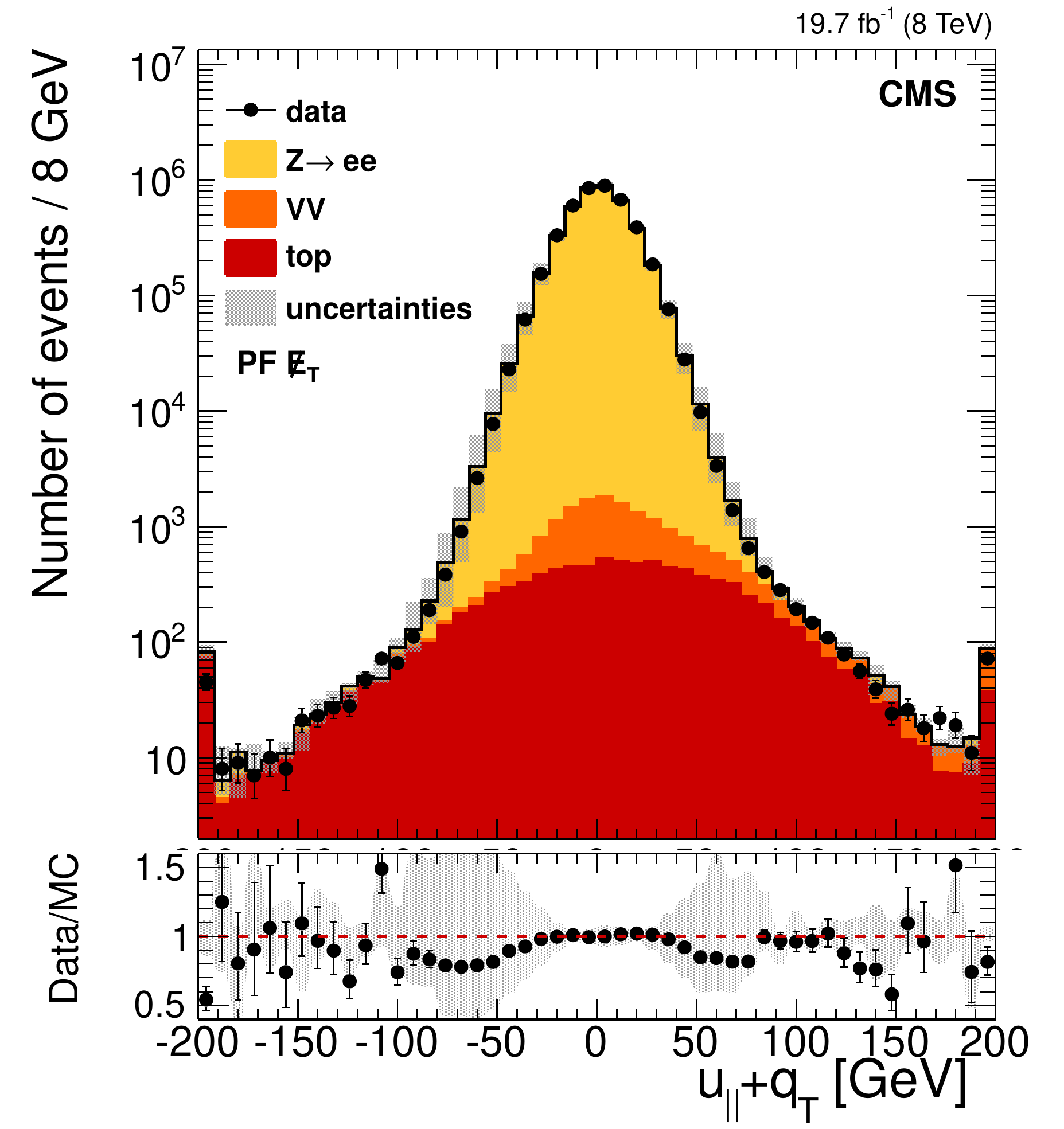}
\includegraphics[width=0.32\textwidth]{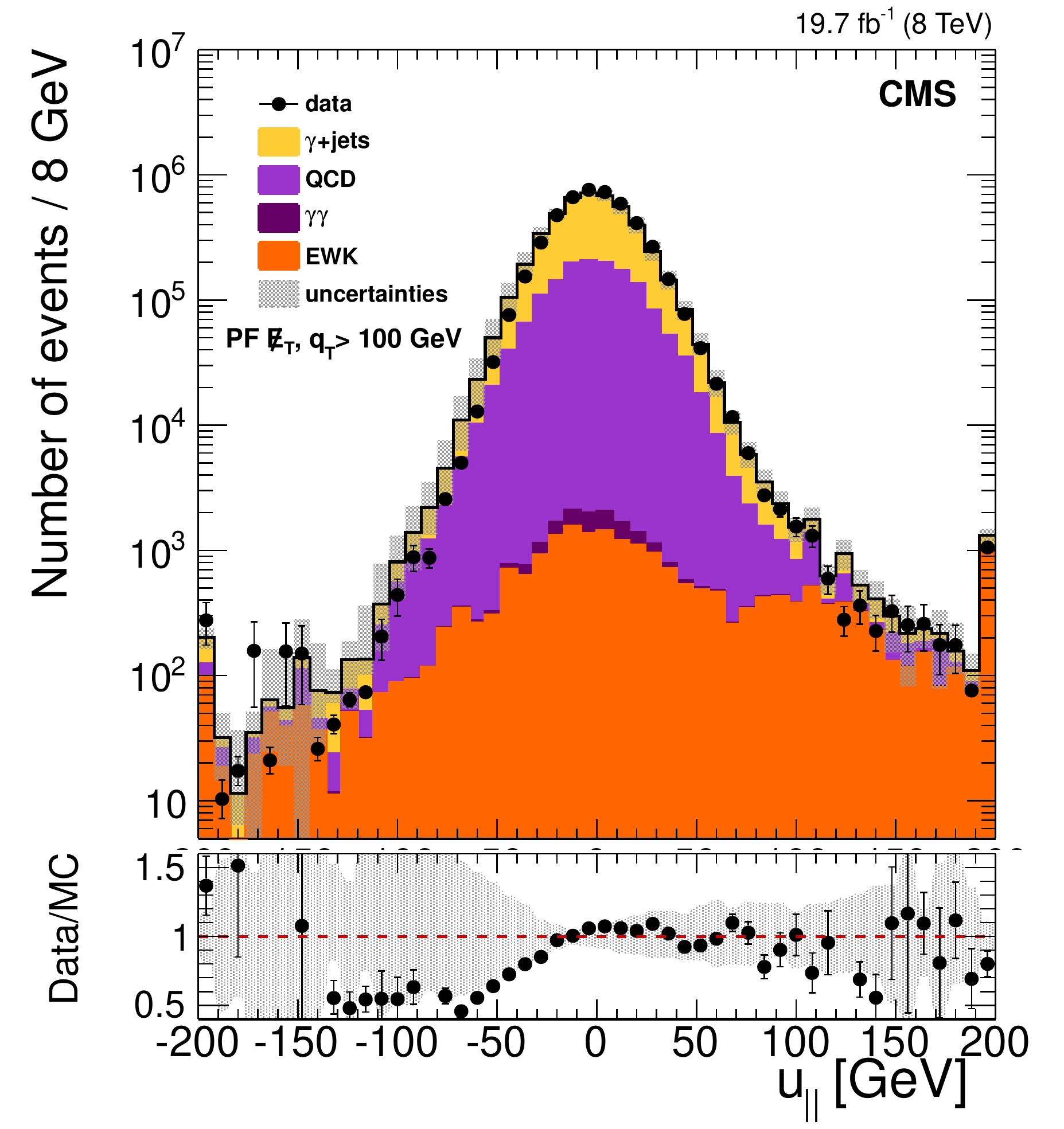}

\caption{Distributions of \uperp (top) and \redupara (bottom) for \pfvecmet for \Zmm\ (left), \Zee\ (middle), and direct-photon events (right);
The points in the lower panel of each plot show the data/MC ratio, including the
statistical uncertainties of both data and simulation;
the grey error band displays the systematic uncertainty of the simulation. The first (last) bin contains the underflow (overflow) content.
}
\label{fig:recoil_perp_gammajet}
\end{figure}

The decomposition of the hadronic recoil momentum into \uperp and \upar components
provides a natural basis in which to evaluate \pfvecmet characteristics.
Distributions of \uperp are shown in
Fig.~\ref{fig:recoil_perp_gammajet} for \mbox{\Zmm}, \mbox{\Zee},
and direct-photon events. The component \uperp is expected to be
centred at zero by construction, and to be symmetric as it
arises primarily from random detector noise and the underlying event.
Distributions of \redupara{} are also shown in
Fig.~\ref{fig:recoil_perp_gammajet}.  Again by construction,
\upara is balanced with \qt, thus making \redupara centred
around zero and approximately symmetric.
The increased uncertainty in the \redupara{} and \uperp{} distributions
around $\pm70$\GeV is due to the jet energy resolution uncertainty.

\begin{figure}[htb]
\centering
\includegraphics[width=0.75\textwidth]{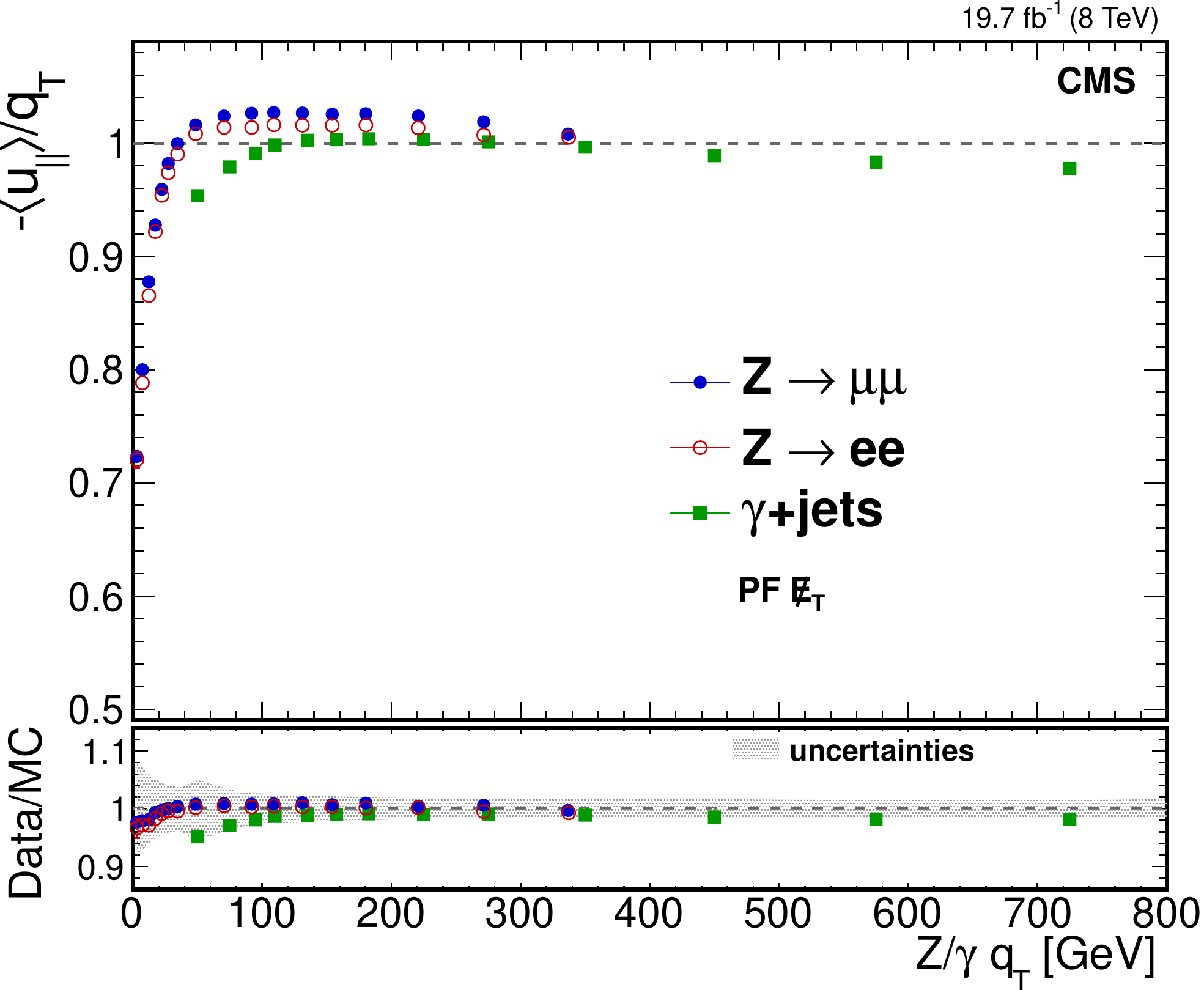}
\caption{
  Response curves for \pfvecmet in events with a \Z{}-boson or direct photon.
  Results are shown for $\Zmm$ events (full blue circles), $\Zee$ events (open red circles),
  and direct-photon events (full green squares).
  The upper frame shows the response in data; the lower frame shows the ratio of data to simulation with
  the grey error band displaying the systematic uncertainty of the simulation, estimated as the maximum of each channel systematic uncertainty.
  The \qt value for each point is determined based on the average \qt value in data contributing to each point.
}
\label{fig:CombResponse}
\end{figure}

The response curves extracted from data, \resp versus \qt, are shown in
Fig.~\ref{fig:CombResponse} for \mbox{\Zmm}, \mbox{\Zee}, and direct-photon
 events.
Deviations from unity indicate a biased hadronic recoil energy scale.
The agreement between data and simulation is reasonable for each channel.
The curves fit to \Z{} data indicate that the \pfvecmet is able to fully recover the
hadronic recoil activity corresponding to a Lorentz boosted \Z{}-boson with $\qt\sim40$
GeV.  Below 40\GeV, the uncorrected unclustered energy contribution (energy not contained within jets or leptons)
starts to be significant compared to the corrected energy of the
recoiling jets, leading to an underestimation of the response.  The curves fit to
\GJ data are 2--3\% lower than those fit to \Z{} data at $\qt<100\GeV$. This effect primarily
stems from the large contribution of QCD multijet events to the $\qt<100\GeV$ region of the
selected photon sample. In these QCD multijet events,
the hadronic recoil of the photon candidate tends to have a higher contamination of gluon jets. As the calorimeter
response to gluon jets is characteristically lower than for quark jets due to difference of jet composition and collimation, the overall average response is reduced for the
photon sample in this region.

\begin{figure}[htb]
\centering
\includegraphics[width=0.45\textwidth]{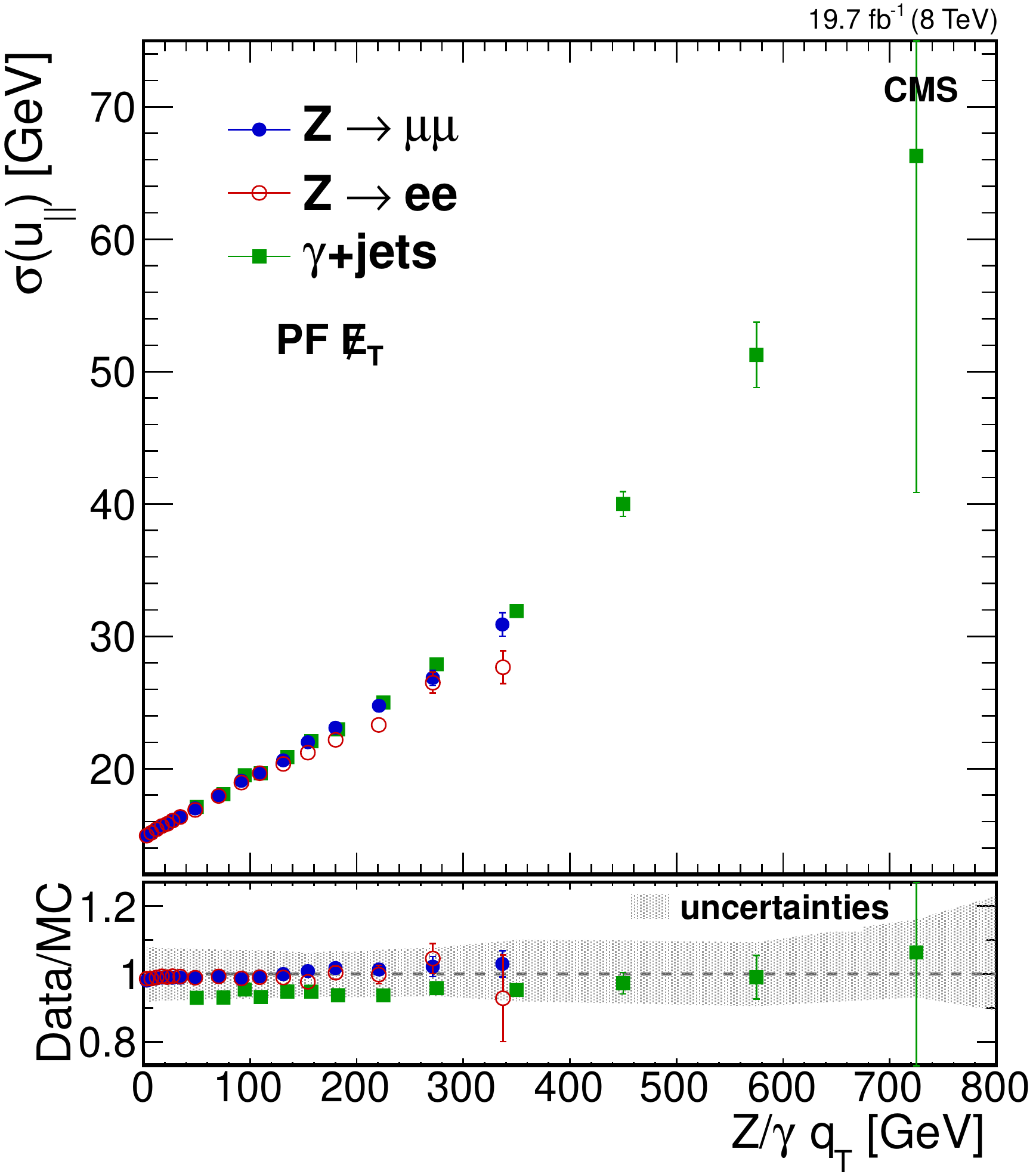}
\includegraphics[width=0.45\textwidth]{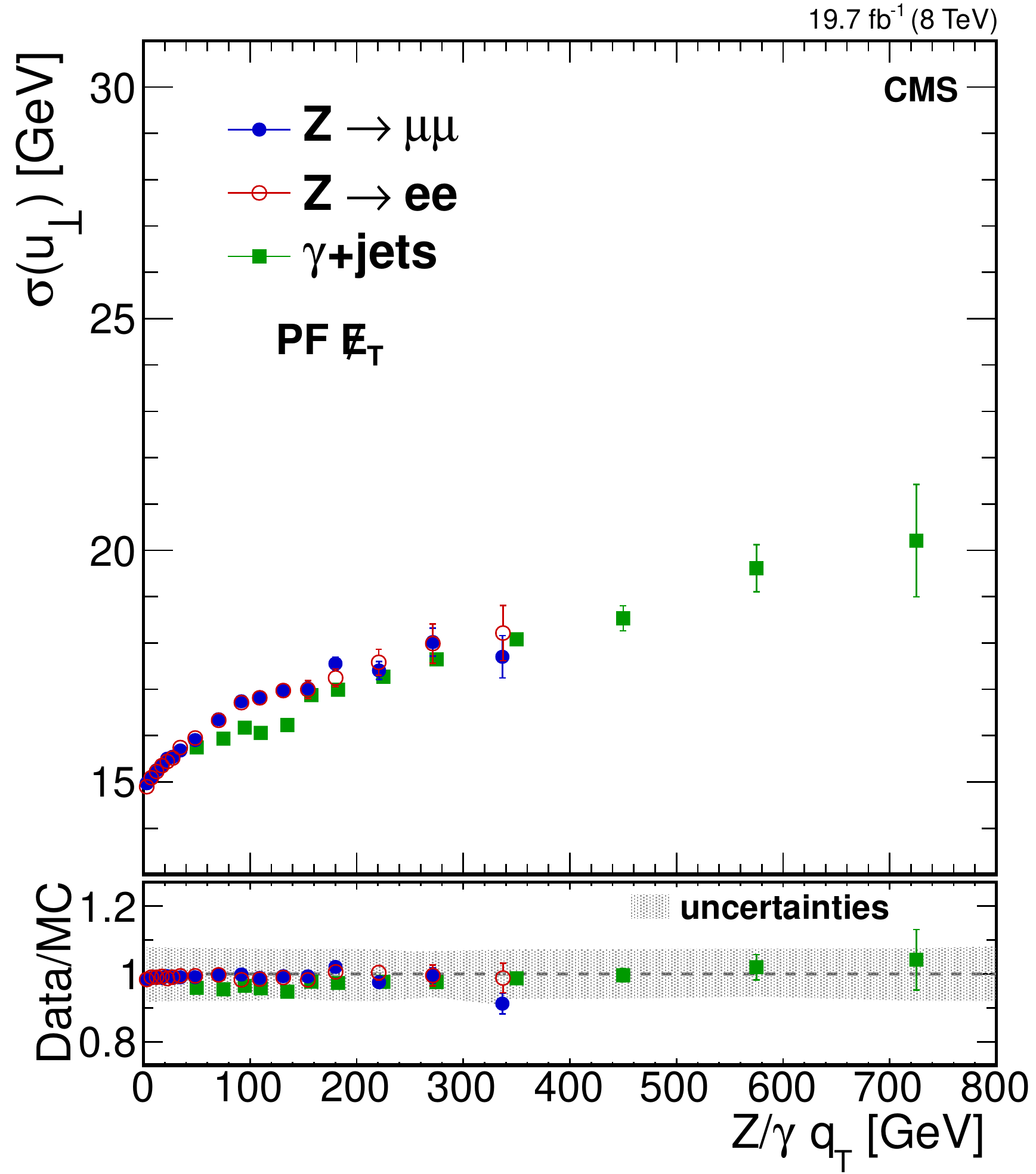}
\caption{
  Resolution curves of the parallel recoil component (left) and perpendicular recoil component (right)
  versus \Z{}/$\gamma$ \qt for \pfvecmet in events with a \Z{}-boson or $\gamma$.
  Results are shown for $\Zmm$ events (full blue circles), $\Zee$ events (open red circles),
  and direct-photon events (full green squares).
  The upper frame of each figure shows the resolution in data;
  the lower frame shows the ratio of data to simulation with
  the grey error band displaying the systematic uncertainty of the simulation, estimated as the maximum of each channel systematic uncertainty.
  The \qt value for each point is determined based on the average \qt value in data contributing to each point.
\label{fig:CombResolutionRMS}}
\end{figure}

The resolution curves, \reso{\upara} and \reso{\uperp} versus \qt, are
shown in Fig.~\ref{fig:CombResolutionRMS}. The resolution increases with
increasing \qt, and the data and simulation curves are in reasonable
agreement for each channel. As the hadronic recoil is produced in the opposite direction of the Z boson or direct photon, \reso{\upara} scales linearly with \qt while \reso{\uperp} is less impacted by the value of \qt.

\begin{figure}[htb]
\centering
\includegraphics[width=0.45\textwidth]{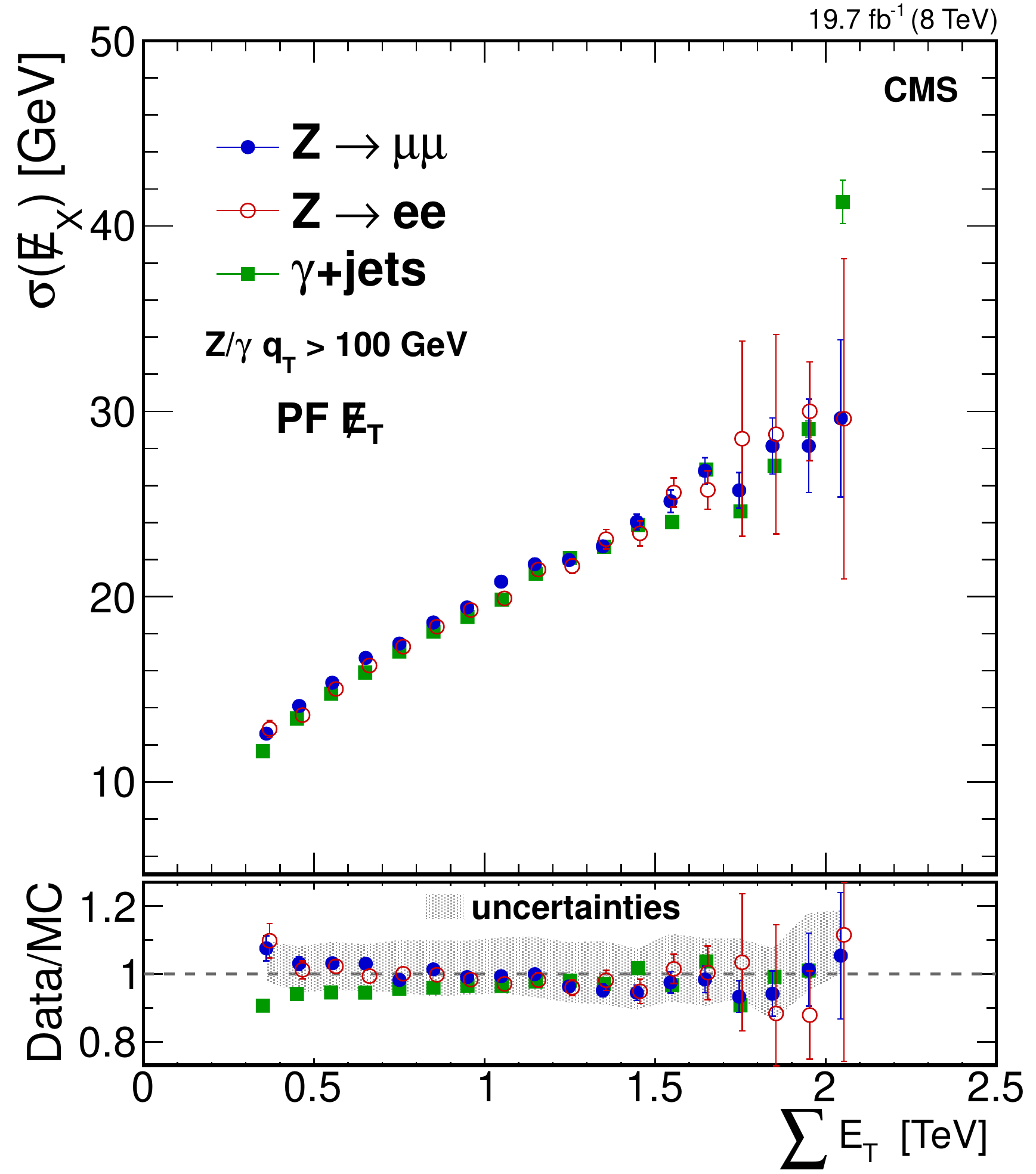}
\includegraphics[width=0.45\textwidth]{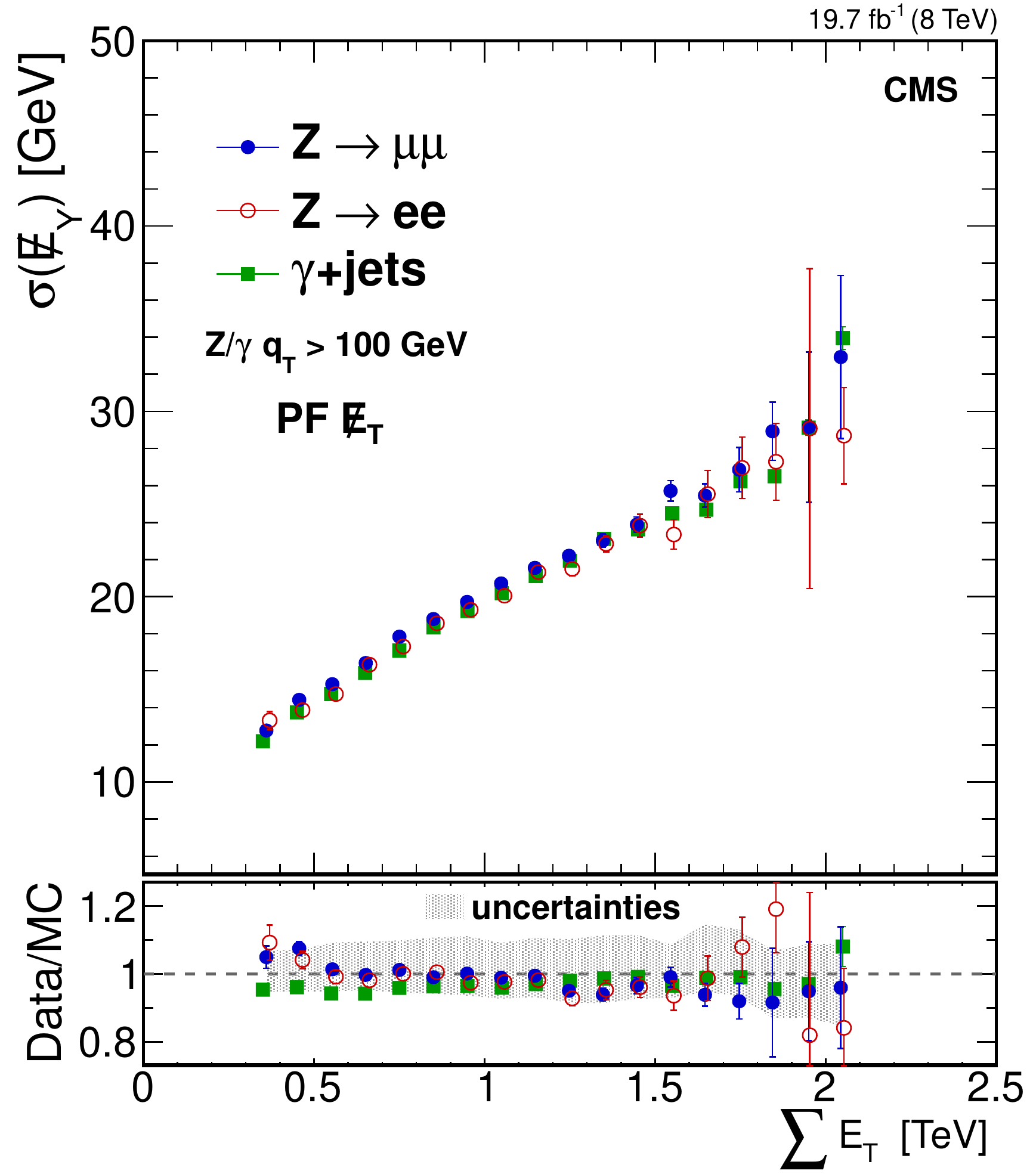}
\caption{
  Resolution of the \pfvecmet projection along the $x$-axis (left) and the $y$-axis (right)
  as a function of \pfsumet in events with a \Z{}-boson or $\gamma$.
  Results are shown for $\Zmm$ events (full blue circles), $\Zee$ events (open red circles),
  and direct-photon events (full green squares).
  The upper frame of each figure shows the resolution in data;
  the lower frame shows the ratio of data to simulation with
  the grey error band displaying the systematic uncertainty of the simulation, estimated as the maximum of each channel systematic uncertainty.
}
\label{fig:CombResolutionSumEt}
\end{figure}

The \cPZ{}-boson and \GJ
\qt spectra  differ from one another, and comparison of resolution curves between the \cPZ{} and \GJ
channels may be affected by their dependence on the \qt{} spectrum.
Thus, for the remaining resolution curves where direct comparisons between
the \cPZ{}-boson and \GJ channels are shown, both
\cPZ{}-boson and \GJ events are required to satisfy $\qt > 100$\GeV, and
event-by-event reweighting of both \cPZ{} data and simulation is applied
to make their \qt spectra similar to that of \GJ data.
Figure~\ref{fig:CombResolutionSumEt} shows the resolution of the
\pfvecmet projections along the $x$ and $y$ axes as a function of
\pfsumet.  The \pfsumet is the scalar sum of \et of all the particles
reconstructed by the particle-flow reconstruction, except for the selected direct photon
or the selected dileptons from the decay of the \cPZ{}-boson candidate.
Resolution curves are found to be in agreement
when comparing different channels and are well described by the simulation.  The
resolution curves for the components of \pfvecmet can be parametrized by a linear
relationship,

\begin{equation} \sigma(\mex,\mey) = \sigma_{0} +
\sigma_{\mathrm{s}}\sqrt{\sumet}, \label{eq:ResoSumEt} \end{equation} where
$\sigma_{0}$ is the intrinsic detector noise resolution and
$\sigma_{\mathrm{s}}$ is the \vecmet resolution stochastic term. Since the fit only contains data with \pfsumet above 300\GeV, the $\sigma_{0}$ parameter is not well constrained in the
fits, and has sizable uncertainties. The uncertainties of the $\sigma_{0}$ parameter are smaller in \GJ data than in \cPZ{} data due to a larger data-sample in the former case. The stochastic term is
$\sigma_{\mathrm{s}}\sim 0.6$ and is compatible for different channels, as shown
in Table~\ref{tab:sumETFits}.

\begin{table}[bt!]
\centering
\topcaption{\label{tab:sumETFits} Parametrization results of the resolution curves for the components of \pfvecmet,
as functions of \pfsumet. The parameter values $\sigma_{0}$ and $\sigma_{\mathrm{s}}$ are obtained from data.
For each parameter, we also present $R_r$, the ratio of values obtained in data and simulation. For the ratios, the
first uncertainty is from the fit, and the second uncertainty corresponds to the propagation of the following into
the parameterization: systematic uncertainties in the jet energy scale, jet energy resolution, lepton/photon energy scale,
and unclustered energy scale, as well as, for direct-photon events only, the systematic uncertainty assigned to the QCD multijet estimation response correction described in Section~\ref{sec:samples}.}
\begin{tabular}{l|cc|cc}

\multicolumn{1}{c}{\multirow{2}{*}{Channel}} & \multicolumn{4}{c}{\mex component} \\
\cline{2-5}
 \multicolumn{1}{c}{}& $\sigma_{0}$ (GeV) & $R_r=\sigma_0(\text{data})/\sigma_0(\mathrm{MC})$
 & $\sigma_{\mathrm{s}}$ (GeV$^{1/2}$) & $R_r=\sigma_\mathrm{s}(\text{data})/\sigma_\mathrm{s}(\mathrm{MC})$ \\
\hline
 \GJ & 0.70 $\pm$ 0.01 & 2.37 $\pm$ 1.11 $\pm$ 0.17 & 0.60 $\pm$ 0.01 & 0.99 $\pm$ 0.05 $\pm$ 0.06 \\
 \Zee & 0.84 $\pm$ 0.46 & 0.83 $\pm$ 0.16 $\pm$ 0.00 & 0.60 $\pm$ 0.02 & 1.01 $\pm$ 0.02 $\pm$ 0.07 \\
 \Zmm  & 1.37 $\pm$ 0.34 & 0.51 $\pm$ 0.30 $\pm$ 0.00 & 0.59 $\pm$ 0.01 & 1.05 $\pm$ 0.02 $\pm$ 0.08 \\
\hline

\multicolumn{1}{c}{\multirow{2}{*}{}} & \multicolumn{4}{c}{\mey component} \\
\cline{2-5}
  \multicolumn{1}{c}{}& $\sigma_{0}$ (GeV) & $R_r=\sigma_0(\text{data})/\sigma_0(\mathrm{MC})$
 & $\sigma_{\mathrm{s}}$ (GeV$^{1/2}$) & $R_r=\sigma_\mathrm{s}(\text{data})/\sigma_\mathrm{s}(\mathrm{MC})$ \\
\hline
 \GJ & 0.76 $\pm$ 0.05 & 2.34 $\pm$ 1.10 $\pm$ 0.35 & 0.60 $\pm$ 0.01 & 0.99 $\pm$ 0.05 $\pm$ 0.04 \\
 \Zee & 1.30 $\pm$ 0.45 & 0.70 $\pm$ 0.76 $\pm$ 0.09 & 0.58 $\pm$ 0.02 & 1.04 $\pm$ 0.06 $\pm$ 0.08 \\
 \Zmm  & 1.47 $\pm$ 0.33 & 0.48 $\pm$ 0.26 $\pm$ 0.00 & 0.59 $\pm$ 0.01 & 1.07 $\pm$ 0.02 $\pm$ 0.09 \\
\end{tabular}

\end{table}

\begin{figure}[htb]
\centering
\includegraphics[width=0.45\textwidth]{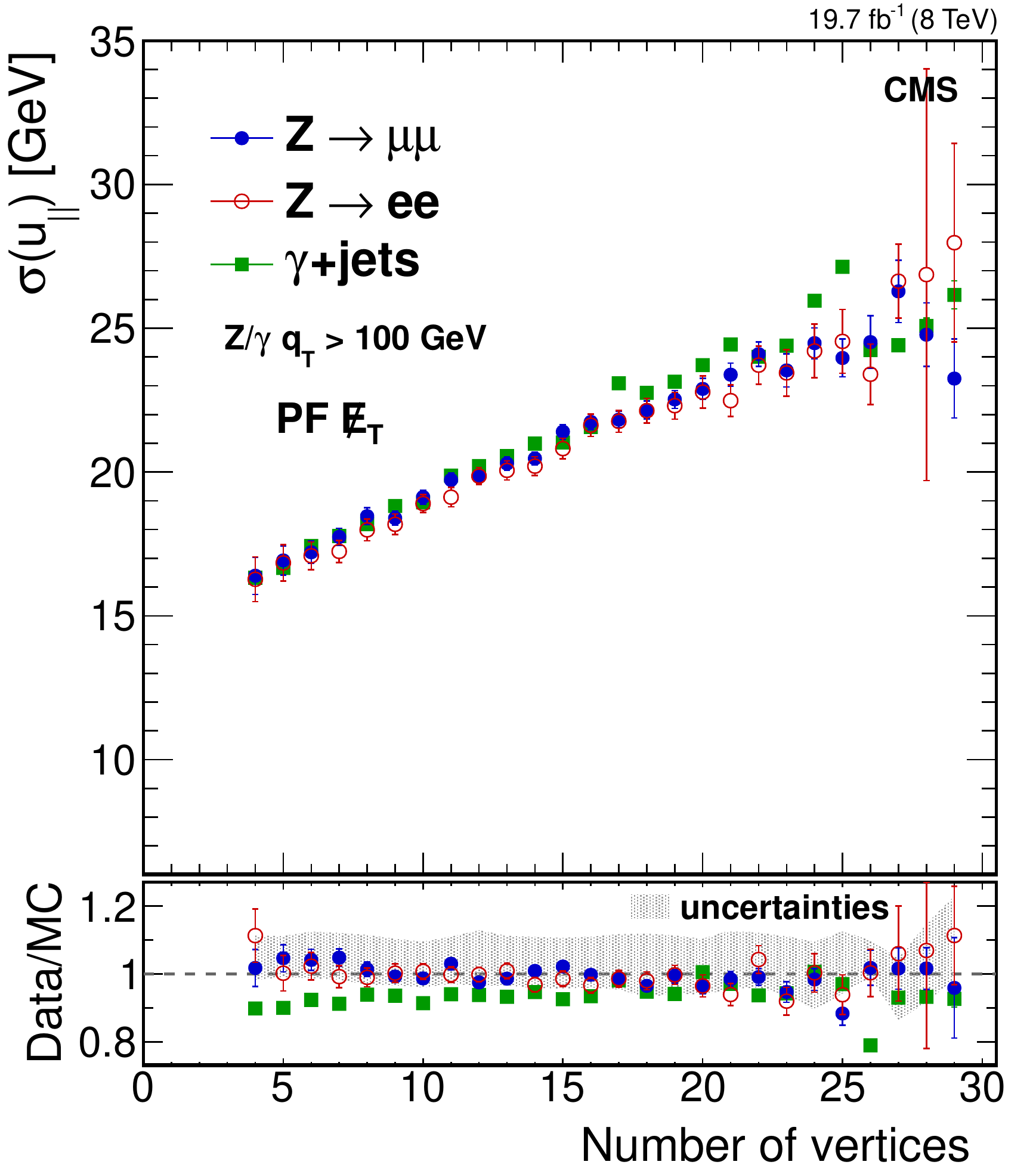}
\includegraphics[width=0.45\textwidth]{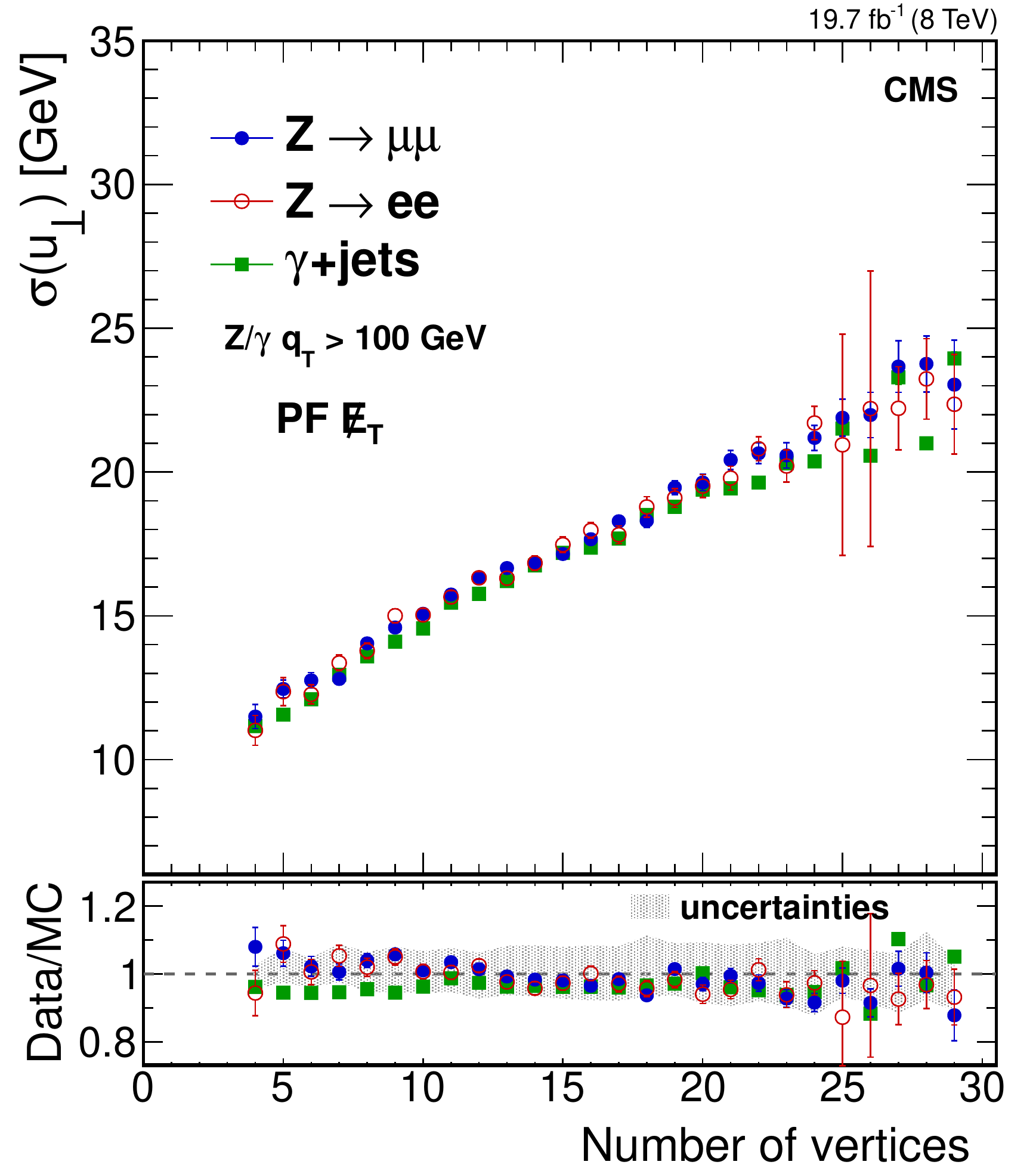}
\caption{
  Parallel (left) and perpendicular (right) recoil component resolution curves
  versus the number of reconstructed vertices for \pfvecmet\ in events with a \Z{}-boson or $\gamma$.
  Results are shown for $\Zmm$ events (full blue circles), $\Zee$ events (open red circles),
  and direct-photon events (full green squares).
  The upper frame of each figure shows the resolution in data;
  the lower frame shows the ratio of data to simulation with
  the grey error band displaying the systematic uncertainty of the simulation, estimated as the maximum of each channel systematic uncertainty.}
\label{fig:CombResoNVtx}
\end{figure}

Figure~\ref{fig:CombResoNVtx} shows the resolution curves \reso{\upara}
and \reso{\uperp} versus the number of primary vertices
$N_\text{vtx}$, for both \cPZ{}-boson channels and the \GJ channel. The
offset of the curve is related to the resolution in \cPZ{} or \GJ events
without pileup and the dependence on $N_\text{vtx}$
indicates how much the pileup degrades the \vecmet resolution. Since the
hard-scatter interaction and each additional collision are uncorrelated,
these resolution curves can be parametrized by the function,

\begin{equation}
  f(N_\text{vtx}) = \sqrt{ \sigma_\mathrm{c}^2 + \frac{N_\text{vtx}}{0.7}\times\sigma_\mathrm{PU}^2 },
\end{equation}

where $\sigma_\mathrm{c}$ is the resolution term induced by the
hard-scatter interaction and $\sigma_\mathrm{PU}$ is the resolution term
induced on average by one additional pileup collision. The factor 0.7
accounts for the fact that  only approximately 70\% of pp interactions produce a
reconstructed vertex isolated from other vertices.
Results of the parameterizations are given in
Table~\ref{tab:NVtxFits}. From there, one can see that
different channels are compatible with each other, and that the
simulation offers a good description of the performance obtained in data.
For each  additional pileup interaction, the \pfvecmet
resolution is degraded by around 3.3--3.6\GeV in quadrature. As a pileup interaction is isotropic,
 the \pfvecmet response is not impacted by the number of additional pileup interaction in the event.

\begin{table}[bthp]
\centering
\topcaption{\label{tab:NVtxFits} Parametrization results of the resolution curves for the \upara{} and \uperp{}
components calculated with the \pfvecmet as functions of $N_\text{vtx}$. The parameter values $\sigma_\mathrm{c}$ and $\sigma_{\mathrm{PU}}$
are obtained from data. For each parameter, we also present $R_r$, the ratio of values obtained in
data and simulation. For the ratios, the first uncertainty is from the fit, and the second uncertainty corresponds
to the propagation of the following into the parameterization: systematic uncertainties in the jet energy scale,
jet energy resolution, lepton/photon energy scale, and unclustered energy scale, as well as, for photon events only, the systematic uncertainty assigned to the QCD multijet estimation response correction described in Section~\ref{sec:samples}. }

\begin{tabular}{l|cc|cc}
\multicolumn{1}{c}{\multirow{2}{*}{Channel}} & \multicolumn{4}{c}{\upara component} \\

\cline{2-5}
 \multicolumn{1}{c}{} & $\sigma_\mathrm{c}$ (\GeVns{}) & $R_r=\sigma_\mathrm{c}(\text{data})/\sigma_\mathrm{c}(\mathrm{MC})$
 & $\sigma_{\mathrm{PU}}$ (GeV) & $R_r=\sigma_\mathrm{PU}(\text{data})/\sigma_\mathrm{PU}(\mathrm{MC})$ \\
\hline
\GJ & 13.70 $\pm$ 0.05 & 1.13 $\pm$ 0.03 $\pm$ 0.01 & 3.57 $\pm$ 0.01 & 1.02 $\pm$ 0.04 $\pm$ 0.10 \\
 \Zee & 13.89 $\pm$ 0.36 & 0.94 $\pm$ 0.05 $\pm$ 0.03 & 3.36 $\pm$ 0.08 & 1.06 $\pm$ 0.05 $\pm$ 0.09 \\
 \Zmm  & 14.25 $\pm$ 0.26 & 0.95 $\pm$ 0.03 $\pm$ 0.06 & 3.37 $\pm$ 0.06 & 1.07 $\pm$ 0.04 $\pm$ 0.11 \\
\hline
\multicolumn{1}{c}{\multirow{2}{*}{}} & \multicolumn{4}{c}{\uperp component} \\
\cline{2-5}
 \multicolumn{1}{c}{} & $\sigma_{c}$ (GeV) & $R_r=\sigma_\mathrm{c}(\text{data})/\sigma_\mathrm{c}(\mathrm{MC})$
 & $\sigma_{\mathrm{PU}}$ (GeV) & $R_r=\sigma_\mathrm{PU}(\text{data})/\sigma_\mathrm{PU}(\mathrm{MC})$ \\
\hline
 \GJ & 7.79 $\pm$ 0.04 & 1.15 $\pm$ 0.05 $\pm$ 0.03 & 3.28 $\pm$ 0.01 & 1.00 $\pm$ 0.03 $\pm$ 0.08 \\
 \Zee & 8.24 $\pm$ 0.34 & 0.72 $\pm$ 0.09 $\pm$ 0.05 & 3.32 $\pm$ 0.05 & 1.10 $\pm$ 0.03 $\pm$ 0.10 \\
 \Zmm  & 8.21 $\pm$ 0.26 & 0.79 $\pm$ 0.07 $\pm$ 0.05 & 3.33 $\pm$ 0.03 & 1.08 $\pm$ 0.03 $\pm$ 0.11 \\
\end{tabular}

\end{table}

\begin{figure}[htb]
\centering
\includegraphics[width=0.32\textwidth]{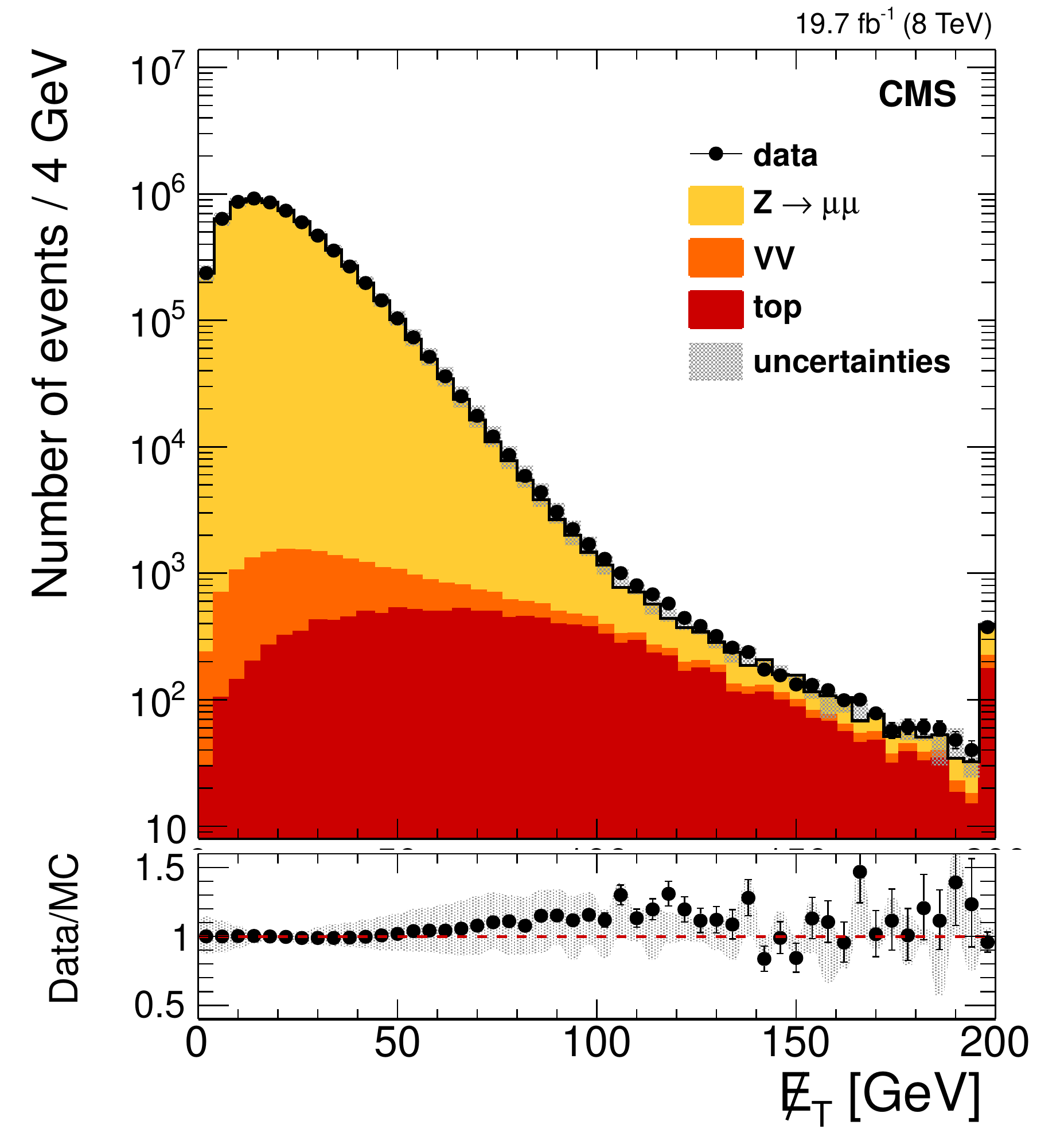}
\includegraphics[width=0.32\textwidth]{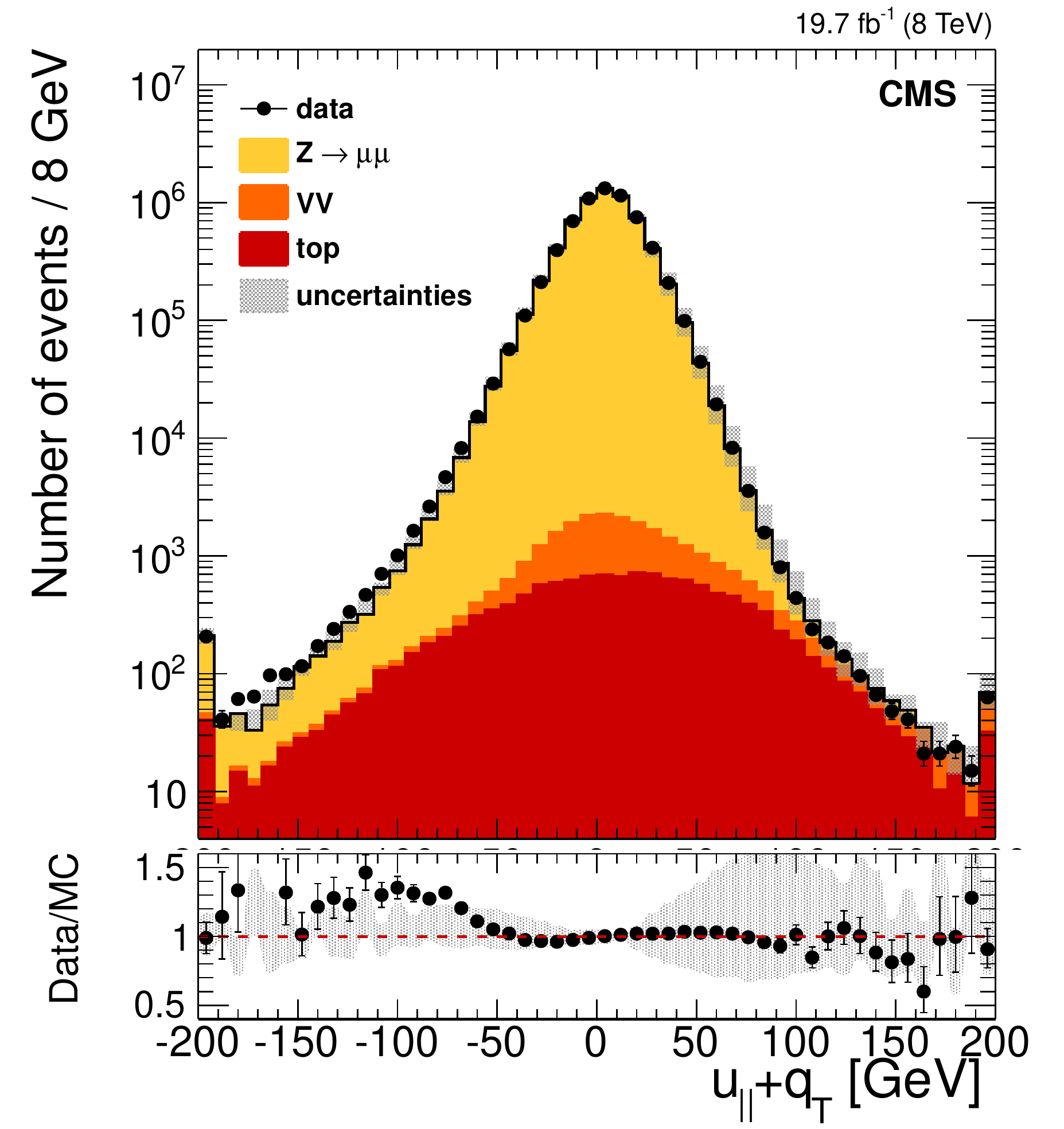}
\includegraphics[width=0.32\textwidth]{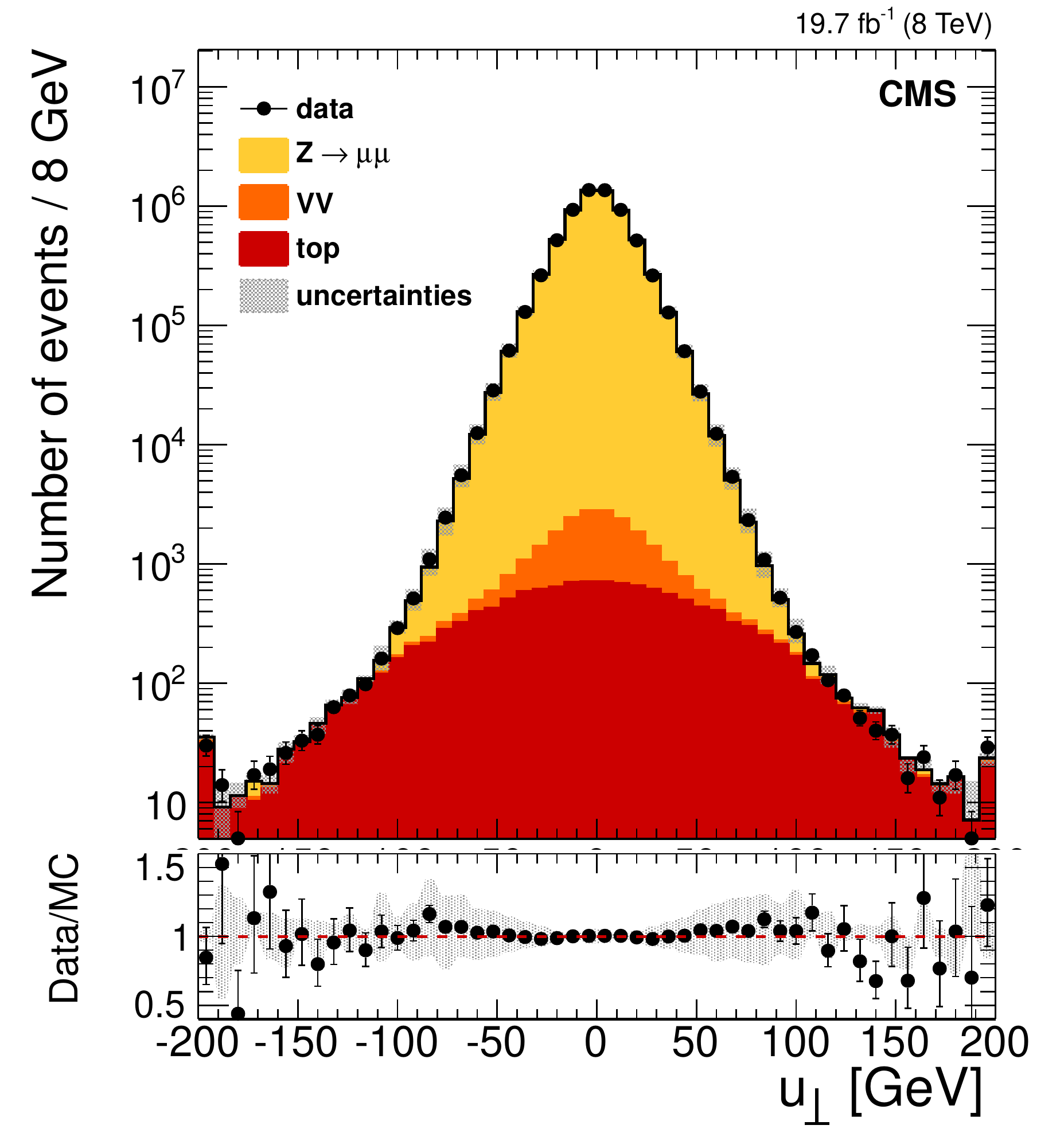}
\caption{
  Calo \met (left), and its parallel (middle) and perpendicular (right) recoil component spectra for \Zmm events.
  The points in the lower panel of each plot show the data/MC ratio, including the
  statistical uncertainties of both data and simulation;
  the grey error band displays the systematic uncertainty of the simulation. The first (last) bin contains the underflow (overflow) content.
}
\label{fig:CaloMET}
\end{figure}

\begin{figure}[htb]
\centering
\includegraphics[width=0.45\textwidth]{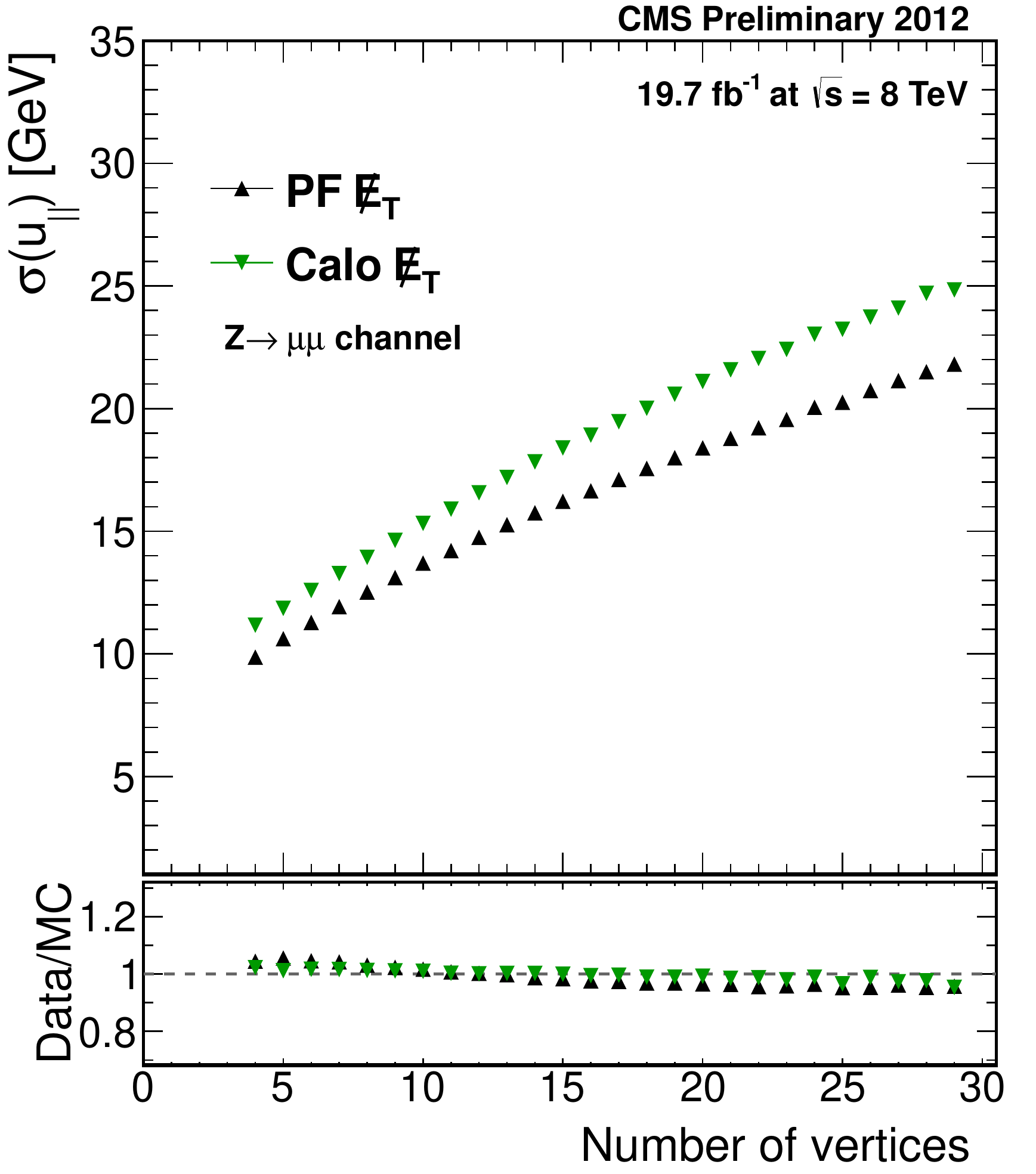}
\includegraphics[width=0.45\textwidth]{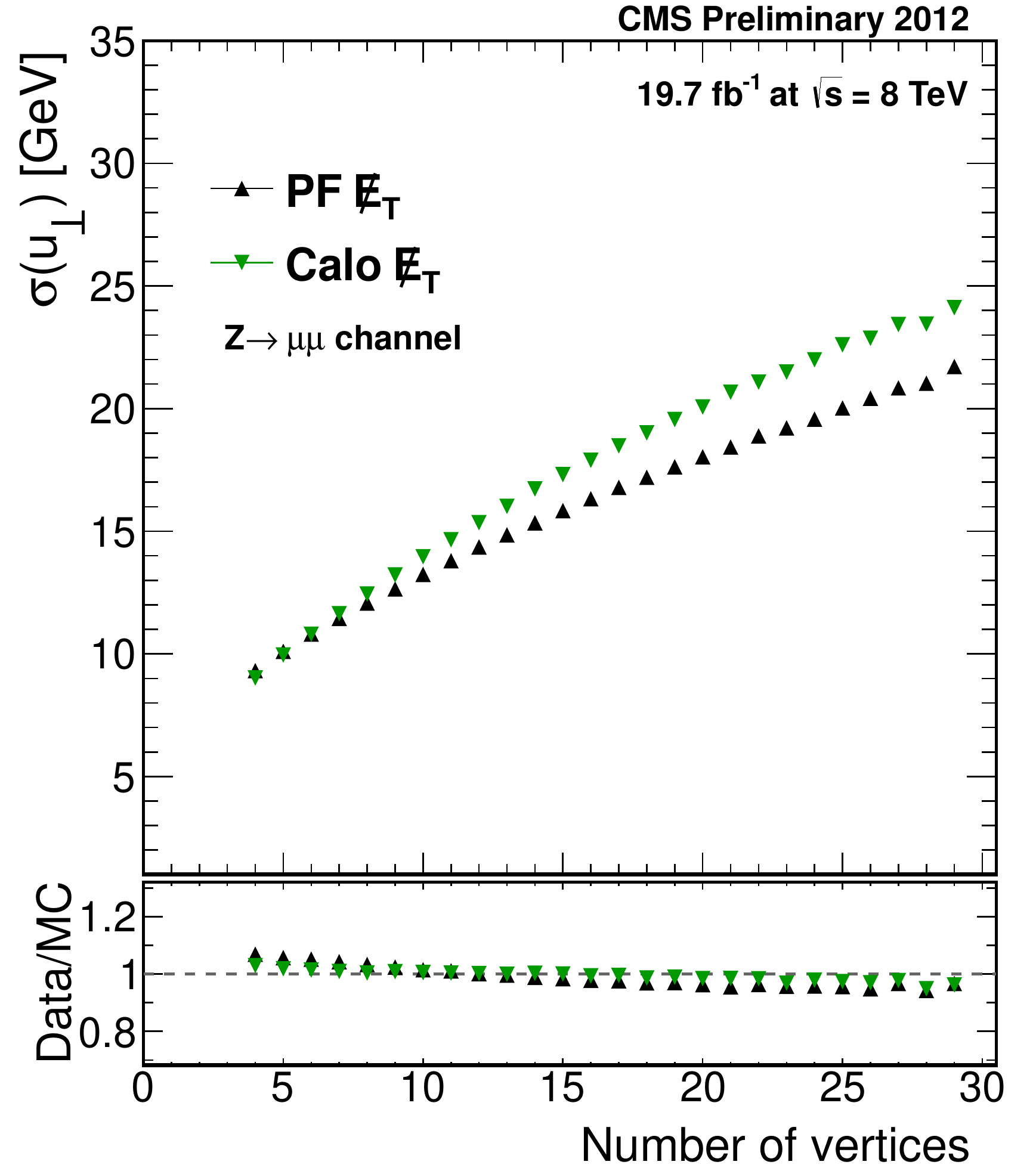}
\caption{
  Resolution curves of the parallel (left) and perpendicular (right) recoil component
  versus the number of reconstructed vertices for Calo~\met (green downward-triangle)
  and \pfvecmet (black upward-triangle) for \Zmm events.
  The upper frame of each figure shows the resolution in data;
  the lower frame shows the ratio of data to simulation.
}
\label{fig:CaloReso}
\end{figure}

The Calo\,\met spectrum, as well as the Calo\,\met recoil components and
are shown in Figs.~\ref{fig:CaloMET}, to be compared with
Figs.~\ref{fig:ZbosonMET} and \ref{fig:recoil_perp_gammajet}.  The
comparison of resolution curves as a function of the number of
reconstructed vertices between \pfvecmet and Calo\,\vecmet, shown in
Fig.~\ref{fig:CaloReso}, demonstrates how the PF reconstruction of \met
has stronger performance in terms of \met resolution dependence on
pileup relative to the \met reconstruction based solely on the
calorimeters.

\section{Pileup-mitigated \texorpdfstring{\bigmet}{MET}}
\label{sec:nopumet}

Since the vast majority of pileup interactions do not have significant
\vecmet and the average value of \vecmet projected on any axis is
zero, the effect of pileup interactions on the \met response is small.
However, as shown in Section~\ref{sec:resolution}, pileup interactions
have a considerable effect on the \vecmet resolution.
Table~\ref{tab:NVtxFits} shows that each pileup interaction adds an
additional 3.3--3.6\GeV of smearing to the \vecmet resolution in
quadrature to both
\uperp and \upar
in \Zmm, \Zee,
and direct-photon events.  In events
where the recoil \pt is small and the number of pileup interactions is around
the mean value of the sample collected during the 2012 run, which corresponds
 to approximately
21 pileup interactions, the contribution to the \vecmet resolution from
pileup interactions is larger than the contribution from the hadronic recoil.

In this section we discuss two algorithms that reduce the effect of pileup interactions on the \vecmet
reconstruction, hereafter referred to as the No--PU \pfvecmet and MVA
\pfvecmet algorithms.  These algorithms divide each event
into two components: particles that are likely to originate from the
primary hard-scattering pp interaction (HS particles) and particles that
are likely to originate from pileup interactions (PU particles).

\subsection{Identification of PU-jets}

Separation of charged PF particles originating from the primary
hard-scattering pp interaction and those from pileup interactions is
best performed by matching them to either the primary vertex or to pileup
vertices.  This information is also used to identify jets originating
primarily from pileup interactions (pileup jets).  Pileup jets often
appear as an agglomeration of lower-\pt sub-jets.  To identify pileup
jets we use a multivariate boosted decision tree (BDT) algorithm that
uses jet shape variables and vertex information and is referred to as the
``MVA pileup jet identification discriminator'' (MVA pileup jet
ID)~\cite{CMS-PAS-JME-13-005}.  Both No--PU and MVA \pfvecmet algorithms utilize
the MVA pileup jet ID.

Details of the No--PU and MVA \pfvecmet algorithms and their
performance in \Zmm, \Zee, and \GJ events are presented in the following sections.
These algorithms provide a crucial improvement to physics analyses
sensitive to low or moderate \met values, such as Higgs boson searches in the
$\tau$-lepton final states~\cite{Chatrchyan:2014nva}.

\subsection{The No--PU PF \texorpdfstring{\bigmet}{MET} algorithm}

The No--PU \pfvecmet algorithm computes the transverse momentum
imbalance by separately weighting contributions from the HS and PU
particles. In contrast to the global pile-up correction included in
Eq.~\ref{eq:type0}, this algorithm therefore treats individual
particles.

The particles that are classified as HS particles are:
\begin{itemize}
\item ``leptons'' (electrons/photons, muons, and hadronic tau decays),
\item particles within jets of $\pt > 30$\GeV that pass the
  MVA pileup jet ID (HS-jets),
\item charged hadrons associated to the hard-scatter vertex
  (unclustered HS-charged hadrons), by matching the associated tracks to the reconstructed vertex of the event.
\end{itemize}
Particles that are considered to be PU particles are:
\begin{itemize}
\item charged hadrons that are neither within jets of $\pt > 30$\GeV
  nor associated to the hard-scatter vertex (unclustered PU-charged
  hadrons),
\item neutral particles not within jets of $\pt > 30$\GeV
  (unclustered neutrals),
\item particles within jets of $\pt > 30$\GeV that fail the MVA pileup
  jet ID (PU-jets).
\end{itemize}

HS particles
enter the transverse momentum
balance in the usual way (see Section \ref{sec:metreco}).  The transverse momenta of PU particles
are scaled down in order to reduce the
impact of pileup on the \met resolution.  The scale factor is
based on the ratio of the scalar sum of the transverse momenta of charged particles that
originate from hard-scattering pp collision and are neither associated to leptons nor to jets of $\pt > 30\GeV$
 (unclustered HS-charged hadrons) to the scalar sum of the transverse momenta
of all unclustered charged hadrons in the event,
\begin{equation}
  S_\mathrm{F} = \frac{\sum_\text{HS-charged} \pt}{\sum_\text{HS-charged} \pt + \sum_\text{PU-charged} \pt}.
  \label{eq:SF}
\end{equation}

Based on this scale factor,
the No--PU \pfvecmet is then computed as,
\begin{multline}
  \vecmet  =  -\Bigg[ \sum_\text{leptons} \vecpt +
    \sum_\text{HS-jets} \vecpt +
    \sum_\text{HS-charged} \vecpt \\
       + S_\mathrm{F}  \Big(
    \alpha  \sum_\text{PU-charged} \vecpt +
    \beta  \sum_\text{neutrals} \vecpt +
    \gamma  \sum_\text{PU-jets} \vecpt +
    \delta  \vec{\Delta}_\mathrm{PU} \Big) \Bigg].
  \label{eq:NoPUMEtComputation}
\end{multline}

The $\vec{\Delta}_\text{PU}$ term is added in a similar way as was done for
the pileup correction applied to the PF~\vecmet{} (c.f.
Eq.~\eqref{eq:type0}), which improves the No--PU \pfvecmet\ resolution.
The parameters $\alpha$, $\beta$, $\gamma$, and $\delta$ have been
determined by numerical optimization of the $\vecmet$ resolution using a
sample of simulated \Zmm\ events.  The optimal values found by this
procedure are  $\alpha = 1.0$, $\beta =0.6$, $\gamma = 1.0$, and $\delta = 1.0$.

\subsection{MVA PF \texorpdfstring{\bigmet}{MET} algorithm}

The MVA \pfvecmet algorithm is based on a set of multivariate regressions
that provide an improved measurement of the \vecmet in the presence of
a high number of pileup interactions. The MVA \pfvecmet is computed as a
correction to the hadronic recoil \vut reconstructed from PF
particles. The correction is obtained in two steps. First, we compute a
correction to the direction of \vut by training a BDT to match the true
hadronic recoil direction in simulated events. In the second step,
another BDT is trained to predict the magnitude of the true \vut on a
dataset where we have already corrected the direction of the \vut using
the regression function from the first step. The corrected \vut is then
added to $\vqt$ to obtain the negative MVA \pfvecmet. The regression for
the correction to the recoil angle is trained on a simulated \Zmm data sample. The
training for the recoil magnitude correction uses a  mixture of simulated \Zmm and $\gamma$+jets
events. The simulated $\gamma$+jets sample is added to the training to
ensure a sufficiently large training sample over the whole \qt region.

To construct the MVA \pfvecmet, we compute five \vecmet variables
calculated from PF particles :
\begin{enumerate}

	\item $\vecmet(1)\equiv -\sum_{X_1} \!\vpt$, where $X_1$ is the set of all PF particles (= \pfvecmet without correction);
	
	\item $\vecmet(2)\equiv -\sum_{X_2} \!\vpt$, where $X_2$ is the set of all charged PF particles that have been associated to the
	selected hard-scatter vertex;
	
	\item $\vecmet(3)\equiv -\sum_{X_3} \!\vpt$, where $X_3$ is the set of all charged PF particles that have been associated to the selected hard-scatter vertex and
  	all neutral PF particles within jets that have passed the MVA pileup jet ID;

	\item $\vecmet(4)\equiv -\sum_{X_4} \!\vpt$, where $X_4$ is the set of all charged PF particles that have not been associated to the selected hard-scatter vertex
   	and all neutral PF particles within jets that have failed the MVA pileup jet ID;

	\item $\vecmet(5)\equiv -\sum_{X_5} \!\vpt + \sum_{Y_5} \!\vpt$, where $X_{5}$ is the set of all charged PF particles that have been associated to the selected hard-scatter vertex and
  	all neutral PF particles (also those that have not been clustered into jets), while $Y_{5}$ is the set of all neutral PF particles within jets that have failed
  	the MVA pileup jet ID.

\end{enumerate}

 The choice of these variables is intended
to address five different sub-components of an event, which can be decorrelated
from each other by considering various linear combinations of the $\vecmet(i)$ variables:
\begin{itemize}
\item the charged PF particles from the hard scatter (in $\vecmet(1)$, $\vecmet(2)$, $\vecmet(3)$ and $\vecmet(5)$);
\item the charged PF particles not from the hard scatter (in $\vecmet(1)$ and $\vecmet(4)$);
\item the neutral PF particles in jets passing the MVA pileup jet ID (in $\vecmet(1)$, $\vecmet(3)$ and $\vecmet(5)$);
\item the neutral PF particles in jets failing the MVA pileup jet ID (in $\vecmet(1)$, $\vecmet(4)$ and $\vecmet(5)$) ;
\item the unclustered neutral PF particles (in $\vecmet(1)$ and $\vecmet(5)$).
\end{itemize}

For each of the $\vecmet(i)$ variables, the vector $\vut(i)$ is computed using
the definition from Section~\ref{sec:resolution}. The
BDT regression then takes as inputs the
magnitude and azimuthal angle $\phi$ of all five types of \vut; the scalar
\pt sum of all PF particles for each respective \vecmet
variable; the momentum vectors of the two highest \pt jets in the event;
and the number of primary vertices.

Two versions of MVA \pfvecmet are used in the following studies. The first
one is trained to optimize the \vecmet resolution, and the second one
is trained to reach unity \vecmet response. The latter one is denoted as
the unity training and the related MVA \pfvecmet is called MVA Unity \pfvecmet.
The unity response training is performed in the same sample used for
the non-unity response training. To ensure the uniformity of the MVA
Unity \pfvecmet training as function of \qt,  the  events have an additional
weight in the training such that the reweighted \qt distribution is flat
over the full range.

\begin{figure}[t!hb]
  \centering
  \includegraphics[width=0.32\textwidth]{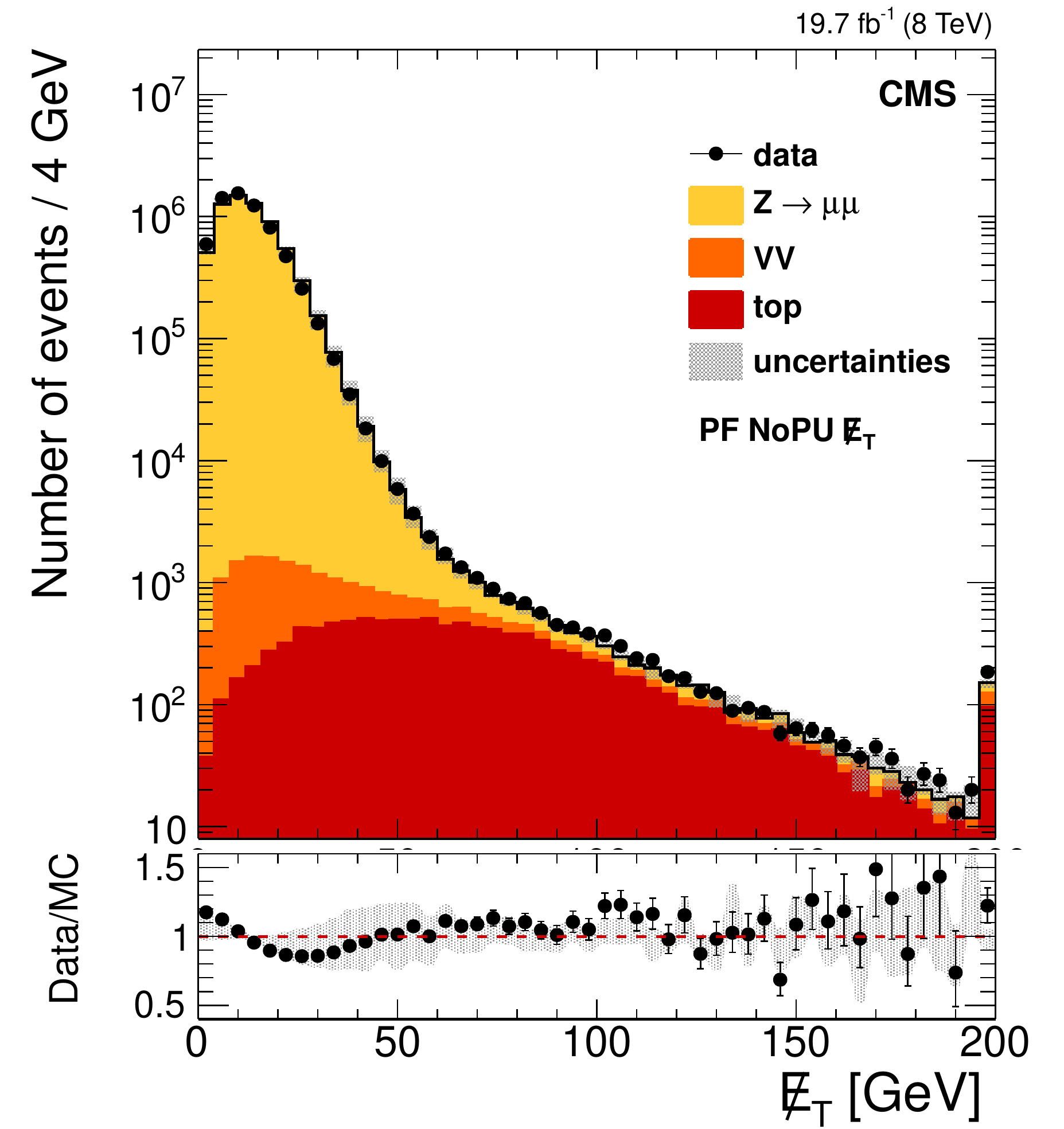}
  \includegraphics[width=0.32\textwidth]{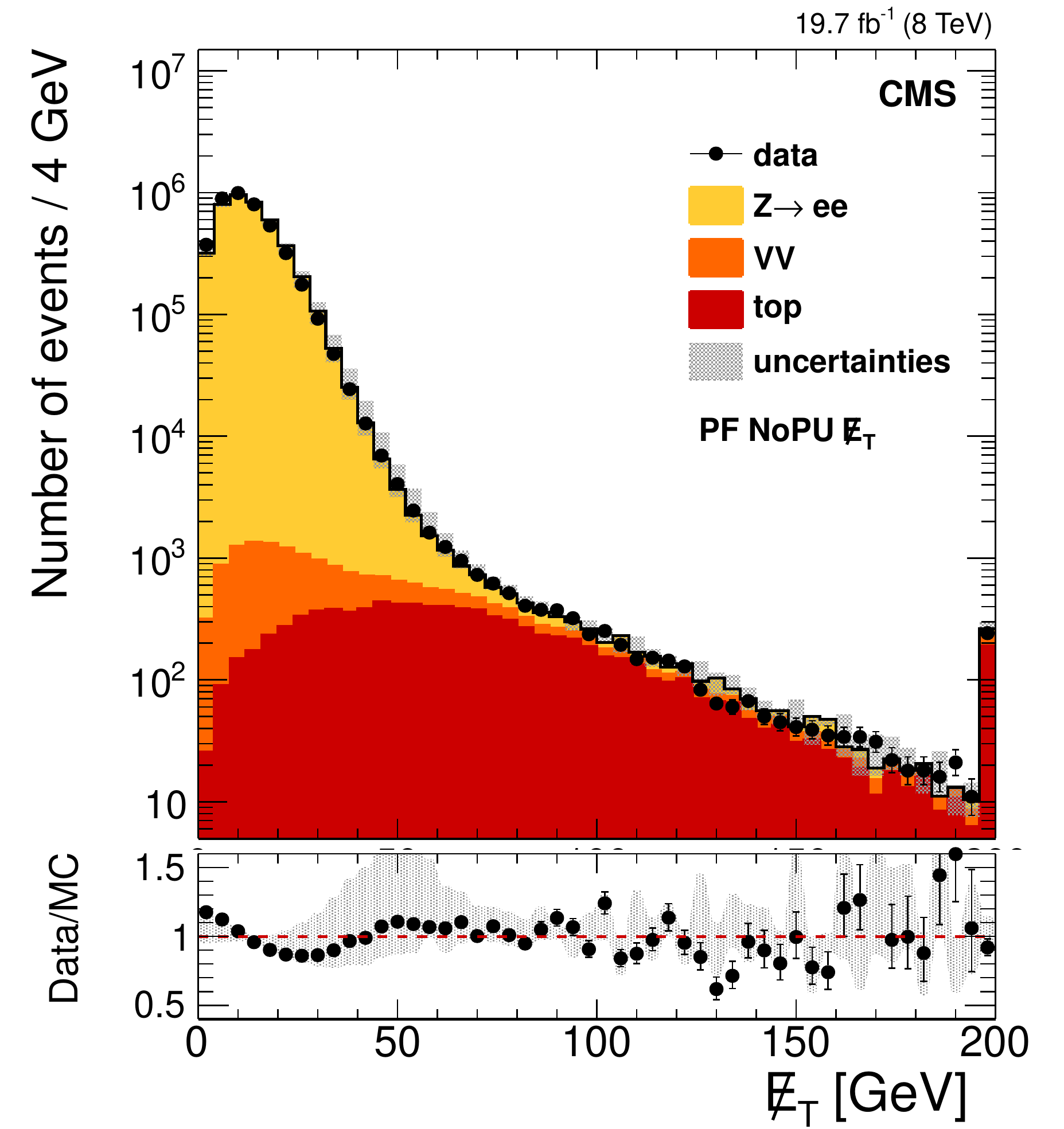}
  \includegraphics[width=0.32\textwidth]{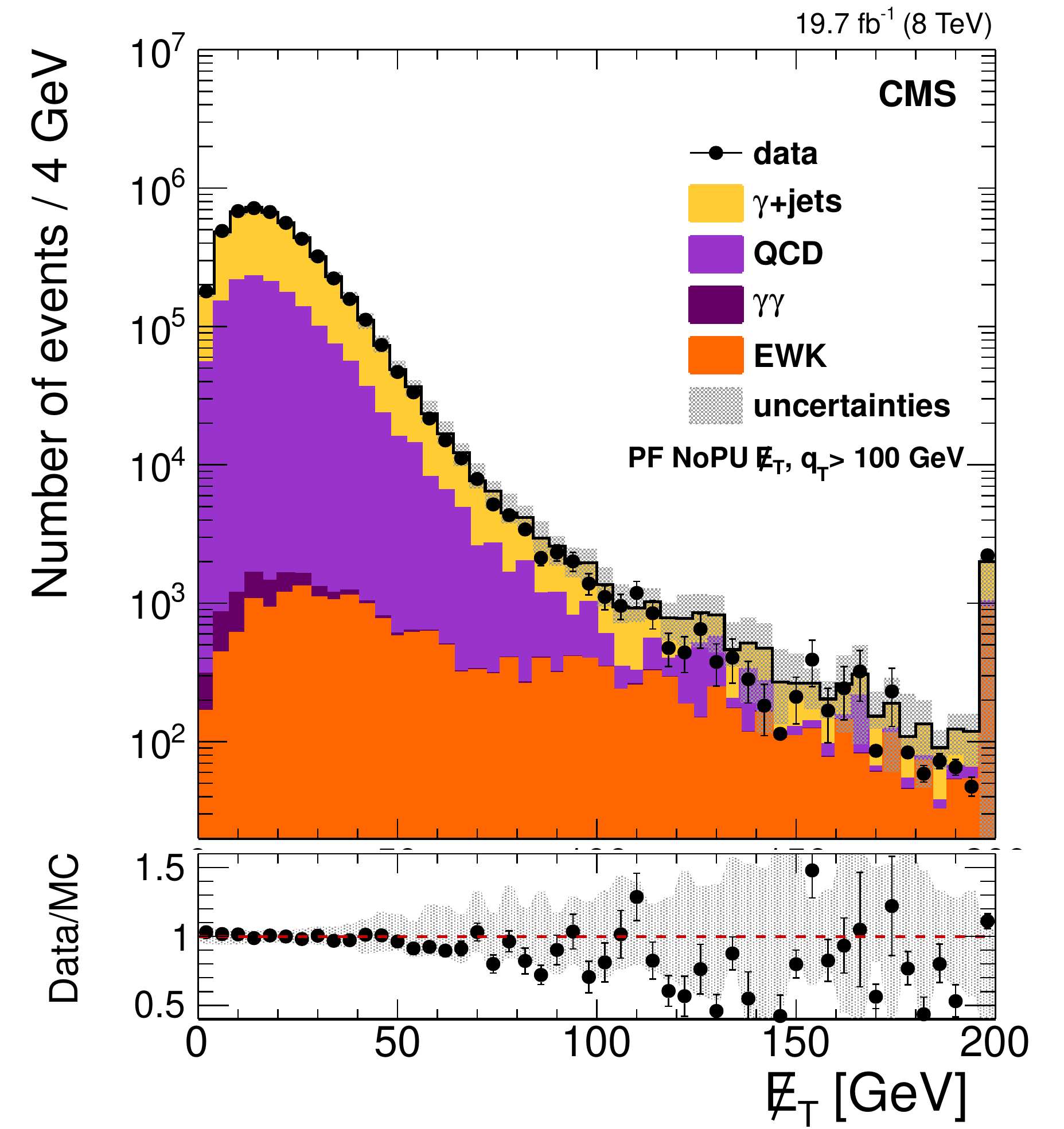}
  \caption{
    No--PU \pfvecmet distributions in \Zmm\ (left), \Zee\ (middle), and \GJ\ (right) events.
    The points in the lower panel of each plot show the data/MC ratio, including the
    statistical uncertainties of both data and simulation;
    the grey error band displays the systematic uncertainty of the simulation. The last bin contains the overflow content.
  }
  \label{fig:NoPU1}
\end{figure}

\begin{figure}[h!tb]
  \centering
  \includegraphics[width=0.32\textwidth]{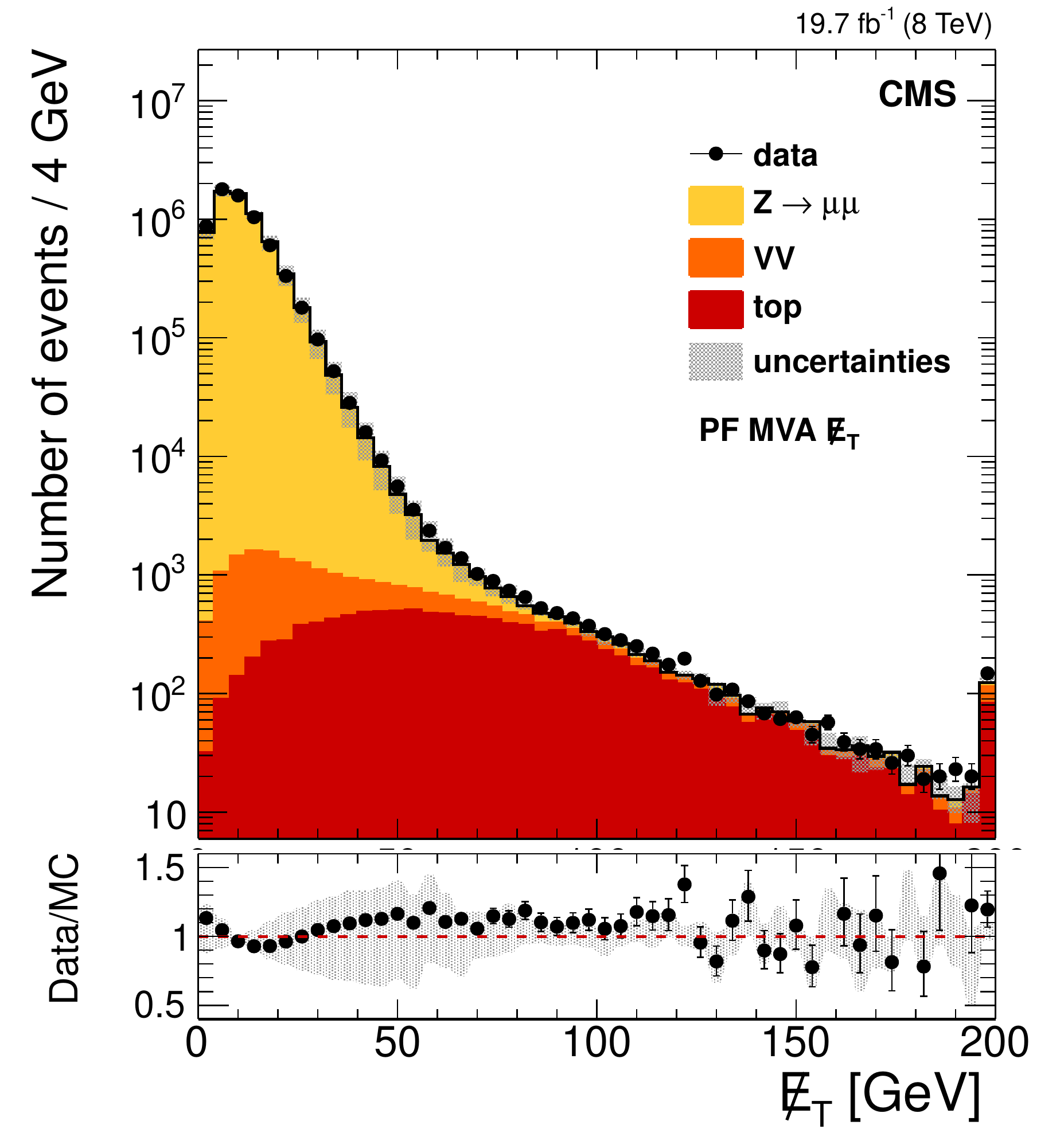}
  \includegraphics[width=0.32\textwidth]{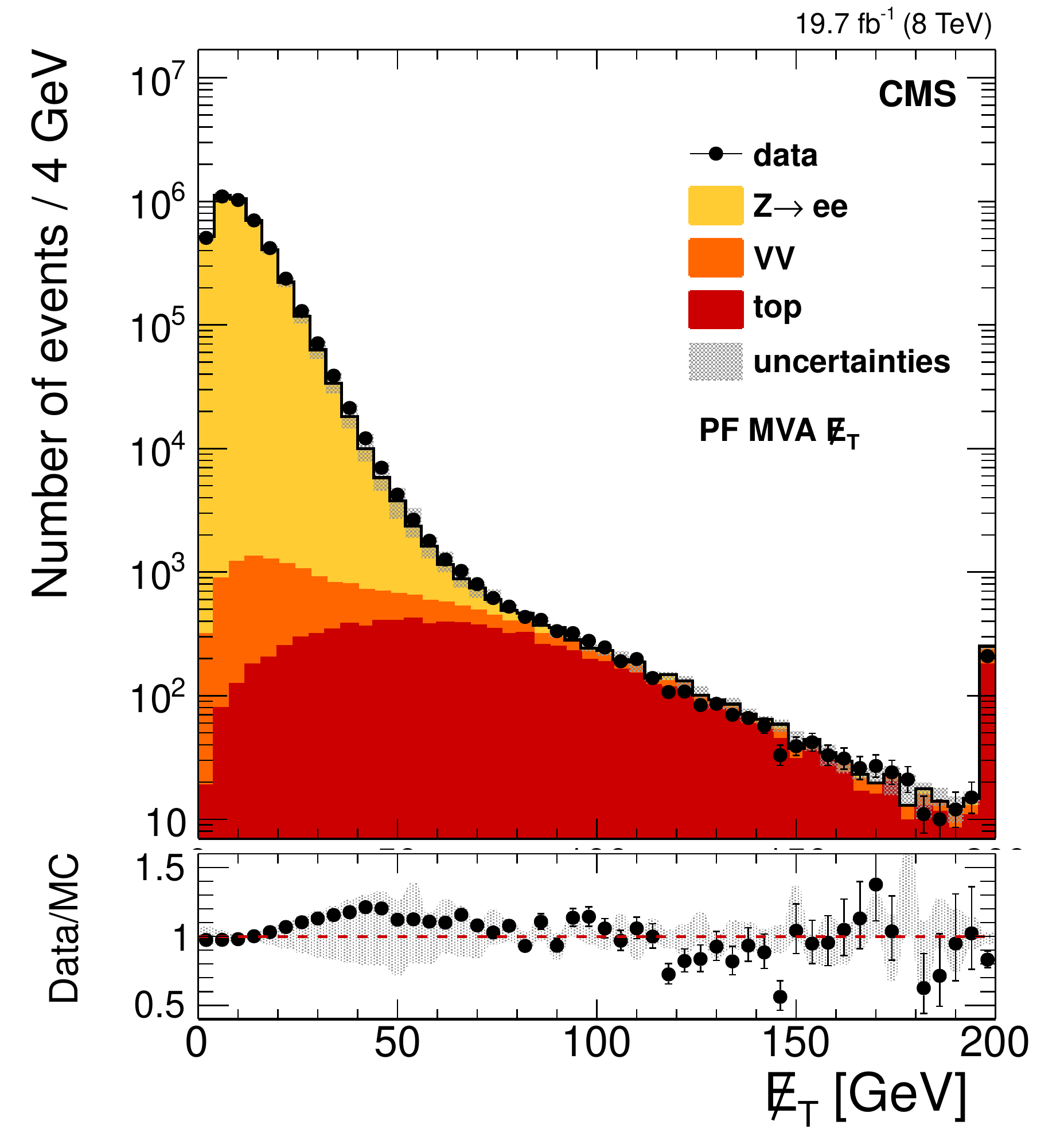}
  \includegraphics[width=0.32\textwidth]{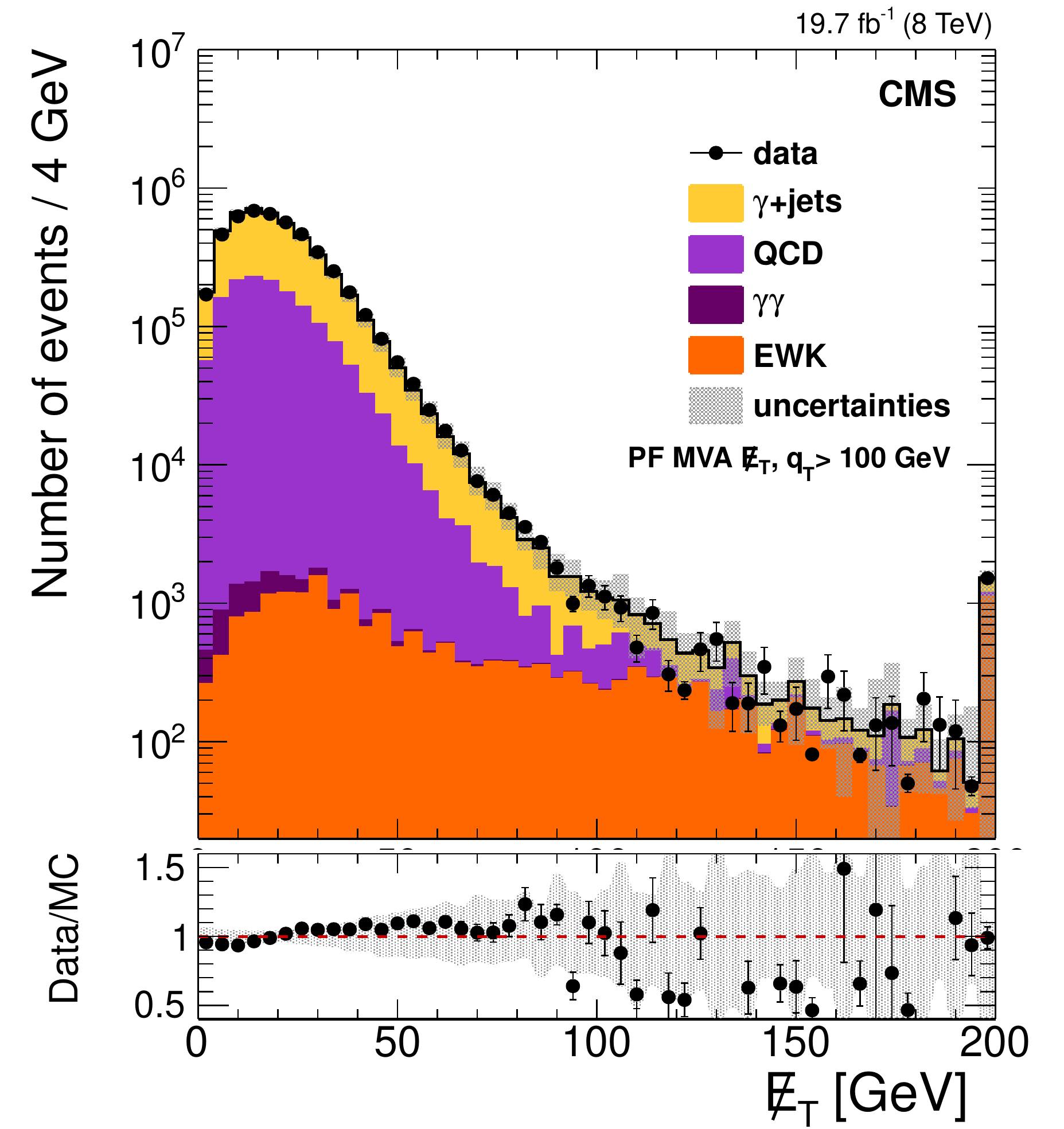}
  \caption{
    MVA \pfvecmet distributions in \Zmm (left), \Zee\ (middle), and \GJ\ (right) events.
    The points in the lower panel of each plot show the data/MC ratio, including the
    statistical uncertainties of both data and simulation;
    the grey error band displays the systematic uncertainty of the simulation. The last bin contains the overflow content.
  }
  \label{fig:MVAMET}
\end{figure}

\begin{figure}[h!tb]
  \centering
  \includegraphics[width=0.32\textwidth]{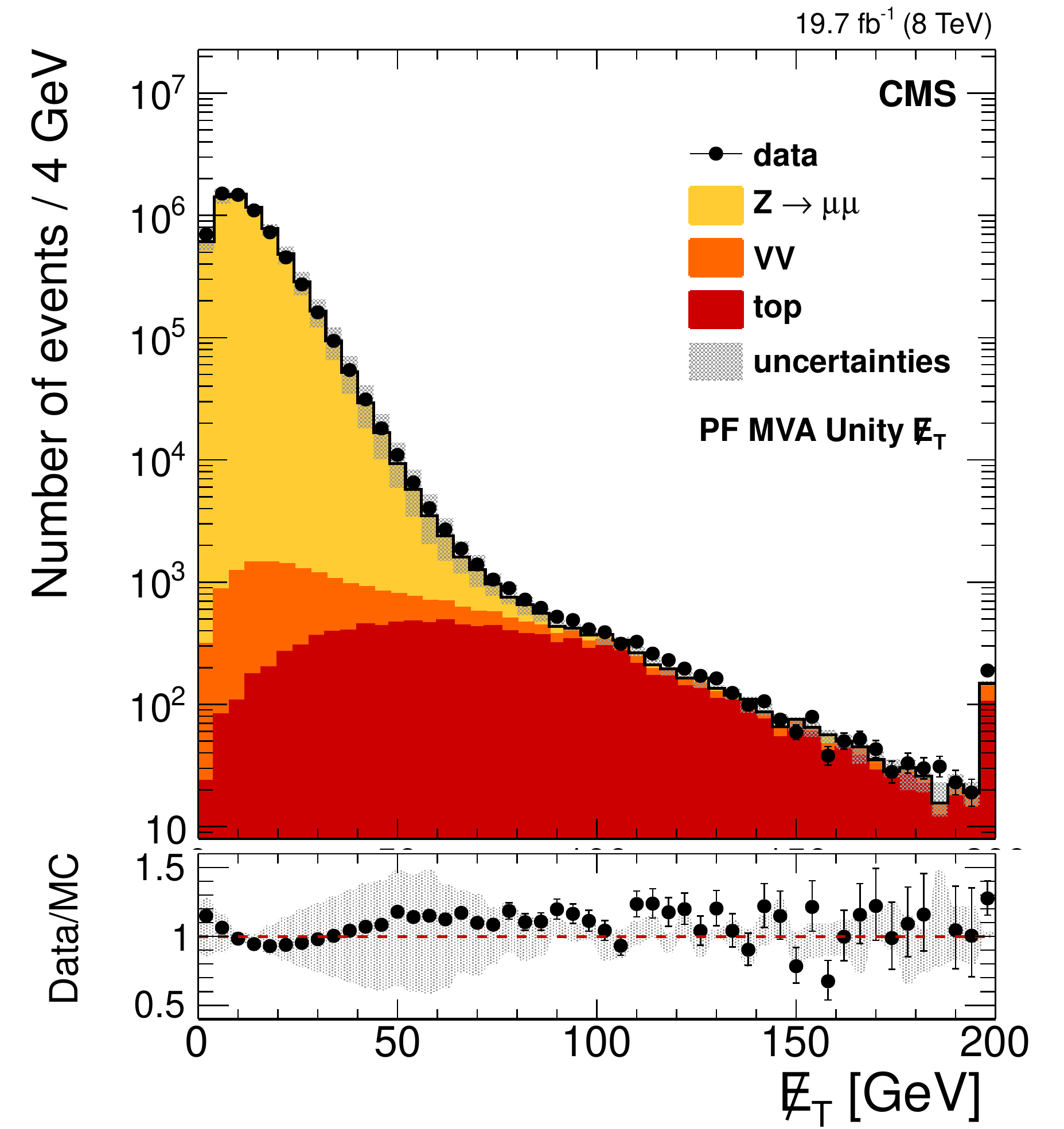}
  \includegraphics[width=0.32\textwidth]{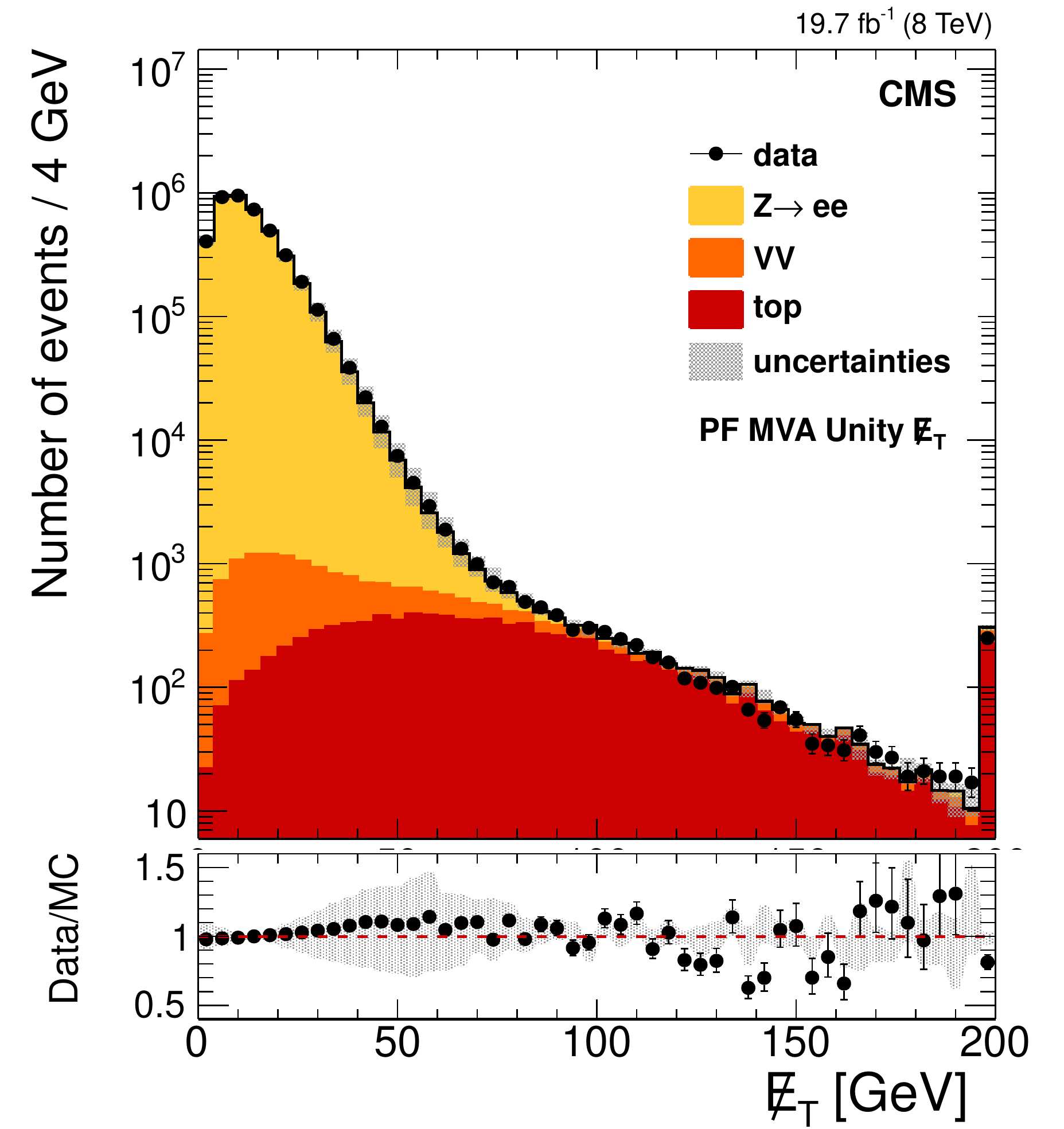}
  \includegraphics[width=0.32\textwidth]{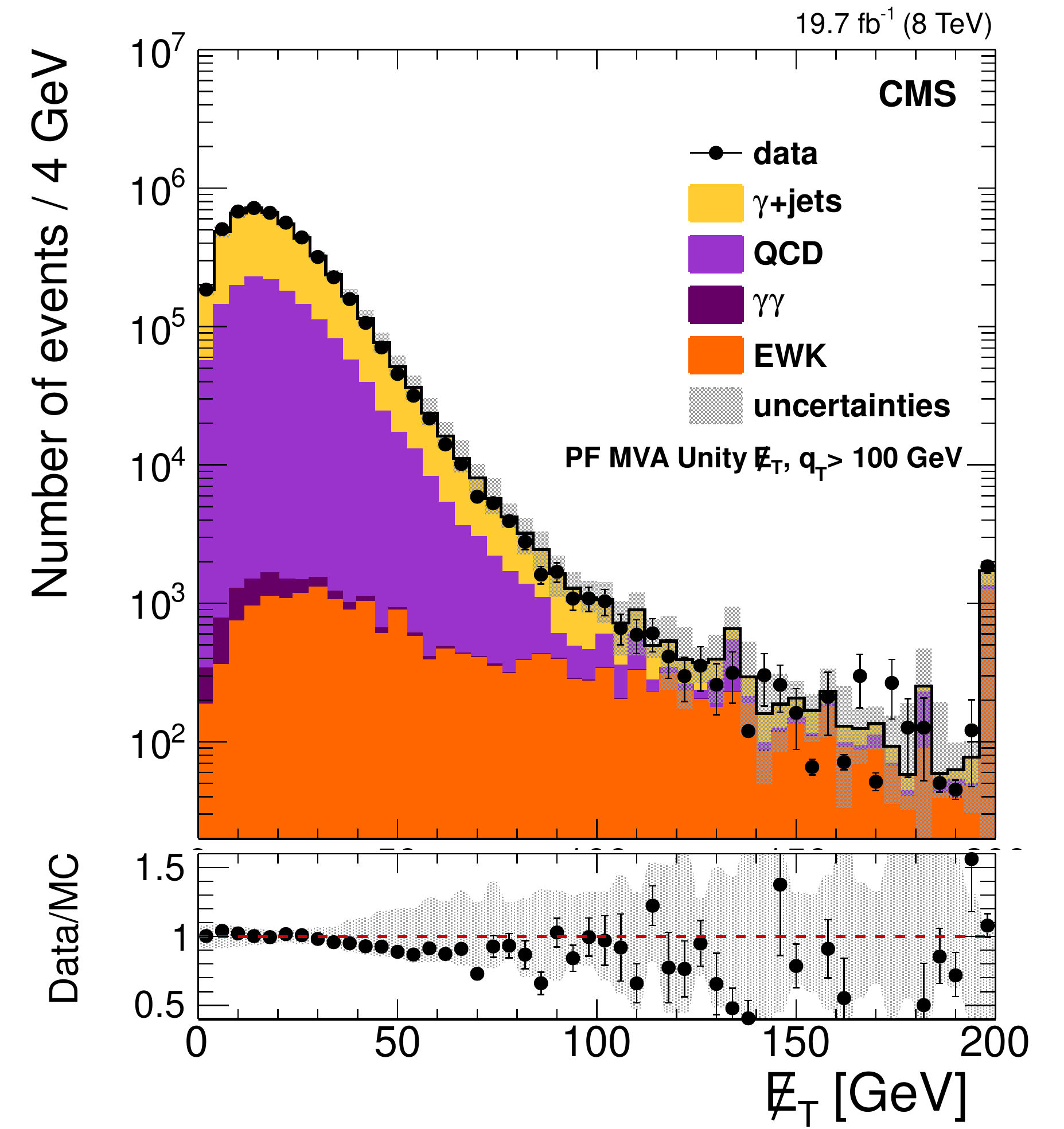}
  \caption{
    MVA Unity \pfvecmet distributions in \Zmm (left), \Zee\ (middle), and \GJ\ (right) events.
    The points in the lower panel of each plot show the data/MC ratio, including the
    statistical uncertainties of both data and simulation;
    the grey error band displays the systematic uncertainty of the simulation. The last bin contains the overflow content.
  }
  \label{fig:MVAMETUnity}
\end{figure}

The No--PU, MVA \pfvecmet, and MVA Unity \pfvecmet distributions for \Zmm,
\Zee, and \GJ\ events are shown in
Figs.~\ref{fig:NoPU1}, \ref{fig:MVAMET}, and \ref{fig:MVAMETUnity},
respectively. Simulation and data are in agreement within the uncertainties.

Some difference between data and simulation can be seen in the region
$\met \leq 70$\GeV. The systematic uncertainty in this region is sizeable,
and is dominated by the uncertainty in the JER~\cite{JETJINST}.
It is found that the JER in simulated events are overestimated by 5\% (up to 20\%) for jets
reconstructed within (outside) the geometric acceptance of the tracking
detectors.  The effect is accounted for by smearing the momenta of jets
in simulated events by the measured difference in the JER.  The uncertainty
on the correction is of a similar size as the correction.  The
difference between data and simulation in the $\met$ distribution is
covered by the present JER uncertainty within one standard deviation.

\subsection{Measurement of No--PU and MVA PF \texorpdfstring{\bigmet}{MET} scale and resolution}
\label{sec:nopu_mva_pfvecmet_performance}

\begin{figure}[t!hb]
  \centering
  \includegraphics[width=0.42\textwidth]{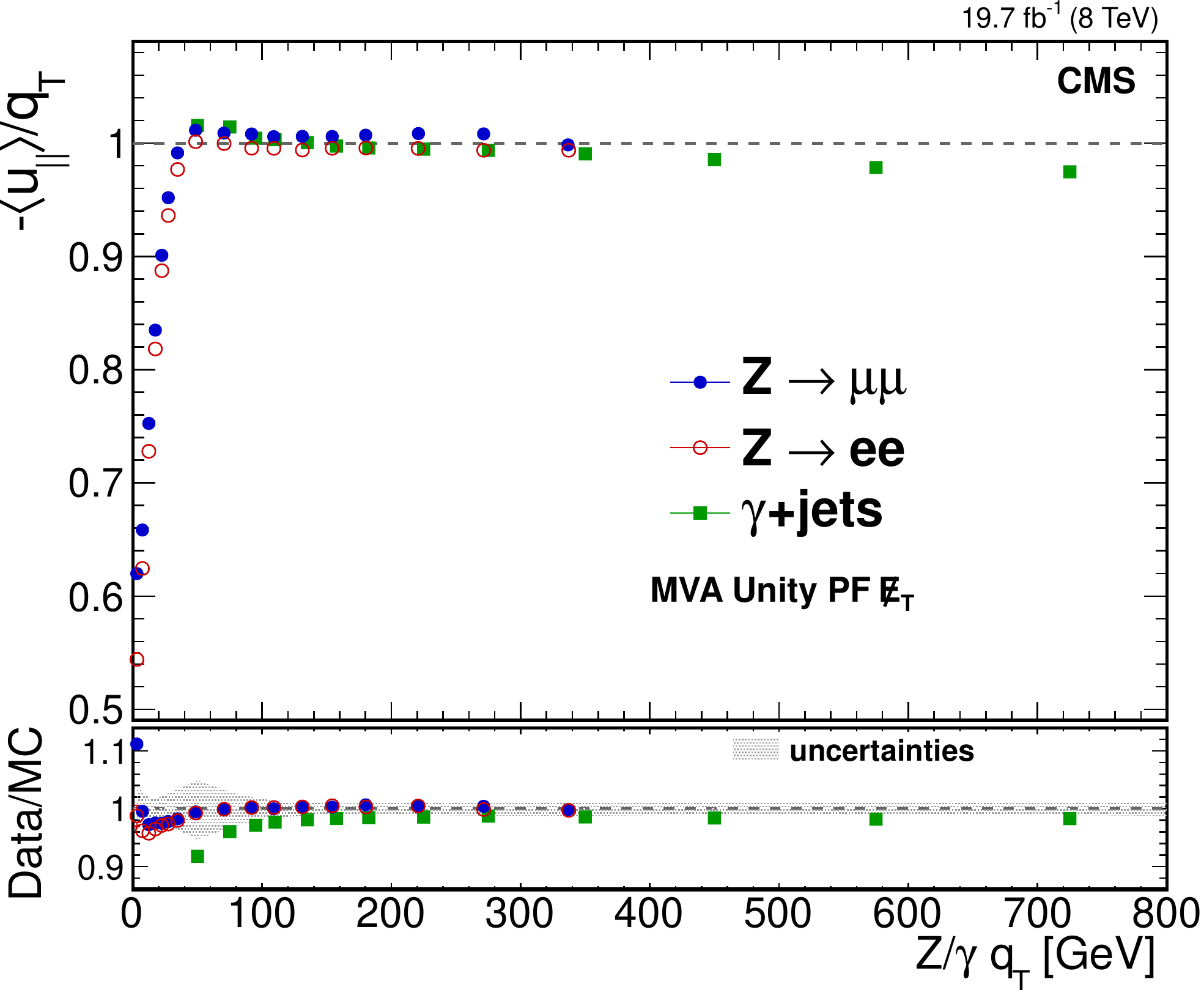}
  \includegraphics[width=0.42\textwidth]{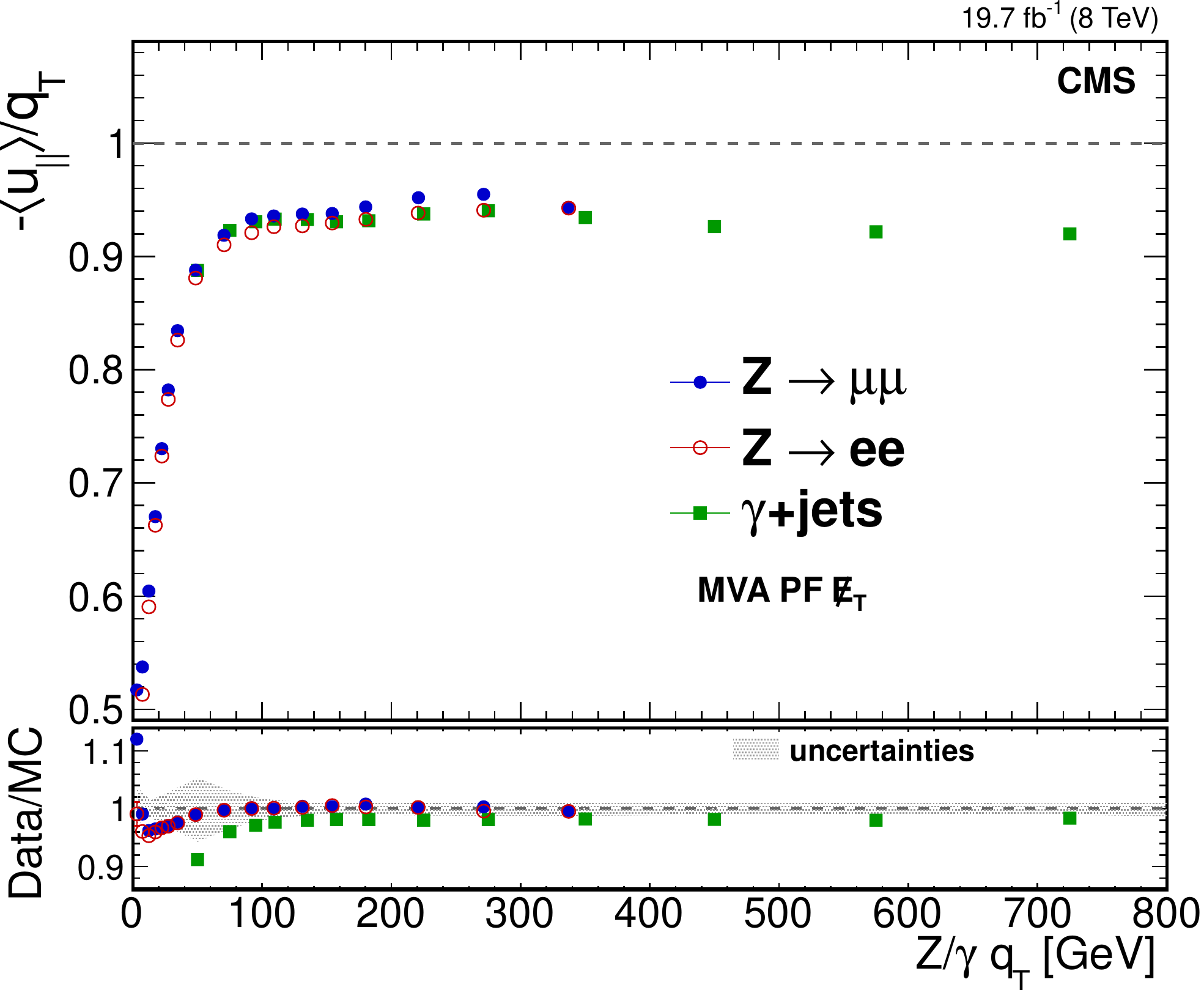}\\
  \includegraphics[width=0.42\textwidth]{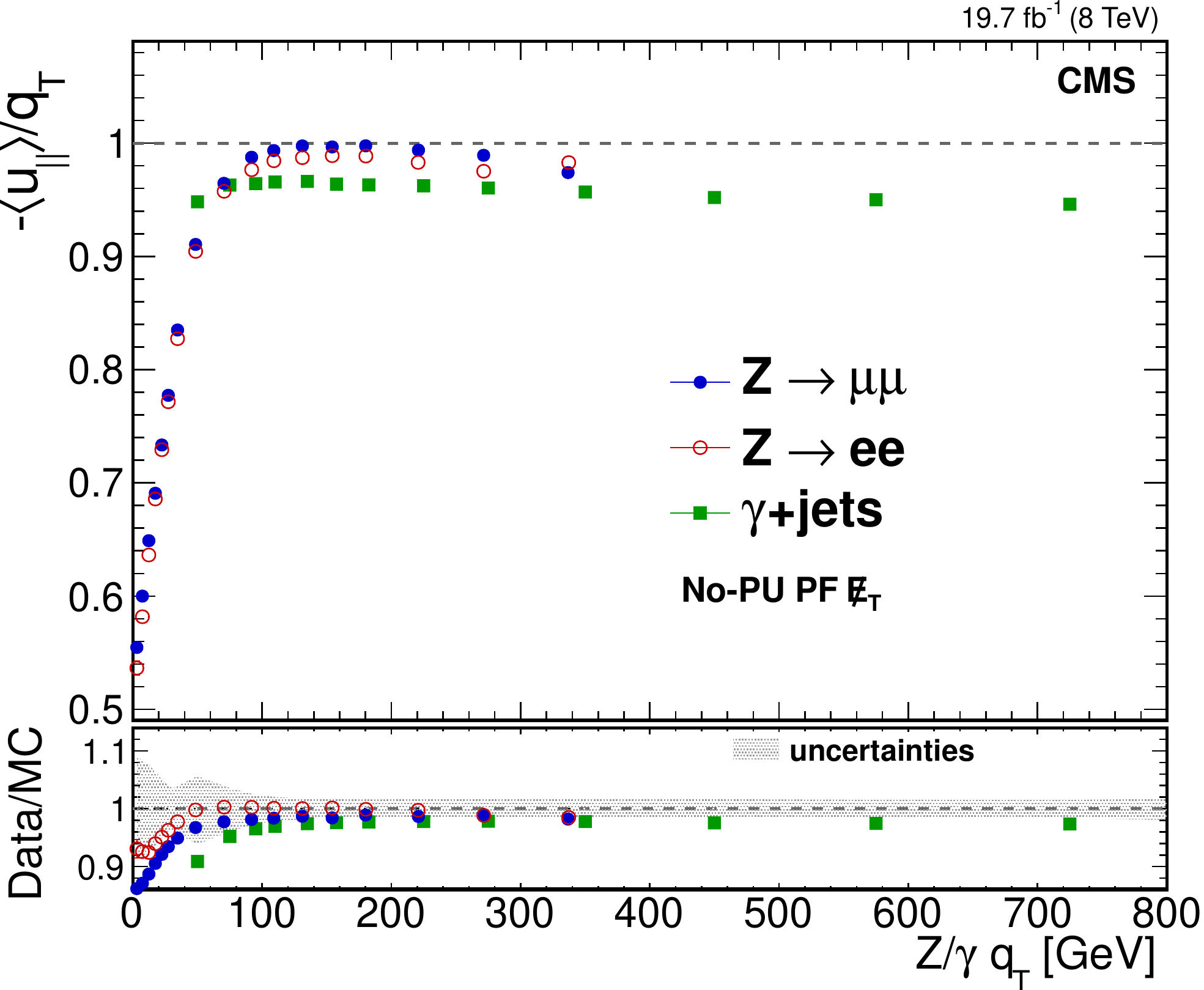}
  \caption{
    Response curves for MVA Unity \pfvecmet (left top),
    MVA \pfvecmet (right top), and No--PU \pfvecmet (bottom),
    in  $\Zmm$ events (full blue circles), $\Zee$ events (open red circles), and direct-photon events (full green squares).
    The upper frame of each figure shows the response in data;
    the lower frame shows the ratio of data to simulation with
    the grey error band displaying the systematic uncertainty of the simulation, estimated as the maximum of each channel systematic uncertainty.
  }
  \label{fig:resp_nopu_mva_qt}
\end{figure}

The response curves of the No--PU \pfvecmet, MVA \pfvecmet, and MVA Unity
\pfvecmet algorithms for \Zmm, \Zee, and \GJ events are shown in
Figure~\ref{fig:resp_nopu_mva_qt}.  Data and simulated distributions
show good agreement, except at the lowest \qt where the recoil
direction is not well defined and becomes sensitive to small
discrepancies in the simulation of low \pt particles. The No--PU \pfvecmet response approaches unity
slower than the standard \pfvecmet (Fig. \ref{fig:CombResponse}) for \Zmm and \Zee events. This is due to events in which a
sizeable fraction of particles originating from the hard scatter
interaction do not carry an electric charge. The response stays below unity for \GJ events. The parameter $\beta = 0.6$ in
Eq.~\eqref{eq:NoPUMEtComputation} has been optimized to yield the best
\vecmet resolution.  Its effect is that the contribution of neutral
particles to the \pfvecmet
computation, which are difficult to separate into distinct contributions
from the hard scatter interaction and pileup, is underestimated by 40\% on average. The MVA \pfvecmet
response is around 0.9 even at high \qt, since the BDT is
trained to achieve the best \vecmet resolution, even if at the expense
of worse response. In contrast, the MVA Unity \pfvecmet reaches a unity
response, due to the dedicated training to achieve the best resolution
given the condition of having unity response.

One conclusion of our studies is that there is a general conflict of
objectives between achieving the best \pfvecmet resolution and reaching
a response close to unity. In order to make the resolution
insensitive to pileup, one needs to scale down the contribution to the
\pfvecmet computation of ``unclustered'' particles and low-\pt jets, both of which
are abundantly produced in minimum bias interactions.  This
procedure inevitably reduces the response at low \qt.

\begin{figure}[htb]
  \centering
  \includegraphics[width=0.42\textwidth]{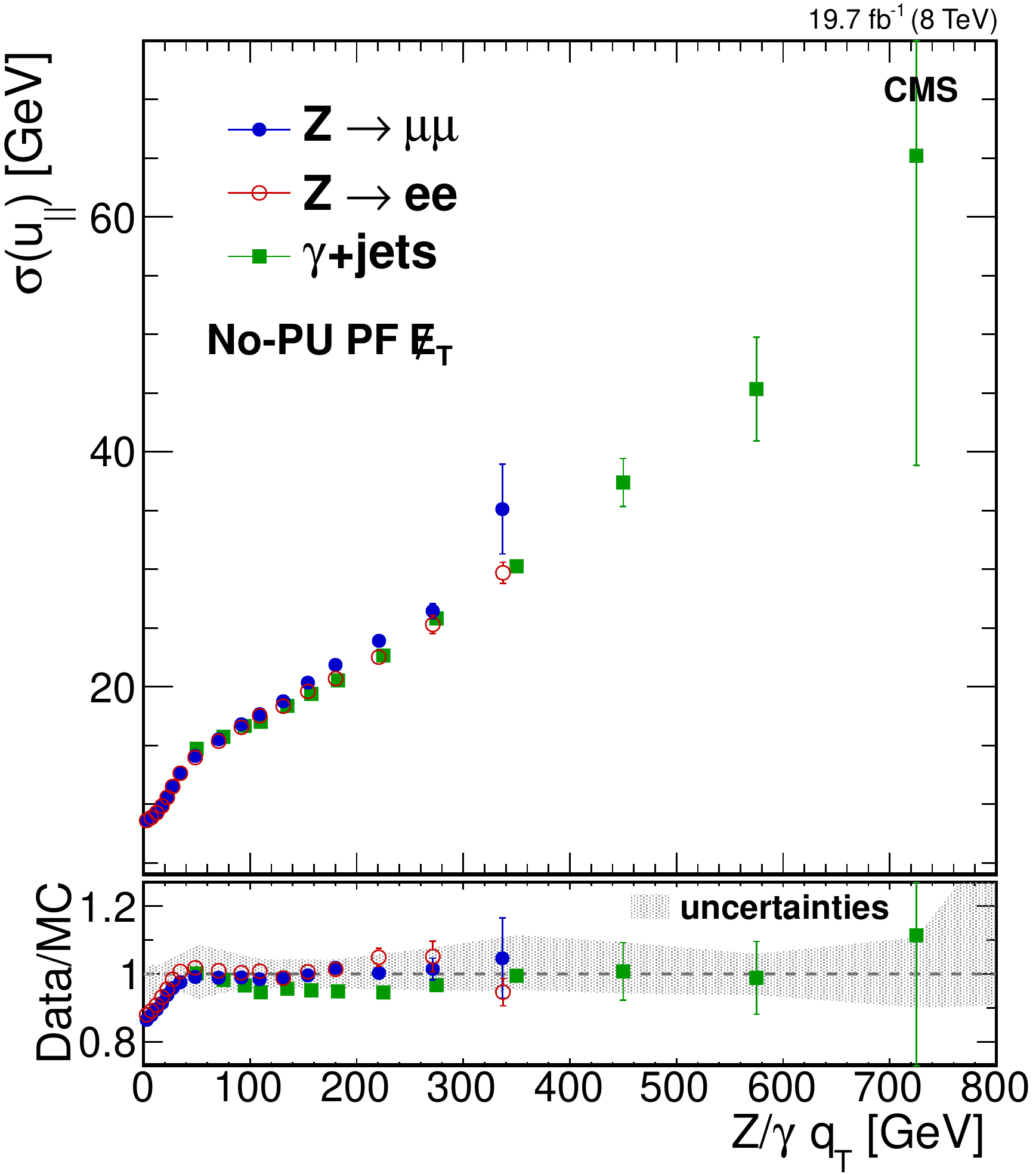}
  \includegraphics[width=0.42\textwidth]{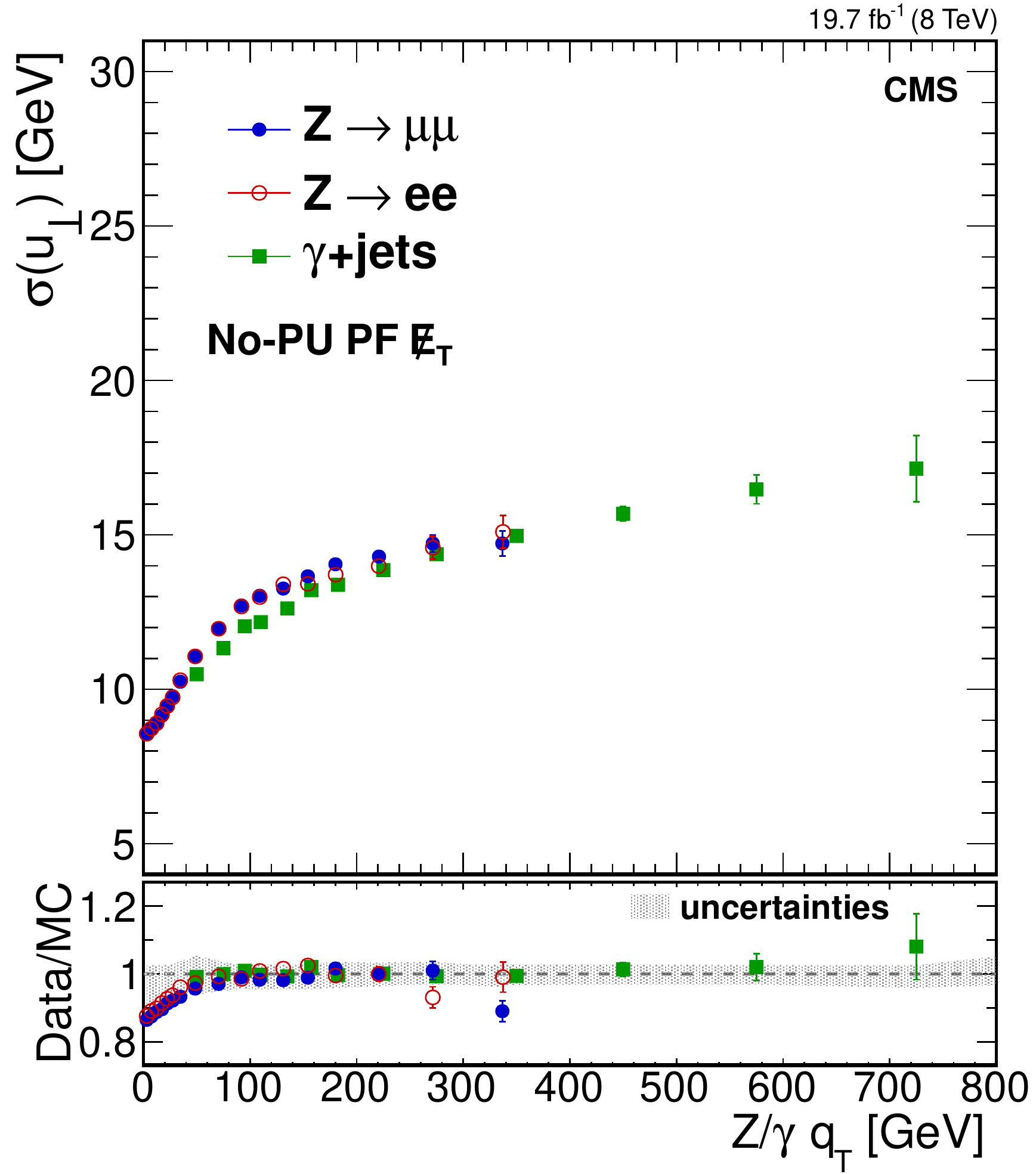}
  \caption{
    Resolution of the parallel (left) and perpendicular (right) recoil component
    as a function of $\qt$ for the No-PU \pfvecmet\ in $\Zmm$ events (full blue circles),
    $\Zee$ events (open red circles), and direct-photon events (full green squares).
    The upper frame of each figure shows the resolution in data;
    the lower frame shows the ratio of data to simulation with
    the grey error band displaying the systematic uncertainty of the simulation, estimated as the maximum of each channel systematic uncertainty.
  }
  \label{fig:reso_nopu_qt}
\end{figure}

\begin{figure}[htb]
  \centering
  \includegraphics[width=0.42\textwidth]{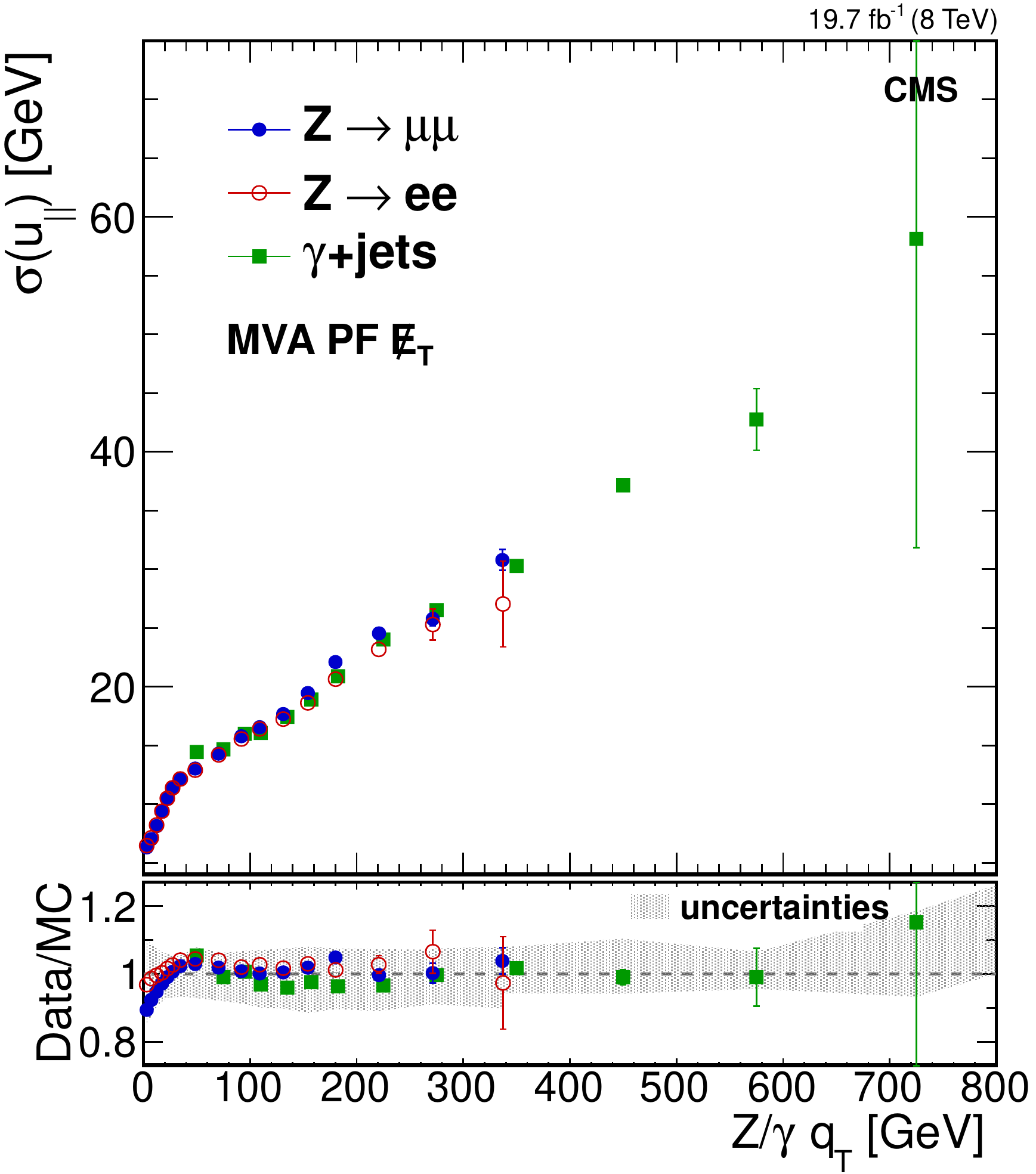}
  \includegraphics[width=0.42\textwidth]{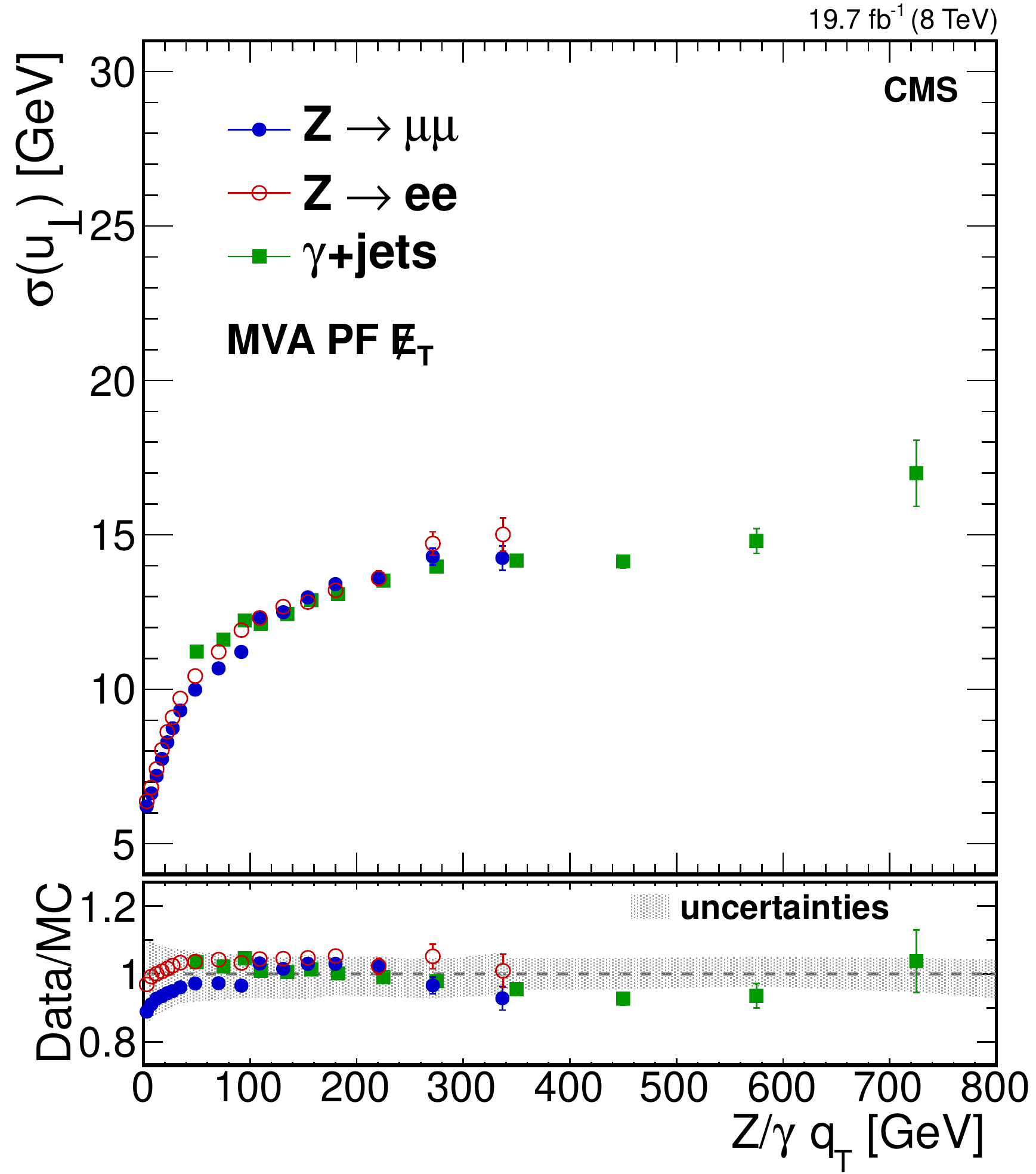}
  \caption{
    Resolution of the parallel (left) and perpendicular (right) recoil component
    as a function of $\qt$ for the MVA \pfvecmet in $\Zmm$ events (full blue circles),
    $\Zee$ events (open red circles), and direct-photon events (full green squares).
    The upper frame of each figure shows the resolution in data;
    the lower frame shows the ratio of data to simulation with
    the grey error band displaying the systematic uncertainty of the simulation, estimated as the maximum of each channel systematic uncertainty.
  }
    \label{fig:reso_mva_qt}
\end{figure}

\begin{figure}[htb]
  \centering
  \includegraphics[width=0.42\textwidth]{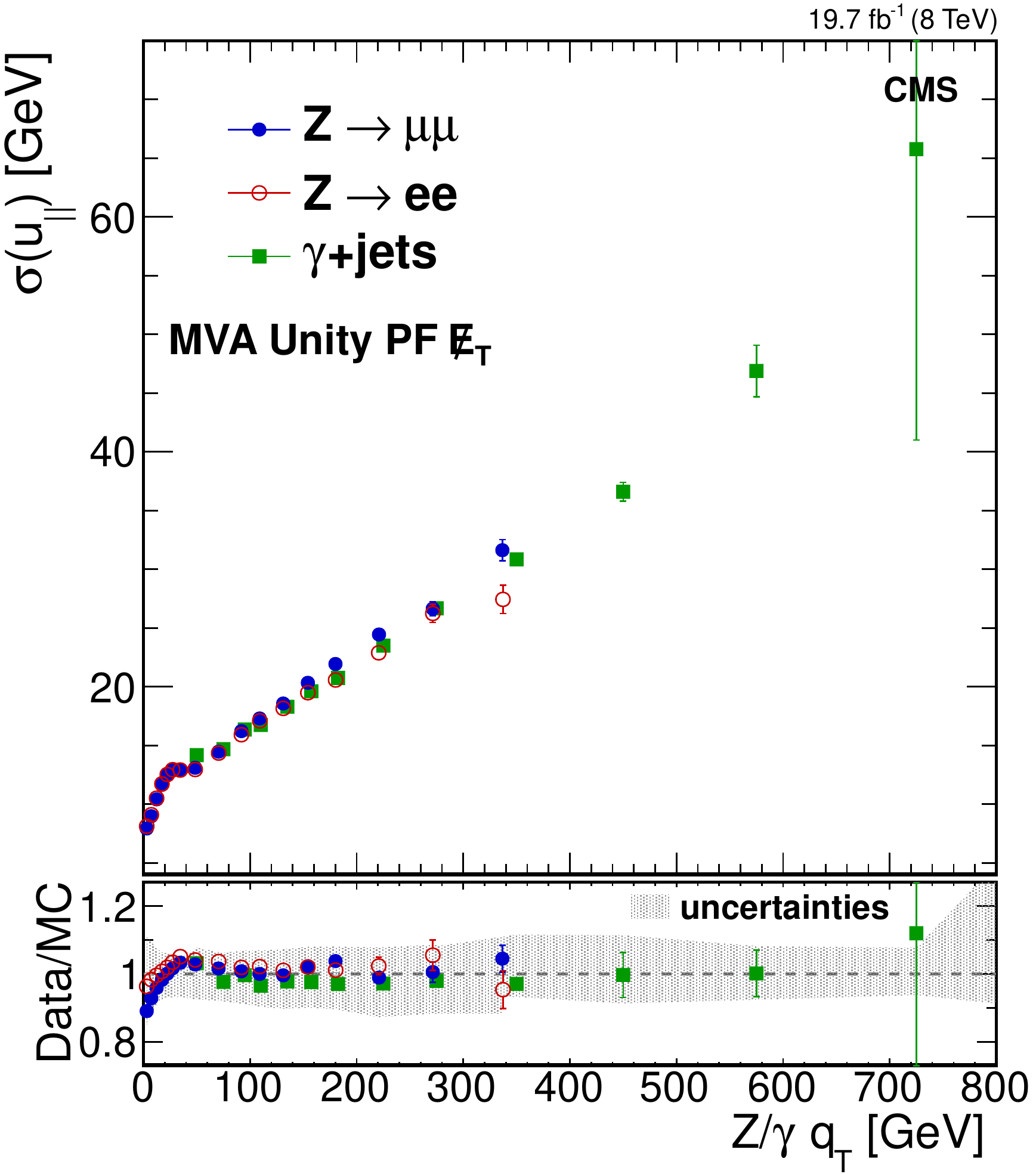}
  \includegraphics[width=0.42\textwidth]{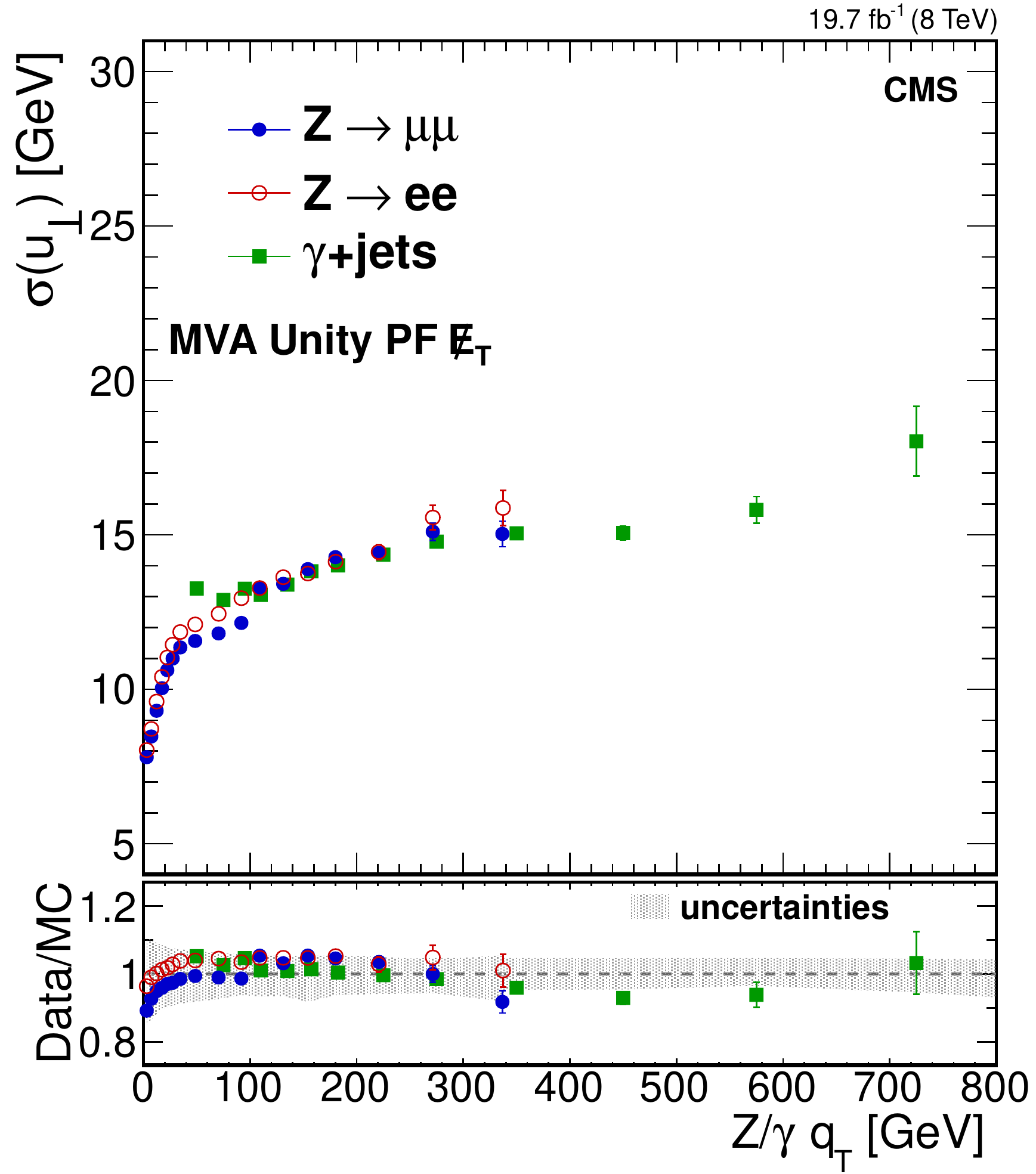}
  \caption{
    Resolution of the parallel  (left) and perpendicular (right) recoil component
    as a function of $\qt$ for the MVA Unity \pfvecmet in $\Zmm$ events (full blue circles),
    $\Zee$ events (open red circles), and direct-photon events (full green squares).
    The upper frame of each figure shows the resolution in data;
    the lower frame shows the ratio of data to simulation with
    the grey error band displaying the systematic uncertainty of the simulation, estimated as the maximum of each channel systematic uncertainty.
  }
    \label{fig:reso_mvaunity_qt}
\end{figure}

\begin{figure}[htb]
  \centering
  \includegraphics[width=0.32\textwidth]{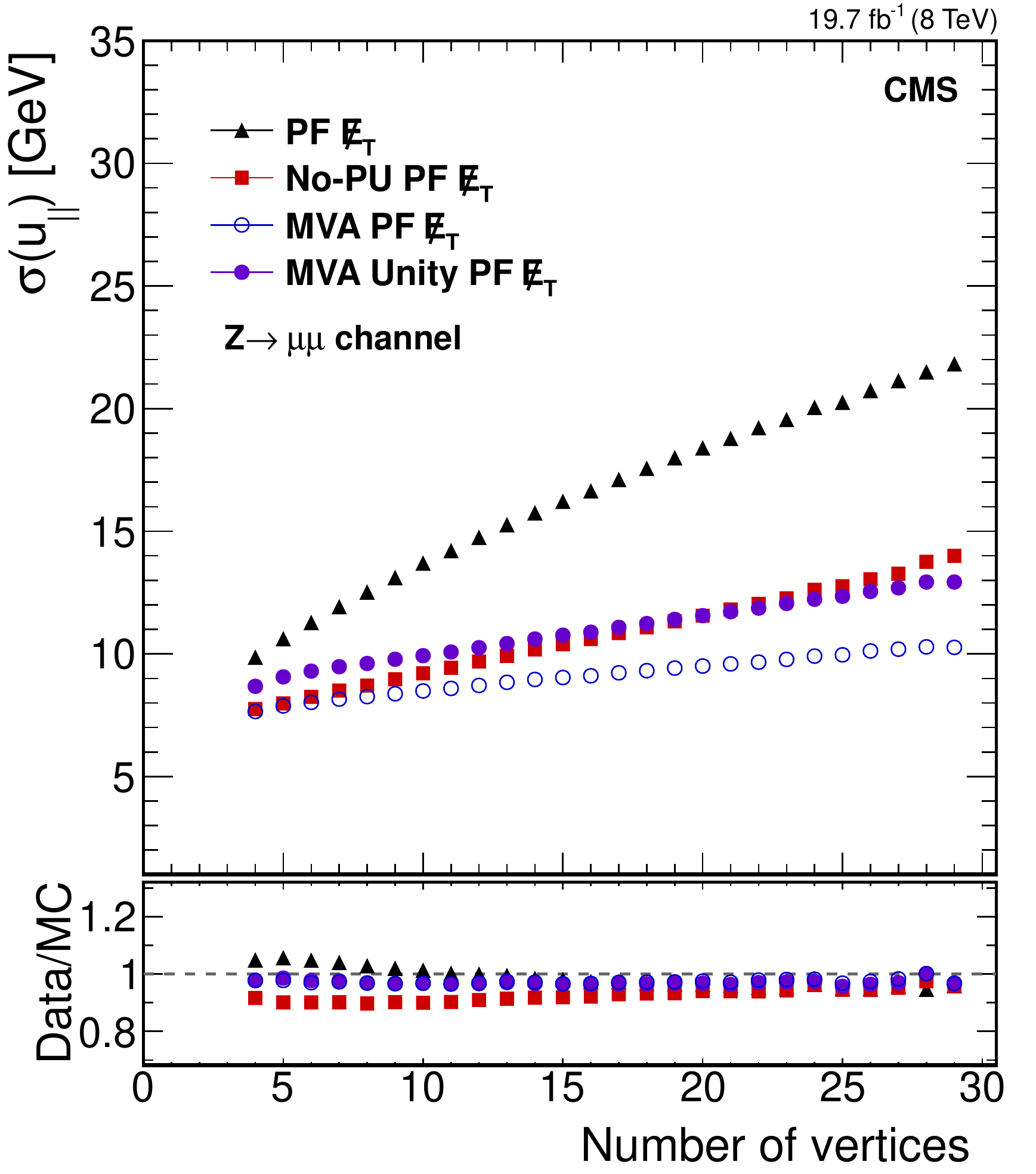}
  \includegraphics[width=0.32\textwidth]{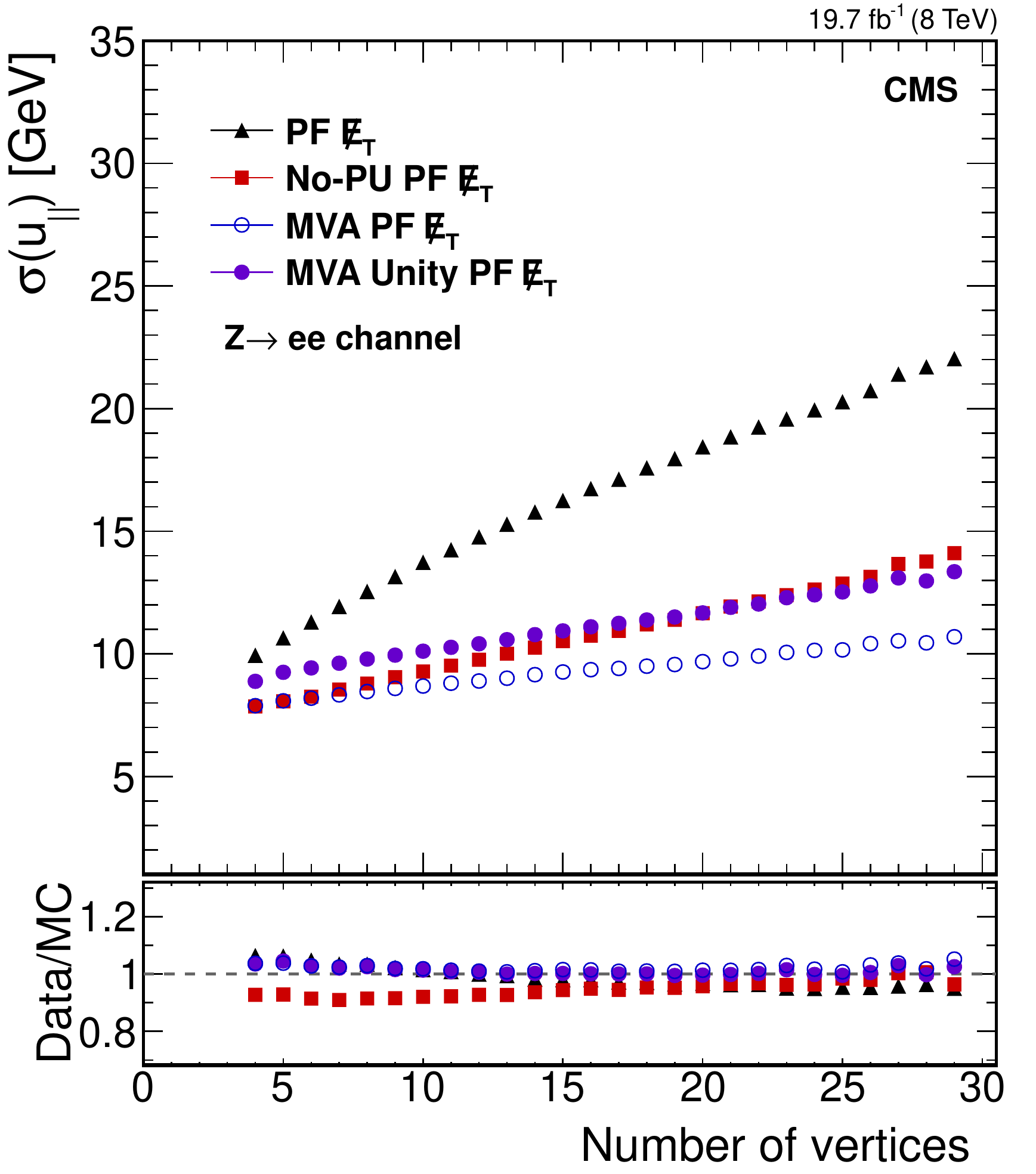}
  \includegraphics[width=0.32\textwidth]{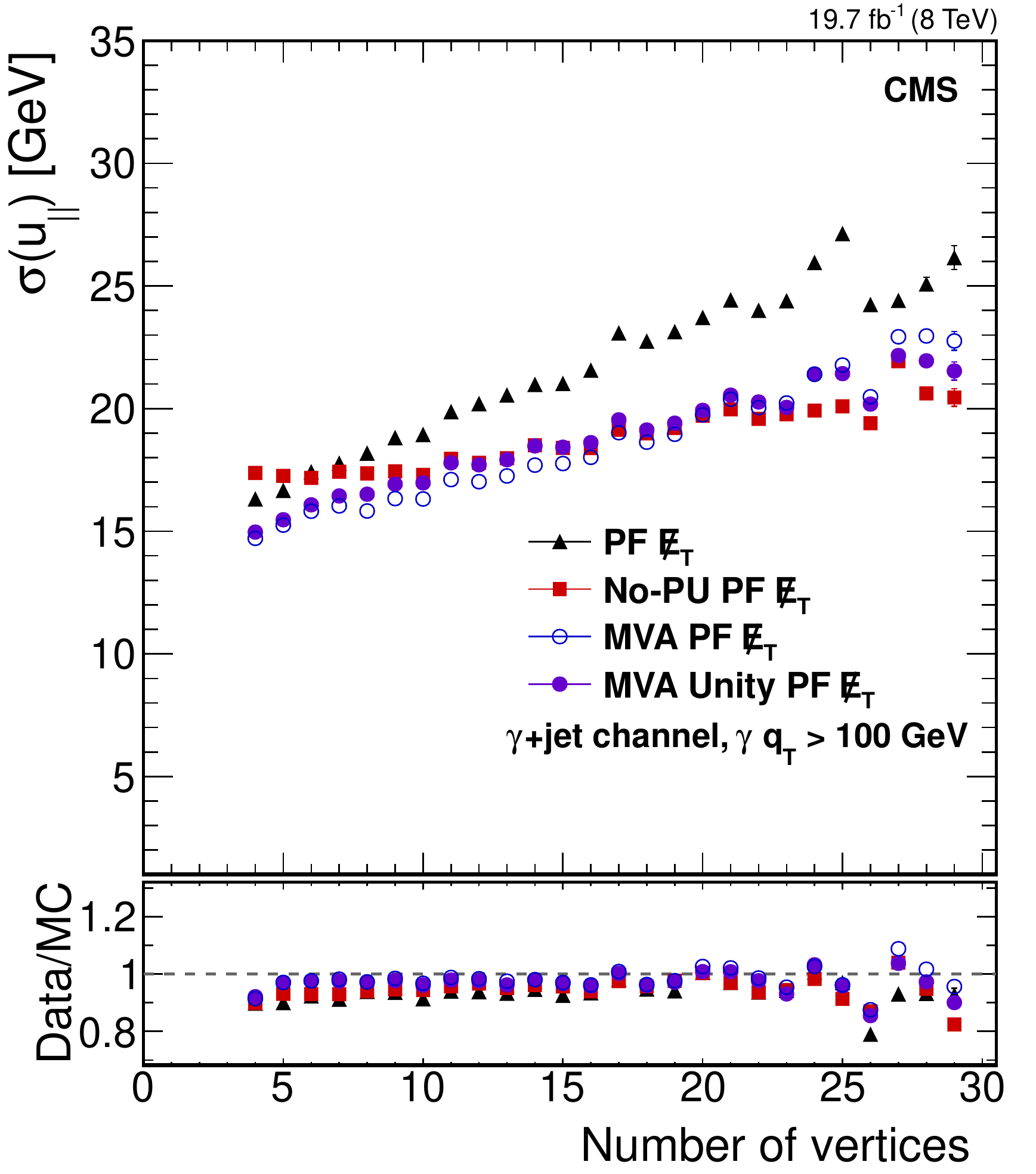} \\
  \includegraphics[width=0.32\textwidth]{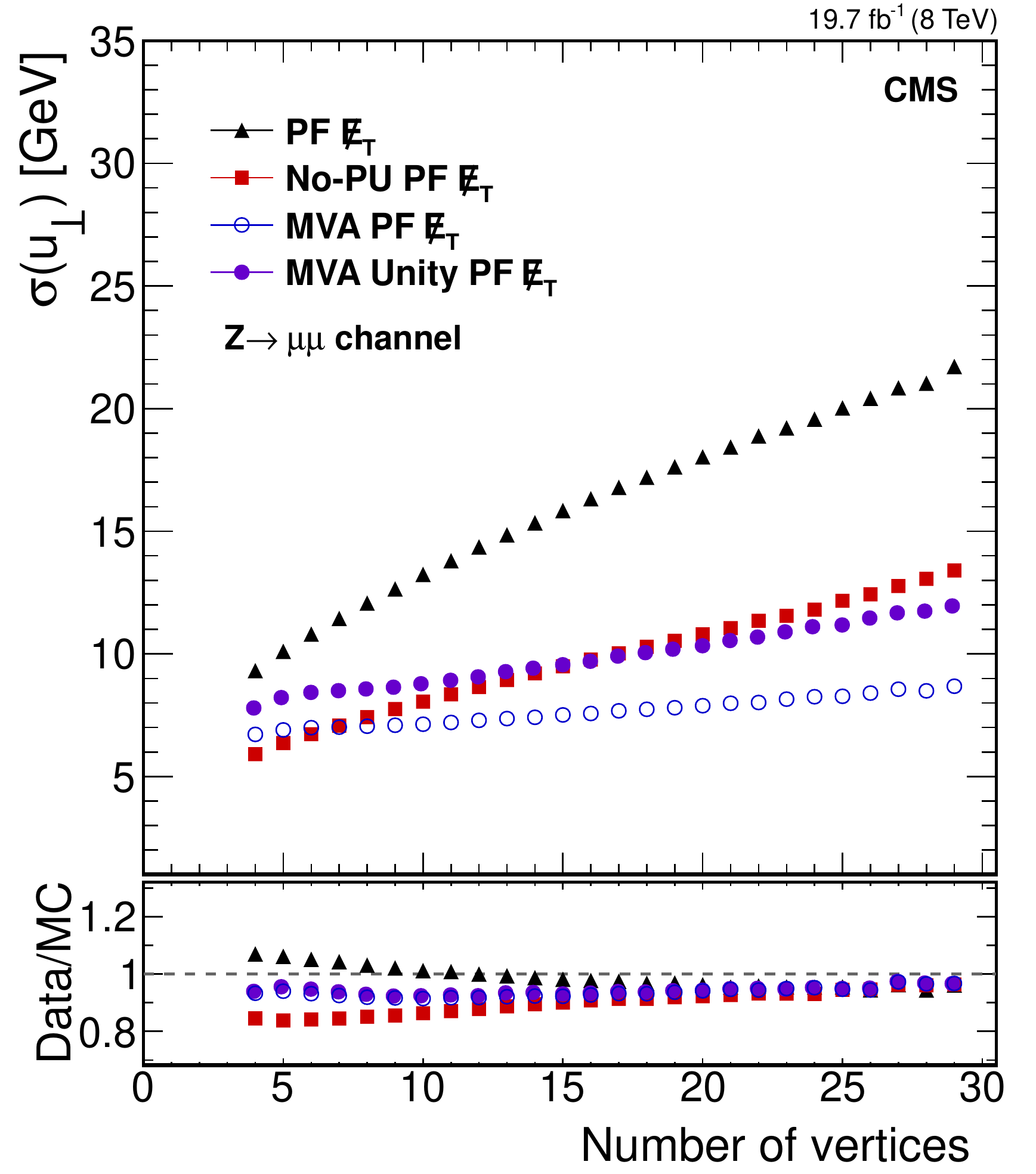}
  \includegraphics[width=0.32\textwidth]{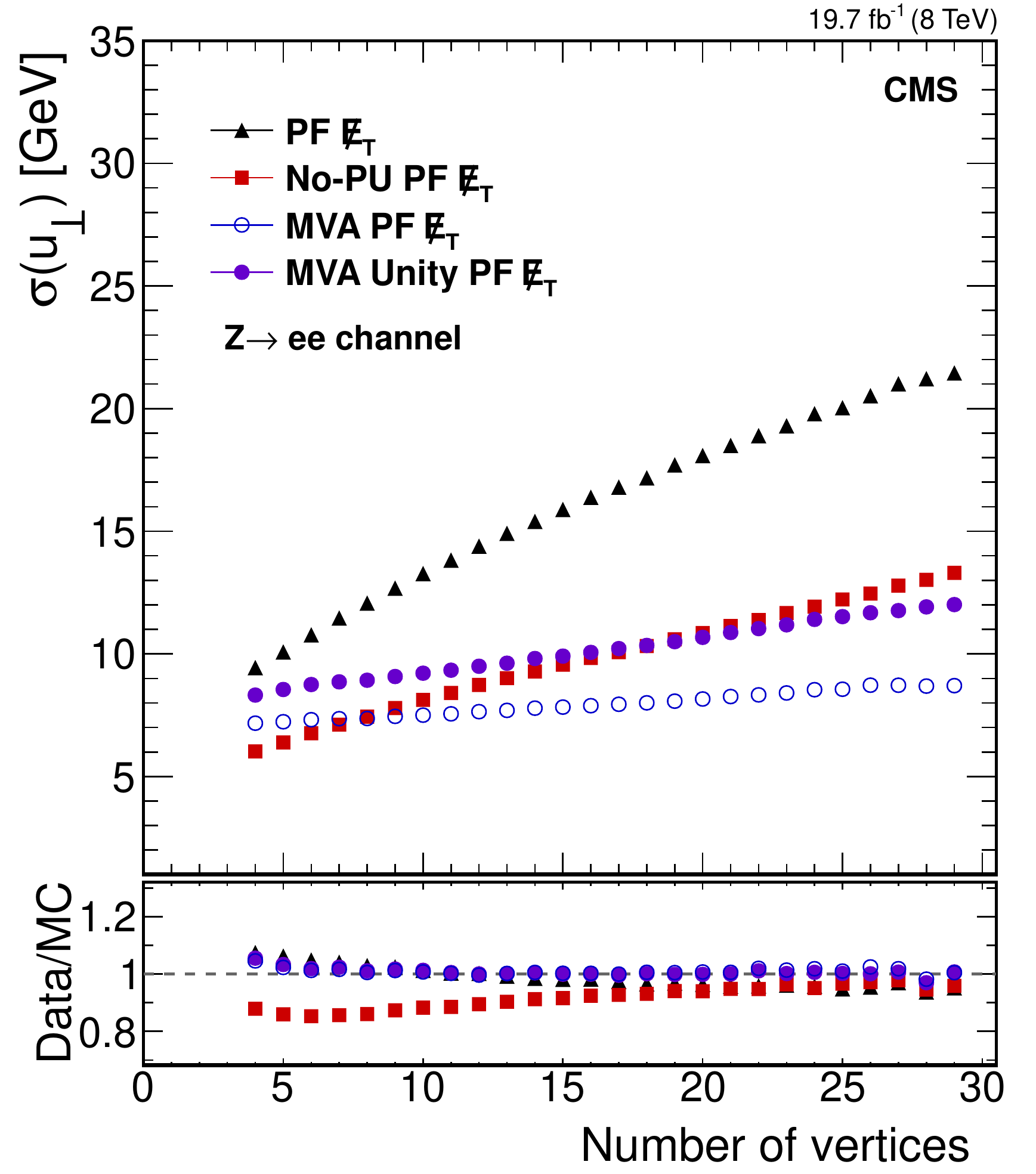}
  \includegraphics[width=0.32\textwidth]{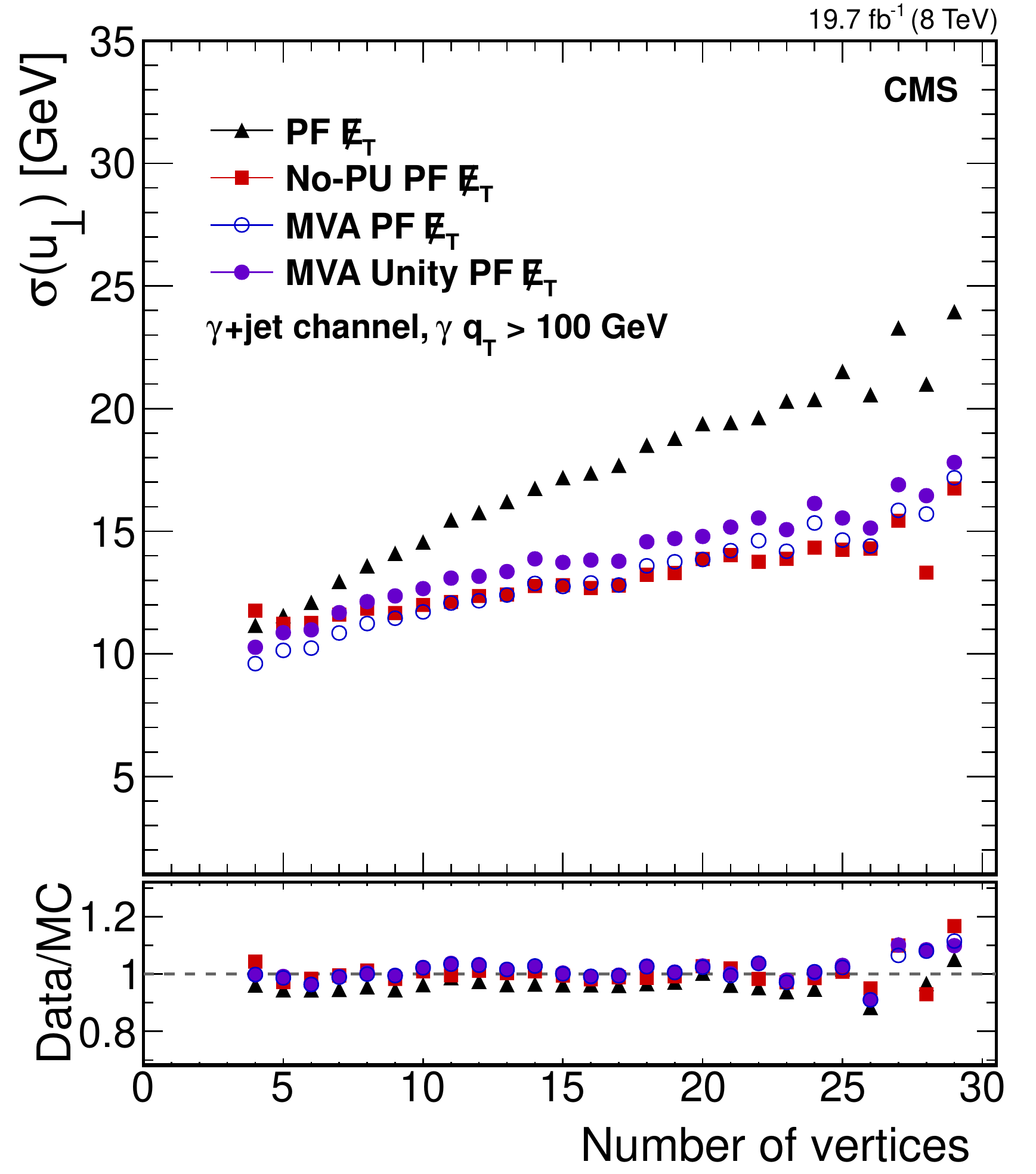}
  \caption{
    Parallel (top) and perpendicular (bottom) recoil component resolution as a
    function of the number of reconstructed vertices for \pfvecmet (black triangles),
    No-PU \pfvecmet (red squares), MVA \pfvecmet (blue open circles),and MVA Unity \pfvecmet (violet full circles)
    in \Zmm\ (left), \Zee\ (middle), and \GJ\ events (right).
    The upper frame of each figure shows the resolution in data;
    the lower frame shows the ratio of data to simulation.
    The \cPZ\ and direct-photon sample curves differ as the photon events are required to satisfy $\qt > 100\GeV$.}
  \label{fig:reso_nvtx}
\end{figure}

The resolution versus boson \qt of the \uperp{} and \upara{} components
are shown in Figs.~\ref{fig:reso_nopu_qt}--\ref{fig:reso_mvaunity_qt}
for the No--PU, MVA, and MVA Unity \pfvecmet. Good agreement is observed
between data and simulation for various algorithms, and between various
channels.  The resolution distributions as a function of $N_\text{vtx}$
are shown in Fig.~\ref{fig:reso_nvtx} and include also, as a reference,
the standard \pfvecmet algorithm shown in Fig.~\ref{fig:CombResolutionRMS}, fully corrected as described in
Section~\ref{sec:metreco}.  The No-PU \pfvecmet and particularly MVA and
MVA Unity \pfvecmet show a significantly reduced dependence of the
resolution on pileup interactions in both data and simulation.  This
reduced pileup dependence can significantly increase the sensitivity of
searches for new physics.  As an example, use of the MVA \pfvecmet
improved the sensitivity of the search for the Higgs boson decaying into
tau-lepton pairs by $\sim$20\% with respect to the
\pfvecmet~\cite{Chatrchyan:2014nva}.

\section{The \texorpdfstring{\bigmet}{MET} significance}
\label{sec:significance}

The ability to distinguish between events with spurious \vecmet and those
with genuine \vecmet is important for  analyses using missing transverse
energy variables. Spurious \vecmet may arise from object
misreconstruction, finite detector resolution, or detector noise. To help
identify such events, we have developed a missing transverse energy
significance variable, which we will denote by ``\vecmet significance", or
simply \metsig. On an event-by-event basis, \metsig evaluates the p-value
 that the observed \vecmet is
inconsistent with a null hypothesis, $\vecmet=0$, given the full event
composition and resolution functions for each object in the event. A high
value of \metsig is an indication that the \vecmet observed in the event is
not well explained by resolution smearing alone, suggesting that the event
may contain unseen objects such as neutrinos or more exotic
weakly interacting particles. A first version of the \vecmet significance
algorithm has been
described in Ref. \cite{METJINST}.

\subsection{Definition of \texorpdfstring{\metsig}{METSig}}

The significance is defined as the log-likelihood ratio,
\begin{equation}
\metsig \equiv 2\ln\left(\frac
{\like(\vet=\sum\vet_{i})}
{\like(\vet=0)}
\right).
\label{eq:metsig:defn}
\end{equation}
The numerator expresses the likelihood of the hypothesis under test that
the true value (\vet) of the missing transverse energy
is equal to the observed value  ($\sum\vet_i$) ,
while the denominator expresses the likelihood of the {\em null
hypothesis}, that the true missing transverse energy is actually zero.
Under the null hypothesis, observation of any non-zero missing transverse
energy is attributed to resolution smearing.

The formulation in Eq.~\eqref{eq:metsig:defn} is completely general and
accommodates any probability distribution functions for the object
resolutions; throughout the bulk of this discussion however, we assume
Gaussian resolutions for measured quantities. This assumption accurately
describes the dominant behavior of energy and momentum measurements in CMS
and greatly simplifies the computation of \metsig as the convolution
integrals underlying the likelihood functions can be done
analytically.  In the Gaussian model, we obtain a simple closed-form solution,
\begin{linenomath}
\begin{equation}
\metsig = \Big(\sum\vet_i \Big)\!^\dag \mathbf{V}^{-1} \Big( \sum\vet_i \Big),
\label{e:metsig-gaussian}
\end{equation}
\end{linenomath}
in which $\mathbf{V}$ is the $2{\times}2$ covariance matrix of the total missing
transverse energy computed by propagating the uncertainties of all objects
in the event or in a defined subset of the event; more details are given in
Ref. \citen{METJINST}. A particularly useful feature of the Gaussian
approximation is that the \metsig, as defined by Eq.~\eqref{e:metsig-gaussian}, is a
$\chi^2$ variable with two degrees of freedom
(one degree of freedom for each component of \vecmet).
For clarity, we note that the term ``significance'' is often used to denote
a linear quantity of the form $x/\sigma_x$ while here it is defined as the
quadratic form $x^2/\sigma^2_x$.

Despite the convenience of Eq.~\eqref{e:metsig-gaussian}, a full treatment of
\vecmet significance must also include non-Gaussian resolutions as these are
known to occur at the percent level in jet measurements. In Section
\ref{s:nongaussian} of this paper we therefore extend the treatment of
\metsig to handle such cases.

\subsection{Jet resolutions}
The \vecmet resolution captured in the covariance matrix $\mathbf{V}$ of
Eq.~\eqref{e:metsig-gaussian} is determined mainly by the momentum resolution of the
hadronic components of the event. For the purpose of \vecmet significance we
separate the hadronic activity into jets with $\pt\ge20\GeV$, which are
reconstructed with the
PF algorithm, and unclustered energy with
$\pt<20\GeV$. The jets are treated as individual objects, each with a
unique resolution function depending on the \pt and $\eta$ of the jet,
while the objects in the unclustered energy are summed vectorially to
produce a single object with $\vpt=\sum_i\vpt^i$, whose resolution is
determined separately. This division separates those
components of the event that carry strong azimuthal information and
contribute distinctively to the topology of the event from those that are
relatively featureless and contribute only to a general broadening of the \vecmet
resolution. Subsequent results are not sensitive to the choice of the 20\GeV
threshold.

The resolution functions of hadronic jets are parametrized with a
Crystal Ball function, which has a core
Gaussian function with additional power-law terms that describe small non-Gaussian
tails~\cite{CrystalBall}. The parameter values are determined initially with samples of
QCD multijet events generated by \PYTHIA v6.4.24 \cite{pythia}, with jets propagated
through the full simulation of the CMS detector; the
reconstructed and generated values of \pt, $\eta$, and $\phi$ are compared
to extract resolution shapes.
A full description of a single jet's Gaussian core resolution is given
by the covariance matrix,
\begin{linenomath}
\begin{equation}
\mathbf{U} = \left( \begin{matrix}
   \sigma_{\pt}^2 & 0 \\
   0 & \pt^2\,\sigma_{\phi}^2
   \end{matrix} \right),
\label{eq:cov}
\end{equation}
\end{linenomath}
in which we assume no correlation between \pt and $\phi$ terms. Both
$\sigma_{\pt}$ and $\sigma_\phi$ are functions of both \pt and $\eta$.
As written, the covariance matrix $\mathbf{U}$ is in the coordinate system
aligned with the jet; in use, all such matrices are rotated by the jet
azimuthal angle $\phi$ into the common CMS
$xy$ basis: $\mathbf{V}={\mathbf R}(\phi) \mathbf{U}\,\mathbf{R}^{-1}(\phi)$.

The widths of the core Gaussian functions obtained from simulation as described above are
retuned with data using the \Zmm\ control sample defined in
Section~\ref{subsec:Zsample}. This is effectively a zero-\met sample
and the observed \vecmet is therefore
expected to derive primarily from jet resolution smearing rather than from
genuine \vecmet. In this sample, jet activity is modest and the \vecmet
characteristics are dominated by the largely isotropic features of the
unclustered energy. The \vecmet significance therefore conforms well to the null hypothesis,
and we use this fact to optimize the Gaussian widths.
Each Gaussian width, $\sigma^{\mathrm{MC}}$, obtained from simulation
is rescaled by an $\eta$-dependent
correction factor: $\sigma(\eta)=a(\eta)\times\sigma^{\text{MC}}$;
the correction factors (in five bins of $\abs{\eta}$) are
determined by a likelihood fit over the \Zmm\ data sample
in which we seek to maximize the null
hypothesis, $\like(\vet = 0)$.  To reduce possible biases stemming from
events with sources of genuine \vecmet, the fit is performed iteratively
with a restriction to exclude high-significance events.

The unclustered energy resolution, $\sigma_{\mathrm{uc}}$, is parametrized by,
\begin{linenomath}
\begin{equation}
  \sigma_{uc}^2 = \SigmaZ^2 + \SigmaS^2\sum_{i=1}^{n}{\abs{\vec{p}_{T_i}}},
  \label{e:pseudo}
\end{equation}
\end{linenomath}
where the summation is over the $n$ low-\pt objects included in the unclustered energy
and $\SigmaZ$ and $\SigmaS$ are free parameters obtained from the same likelihood fit
as described above. Because the best fit normally returns $\SigmaZ=0$ (as one would expect),
we see that the resolution of the unclustered energy
exhibits the general form $\sigma_{\mathrm{uc}}\approx\sqrt{n}\,\sigma_{\XX}$ where the quantity $\sigma^2_\XX$
measures the average contribution of low-\pt objects to the \vecmet covariance.
Its contribution to the \vecmet covariance matrix is taken to be
isotropic,
\begin{linenomath}
\begin{equation}
\mathbf{V}_{uc} = \left( \begin{matrix}
   \sigma_{uc}^2 & 0 \\
   0 & \sigma_{uc}^2
   \end{matrix} \right)= n\sigma_{\XX}^2\,\mathbf{I}
\label{e:isotropic}
\end{equation}
\end{linenomath}
as it is constructed from a large
number of (mostly) uncorrelated, low-\pt objects. The matrix $\mathbf{I}$ in
Eq.~\ref{e:isotropic} is the
identity matrix. In practice, a slight
ellipticity due to fluctuations of the unclustered energy is found in some events but can
 be neglected without degrading the \vecmet significance performance. 

Systematic uncertainties associated with hadronic activity are evaluated using
uncertainties on the jet energy scale (2--10\%) and the energy scale of low
energy particles entering into the unclustered energy (10\%), and are
displayed as gray bands in Figs.~\ref{fig:zeromet_sig}--\ref{fig:realmet_pchi2}.  The systematic uncertainty due to
jet energy resolution and unclustered energy resolution is captured here as well.

Electron and muon resolutions are assumed to be negligible when compared to those for the
hadronic activity in each event, and thus do not enter into the \vecmet
covariance.

\subsection{Characteristics of \texorpdfstring{\bigmet}{MET} significance}

\subsubsection{Events with \texorpdfstring{\bigmet}{MET}$=0$}
As \sig\ is \chisq-distributed, an event sample that
nominally has no genuine \vecmet should be flat in
the \chisq\ probability function for two degrees of freedom, $\pchisq$.
Here, $\pchisq$ is defined such that $1-\pchisq$ is the standard cumulative distribution
function of the \chisq\ statistic for two degrees of freedom.
Both \Zmm\ and dijet
samples from pp collisions are dominated by such events.  The dijet
sample is defined in Section \ref{subsec:Dijetsample}; though heavily populated
by events with two high-\pt jets, it is not restricted by any limit on the maximum number of jets.

\begin{figure}[htb]
\centering
\includegraphics[width=0.45\textwidth, height=65mm]{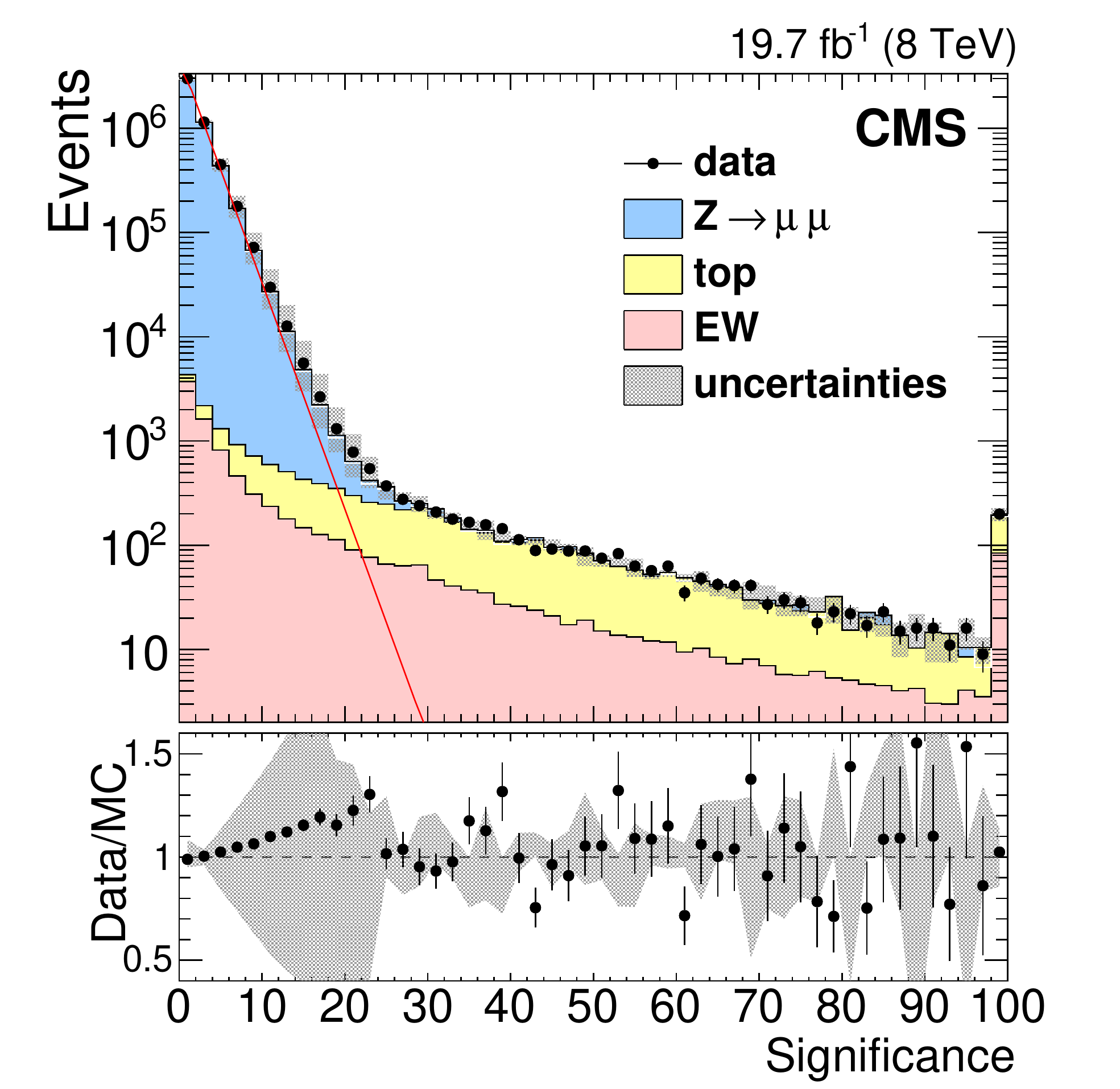}
\includegraphics[width=0.45\textwidth, height=65mm]{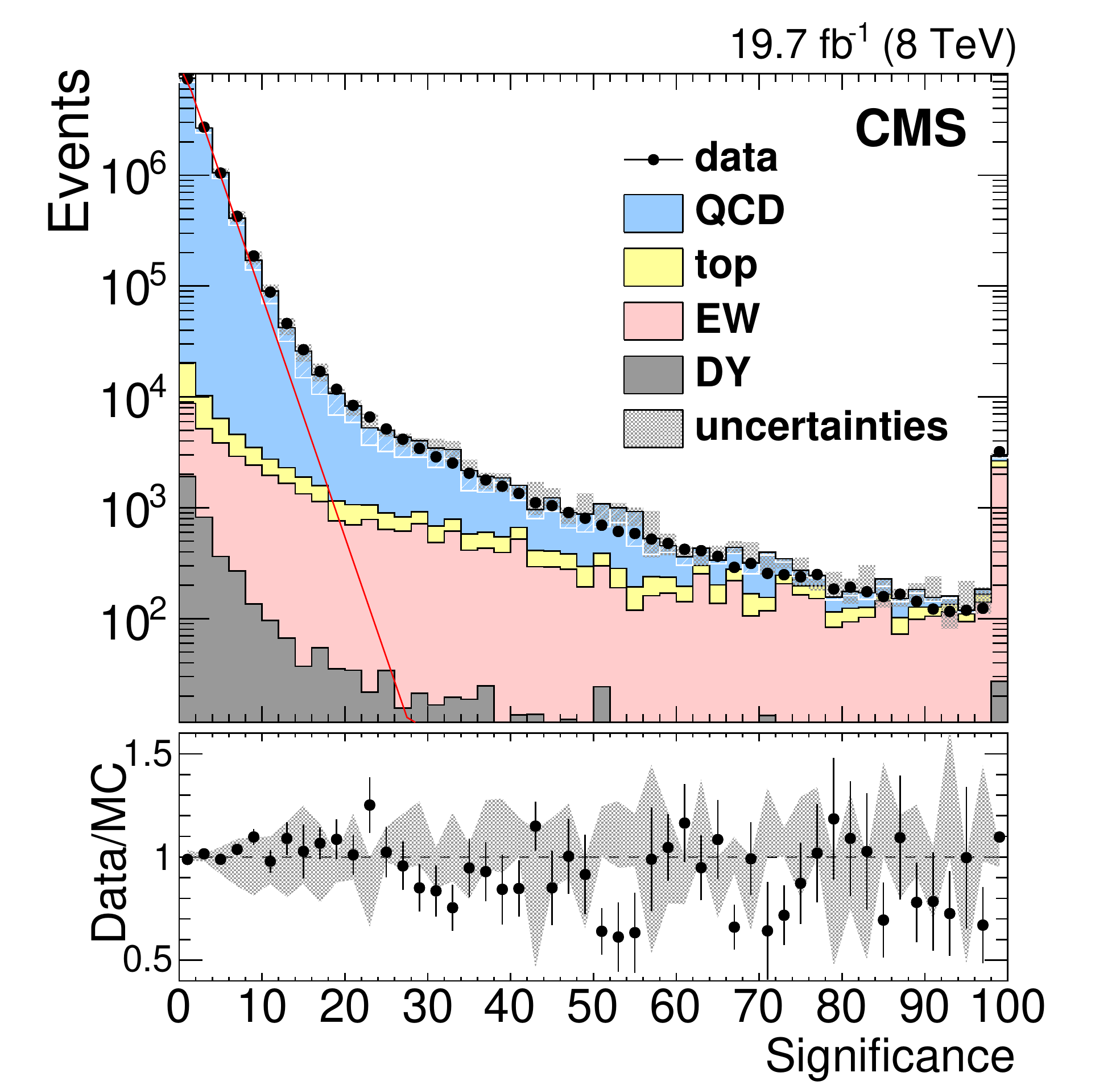}
\caption{
	Distribution of \vecmet significance in the (left) \Zmm\ and (right) dijet samples.
	The red straight line corresponds to a $\chi^2$ distribution of 2 degrees of freedom;
	the white hatched region shows the distribution of events containing genuine non-zero \met.
	The points in the lower panel of each plot show the data/MC ratio, including the
	statistical uncertainties of both data and simulation;
	the grey error band displays the systematic uncertainty of the simulation. The last bin contains the overflow content.
	}
\label{fig:zeromet_sig}
\end{figure}

\begin{figure}[htb]
\centering
\includegraphics[width=0.45\textwidth, height=65mm]{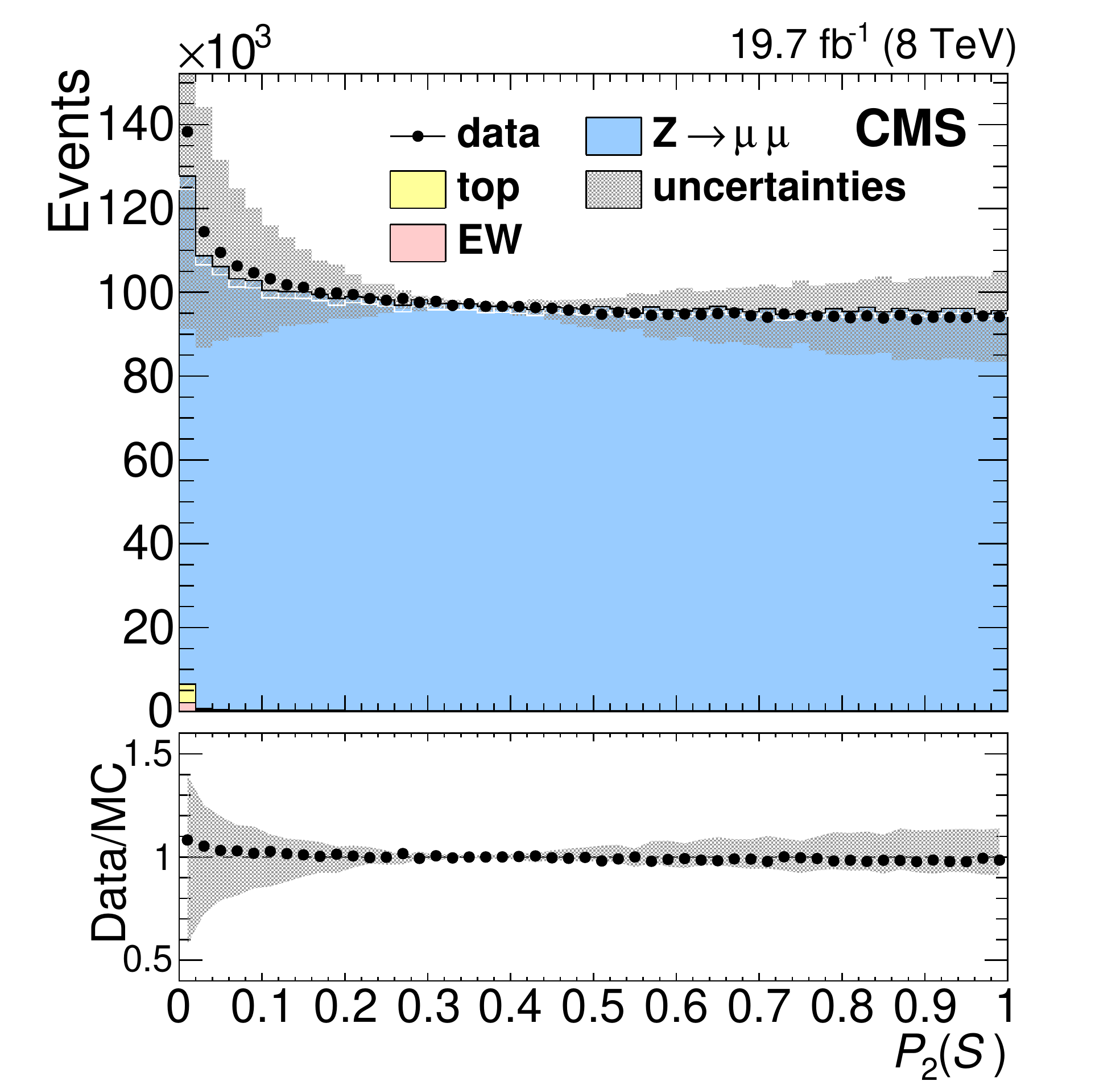}
\includegraphics[width=0.45\textwidth, height=65mm]{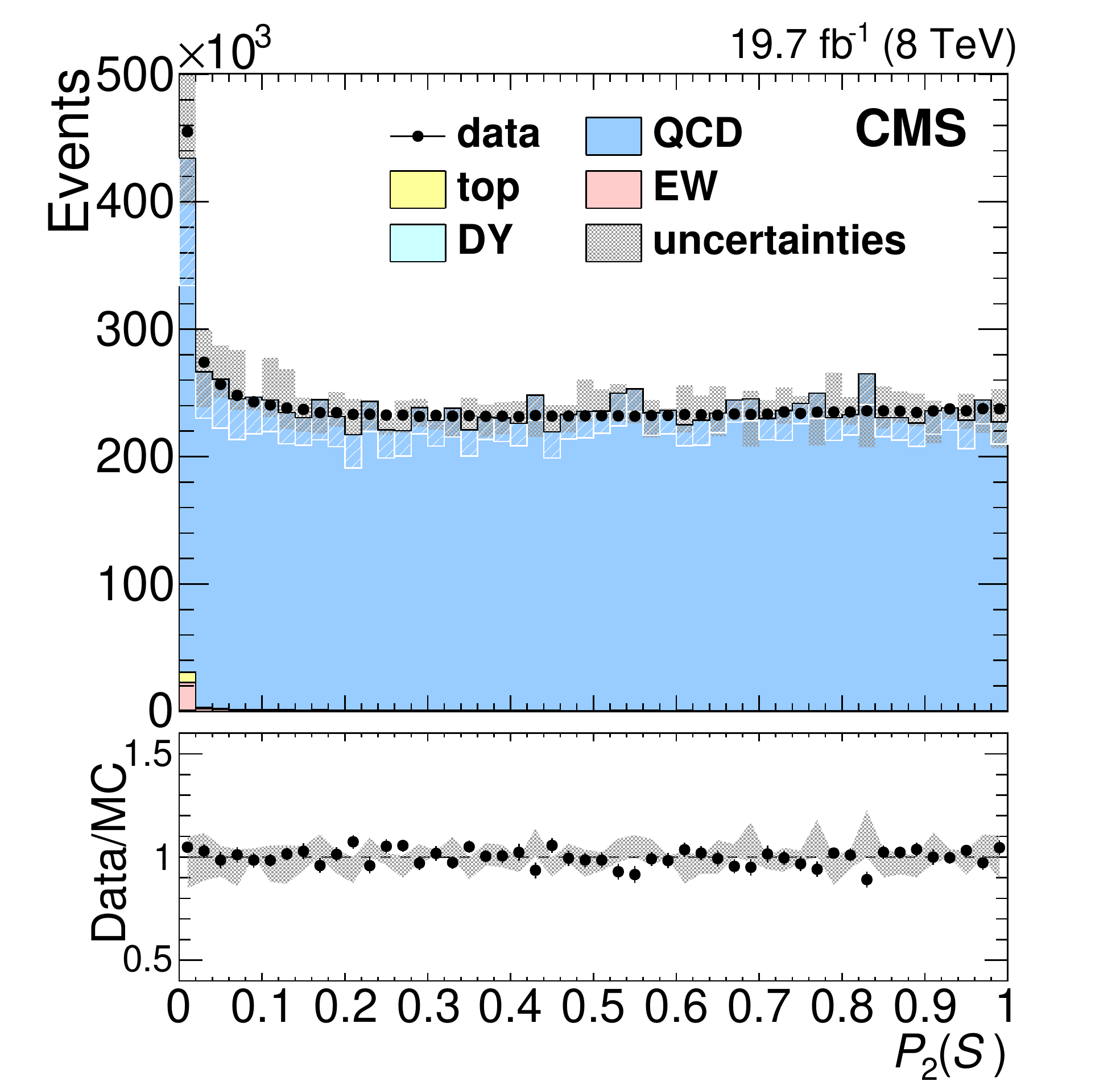}
\caption{
  	Distribution of \pchisq\ in the (left) \Zmm\ and (right) dijet samples.
  	Events that contain a source of genuine \vecmet are represented by the hatched white region.
	The points in the lower panel of each plot show the data/MC ratio, including the
	statistical uncertainties of both data and simulation;
	the grey error band displays the systematic uncertainty of the simulation.
	}
\label{fig:zeromet_pchi2}
\end{figure}

We compare the distributions of \sig\ as well as \pchisq\
in data and simulation for both \Zmm\ and dijet samples in
Figs.~\ref{fig:zeromet_sig} and \ref{fig:zeromet_pchi2}.
The observed spectrum conforms to a $\chi^2$
distribution in the core region, but begins to slightly deviate from a
perfect $\chi^2$ at high values of significance ($\mathcal{S} \gtrsim 9$).
Physics backgrounds containing nonzero true \vecmet (defined here to be
$\met>3\GeV$) are present, but are negligible in comparison to
the dominant zero-\met population. The impact of \Zmm\ events
with true \met due to heavy-quark decays and decays in flight is also found
to contribute to the high-\sig\ region.  Such events only constitute about
1\% of the signal sample in simulated events, however. The general agreement with a
\chisq\ distribution is also apparent in the \pchisq\ spectra,
which are flat over the bulk of events and
show an excess at low values of \pchisq\ (high values of \sig).
It is helpful to keep in mind that
$\pchisq < 0.01$ corresponds to $\mathcal{S} > 9.2$,
$\pchisq < 0.02$ corresponds to $\mathcal{S} > 7.8$,
and
$\pchisq < 0.05$ corresponds to $\mathcal{S} > 6.0$.


\subsubsection{Events with \texorpdfstring{\bigmet}{MET}$\ne0$}
The presence of genuine \vecmet
pushes events to higher values of \sig\ and lower values of \pchisq,
and thus can be used to separate events with genuine \vecmet from
those with only resolution-induced \vecmet. To study the discrimination power
of the significance variable, we use samples of events
containing \PW{}-boson or \ttbar production.
The \Wenu\ channel offers a probe of \vecmet
significance in a scenario dominated by genuine \vecmet,
accompanied by significant zero-\vecmet backgrounds; the
semileptonic \ttbar\ channel similarly provides a genuine \vecmet signal, but with background
events predominantly from higher-\met dileptonic \ttbar\ decays.

\begin{figure}[htb]
\centering
\includegraphics[width=0.45\textwidth]{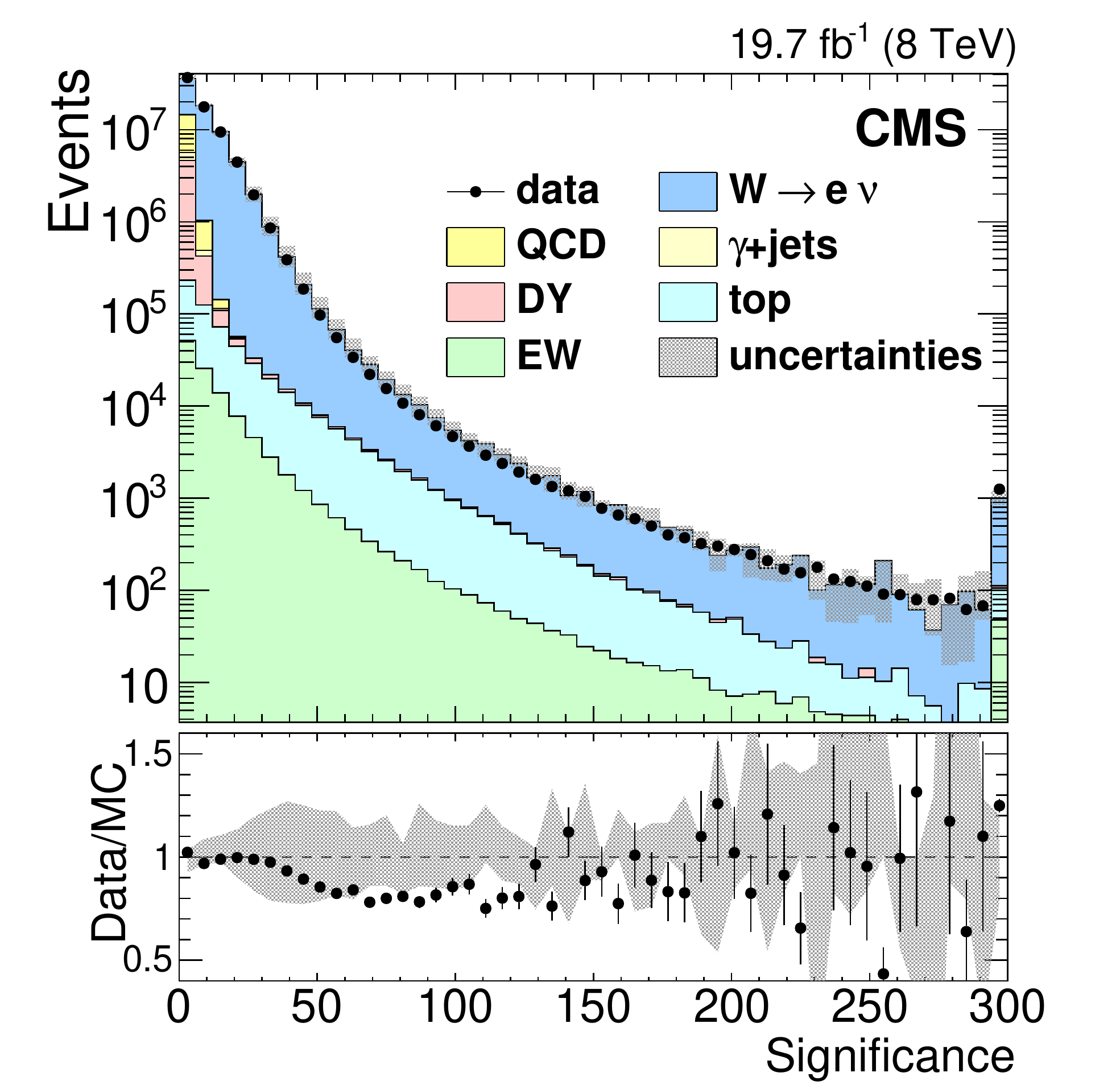}
\includegraphics[width=0.45\textwidth]{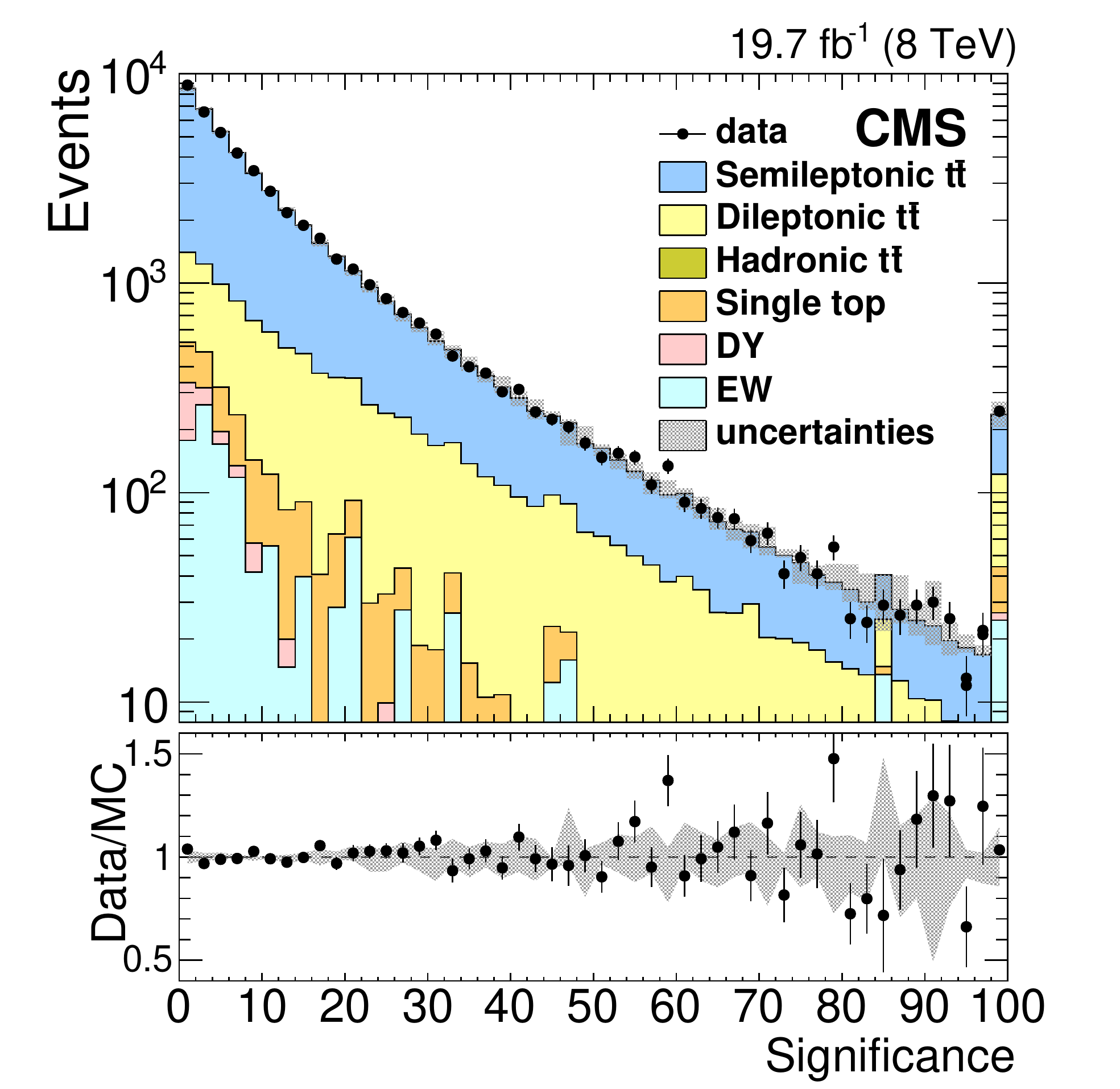}
\caption{
	Distribution of \vecmet significance in the
	(left) \Wenu\ and
	(right) \ttbar\ events. The last bin contains the overflow content. The points in the lower panel of each plot 
        show the data/MC ratio, including the
	statistical uncertainties of both data and simulation;
	the grey error band displays the systematic uncertainty of the simulation.}
\label{fig:realmet_sig}
\end{figure}

\begin{figure}[htb]
\centering
\includegraphics[width=0.45\textwidth]{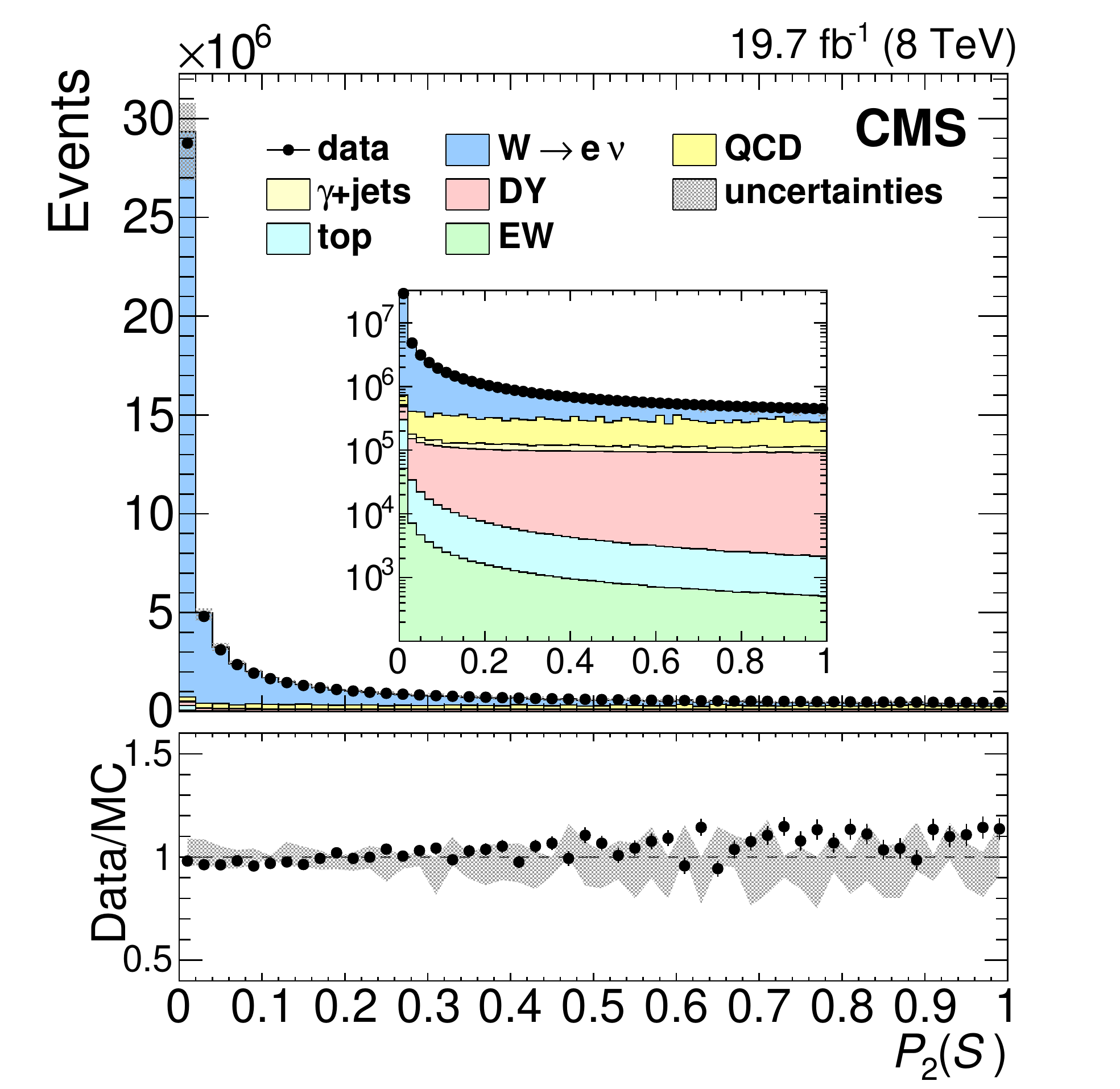}
\includegraphics[width=0.45\textwidth]{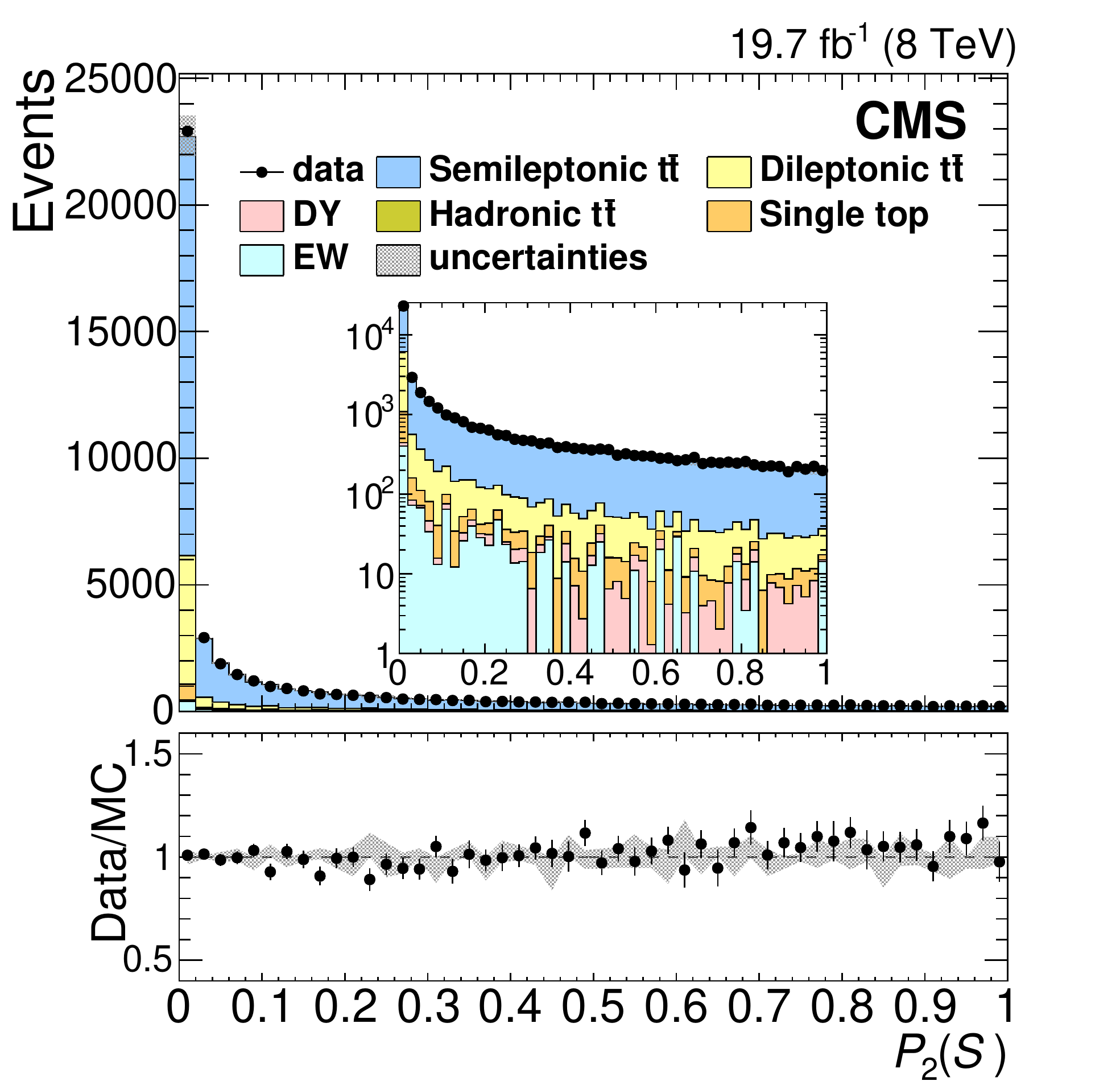}
\caption{
	Distribution of \pchisq\ in the
	(left) \Wenu\ and
	(right) \ttbar\ events.
	The insets show the same data as the main plots, but with a log scale to
	show the background components more clearly. The points in the lower panel of each plot 
        show the data/MC ratio, including the statistical uncertainties of both data and simulation;
	the grey error band displays the systematic uncertainty of the simulation.}
\label{fig:realmet_pchi2}
\end{figure}

The distributions in data and simulation of the \vecmet significance and corresponding \pchisq\
distributions are shown in
Figs.~\ref{fig:realmet_sig}~and~\ref{fig:realmet_pchi2} for both the \Wenu\
and semi-leptonic \ttbar\ events. Some interesting features are apparent
in the composition of simulation events in the significance spectra.
In the \Wenu\ channel, events arising from zero
true \vecmet physics channels, such as QCD and Drell--Yan events, are mostly
found at low values of significance compared to the broad distribution of
non-zero-\vecmet events. Some QCD events show large values of \sig, corresponding to the tail of the distribution observed on Fig.~\ref{fig:zeromet_sig}.
 The semi-leptonic \ttbar\ channel has a significant
non-zero-\vecmet background stemming from dileptonic \ttbar\ decays.  The
dileptonic \ttbar\ spectrum falls more slowly than the semileptonic
\ttbar\ signal in the tail region of \sig.


\subsection{Performance in \texorpdfstring{\Wenu}{W e nu} and semileptonic \texorpdfstring{\ttbar}{ttbar} events}
\label{s:perfwenuttbar}

\begin{figure}[htb]
\centering
\includegraphics[width=0.45\textwidth]{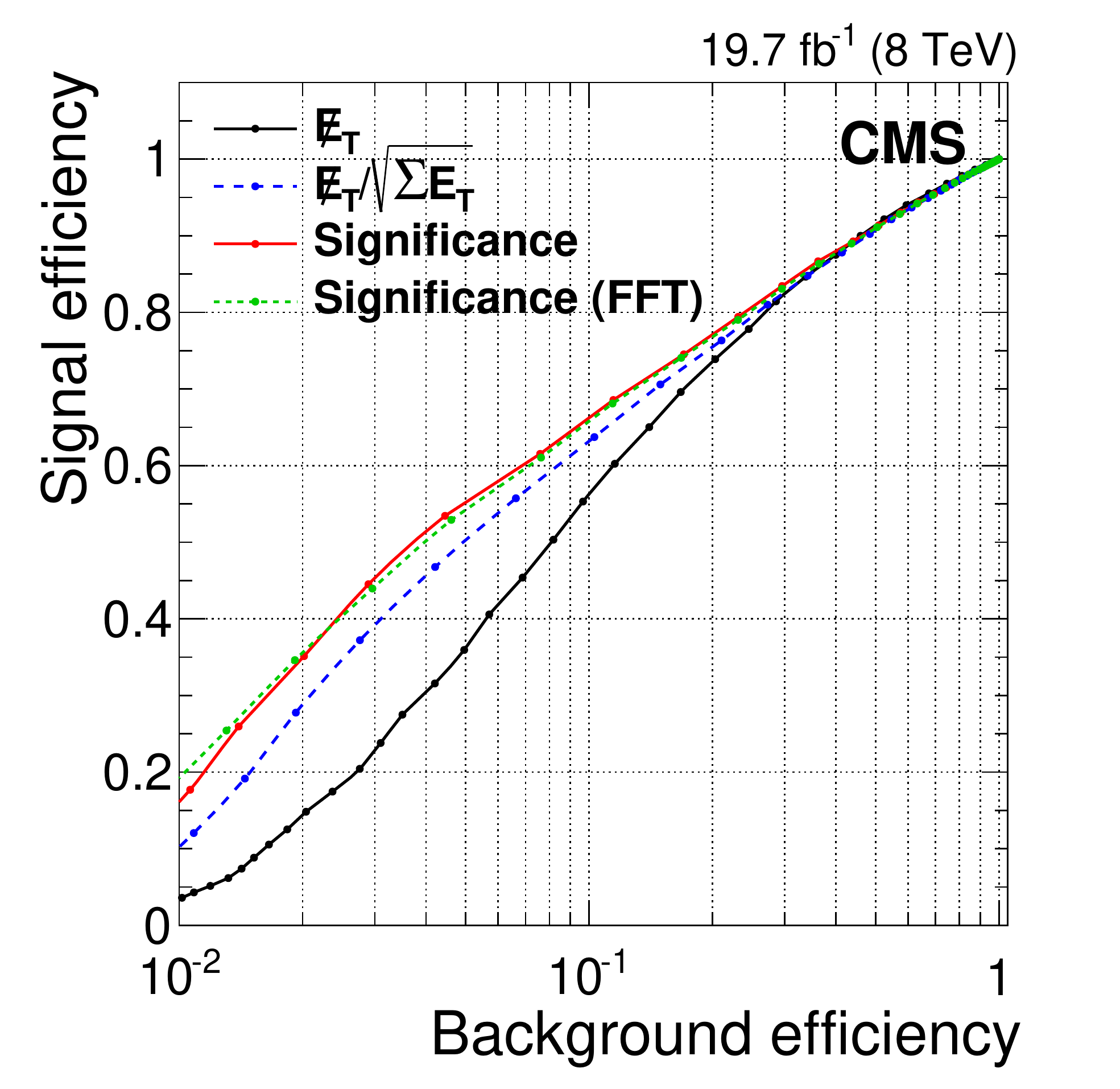}
\caption{
Signal versus background efficiencies for \Wenu\ for various \met-based discriminating variables.
The FFT Significance variable (green dashed line) is discussed in Section~\ref{s:nongaussian}.}
\label{fig:realmet_roc}
\end{figure}

Here we examine the potential gain of introducing the significance variable
into the selection criteria for \Wenu\ and semileptonic \ttbar\ events.
Fig.~\ref{fig:realmet_roc} compares the signal and background
efficiencies for \Wenu\ events in simulation, where increasing thresholds are placed on the
value of \sig, $\pfmet/\sqrt{\sumet}$, and \pfmet.  (The green curve is discussed in Section \ref{s:nongaussian}.)
In the \Wenu\ channel, there is a performance benefit in using \vecmet
significance when compared to simpler background discrimination variables such as
\met alone or the approximate significance variable $\met/\sqrt{\sum\ET}$~\cite{SUS-10-006}.
For example, choosing a
working point with 50\% signal efficiency yields a background efficiency of
8.2\% using \met, 5.1\% using $\met/\sqrt{\sum\ET}$, and 4.0\%
using the significance as a discriminating variable.
For reference, a 50\% signal efficiency working point corresponds to a $\met>40$\GeV requirement.
In the semi-leptonic \ttbar\ channel, \sig\ provides discrimination that is comparable to
\met and $\met/\sqrt{\sum\ET}$.  This reflects the fact that \sig\
is optimized for discriminating events that satisfy the null
hypothesis ($\vec{\epsilon} = 0$) from those that do not.  In the case of
semileptonic \ttbar, the dominant background contribution comes from
dileptonic \ttbar\ decays with large, genuine \met.

We have also evaluated the performance benefit
of modeling individual jet resolutions down to 3\GeV, as in Ref. \citen{METJINST}, as an alternative to
the current threshold of 20\GeV.
Using a lower threshold for individual jets can potentially provide more
detailed information about the low-\pt hadronic activity, but we find that
the performance in the \Wenu\ channel is essentially indistinguishable
when implemented with these two different thresholds, and
therefore use the simpler 20\GeV threshold.

\subsubsection{Pileup}
\label{ss:pileup}
\begin{figure}[htb]
\centering
\includegraphics[width=0.45\textwidth]{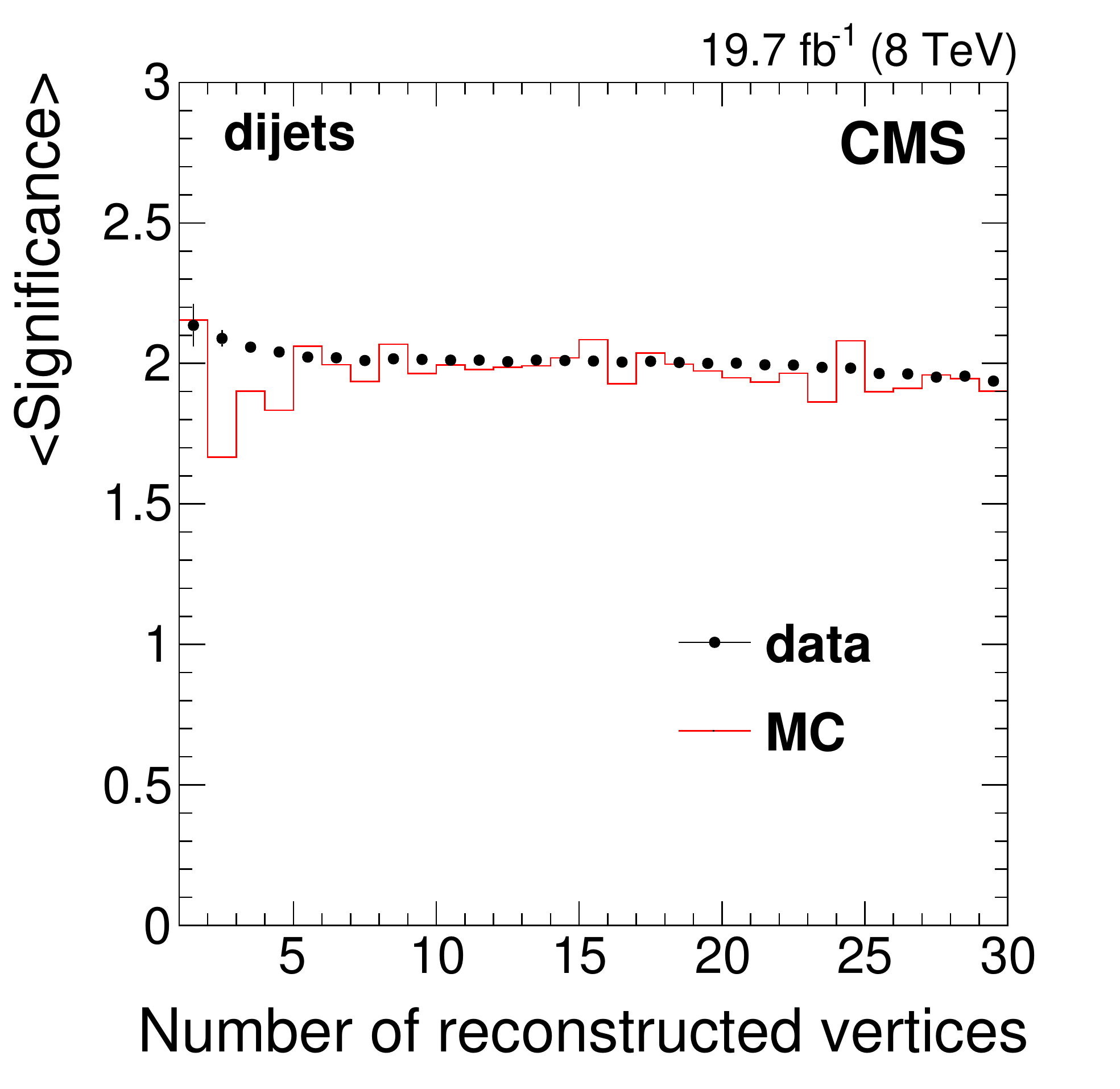}
\includegraphics[width=0.45\textwidth]{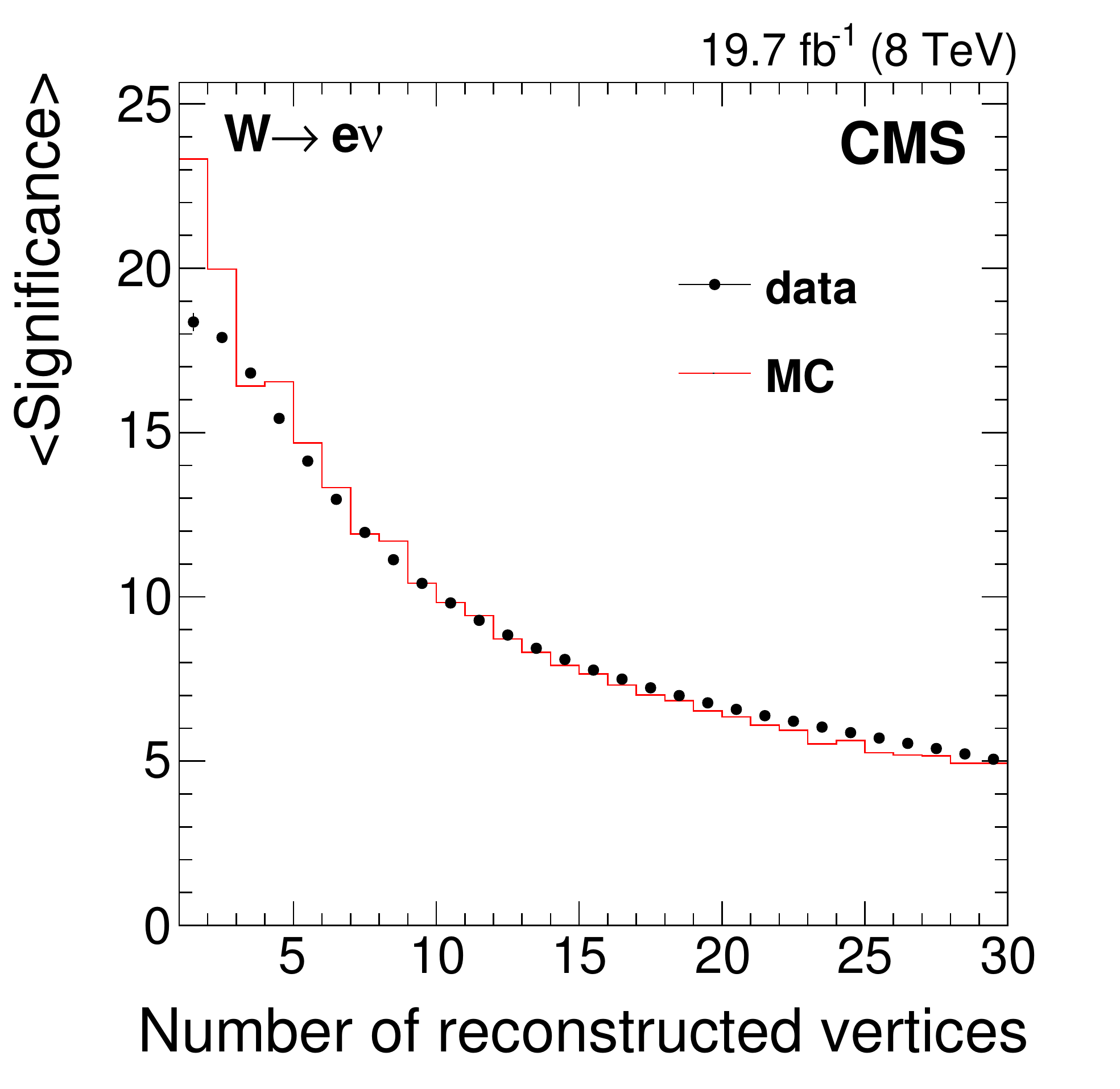}
\caption{
The average \vecmet significance versus the number of reconstructed
vertices for (left) dijet and (right) \Wenu\ event samples.
}
\label{fig:psig_nvert}
\end{figure}

The \vecmet significance variable exhibits simple behavior as a function of the
number of pileup interactions. For event samples such as the \Zmm\ and dijet selections,
in which in most events there is no source of true \met, the \sig\ value remains essentially
constant as the number of primary vertices increases. In samples such as
\Wenu\ and \ttbar, where the average value of \met is
non-zero, a decrease with increasing pileup is seen. This behavior
can be derived formally from the expression for \sig\ given
in Eq.~\eqref{e:metsig-gaussian} with the isotropic model of
unclustered energy given in Eq.~\eqref{e:isotropic} if the additional
covariance due to $n$ pileup vertices is incorporated via the replacement
$\mathbf{V}\to \mathbf{V}_0+n\sigma^2\mathbf{I}$. In this transformation $\mathbf{V}_0$
represents the covariance matrix in the absence of pileup.
It is also
confirmed empirically in Fig.~\ref{fig:psig_nvert}.
As a side point, we note that $\langle{\cal S}\rangle\approx 2$
for the zero-\met events, as one expects for a \chisq\ variable with two
degrees of freedom.

\begin{figure}[htb]
\centering
\includegraphics[width=0.45\textwidth]{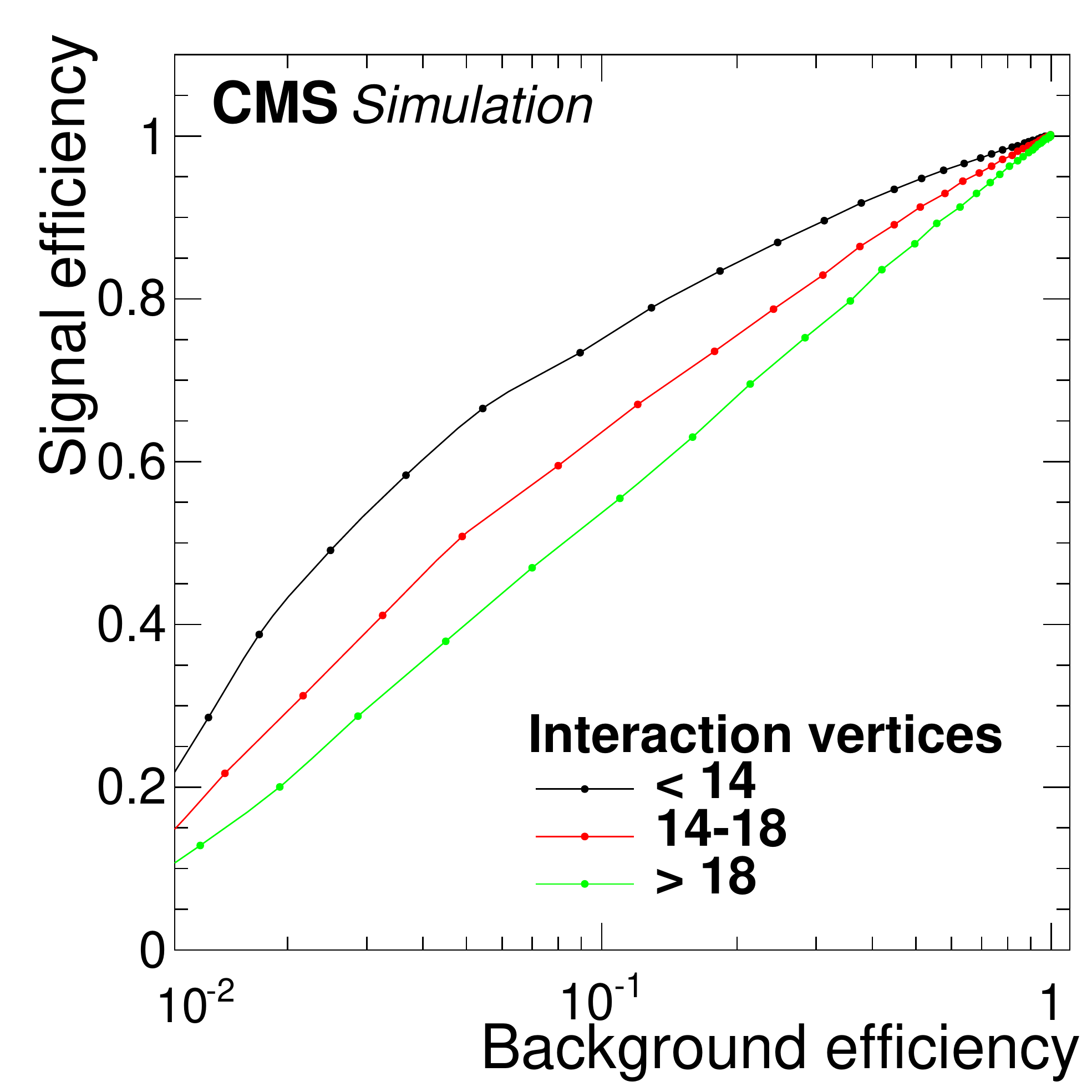}
\includegraphics[width=0.45\textwidth]{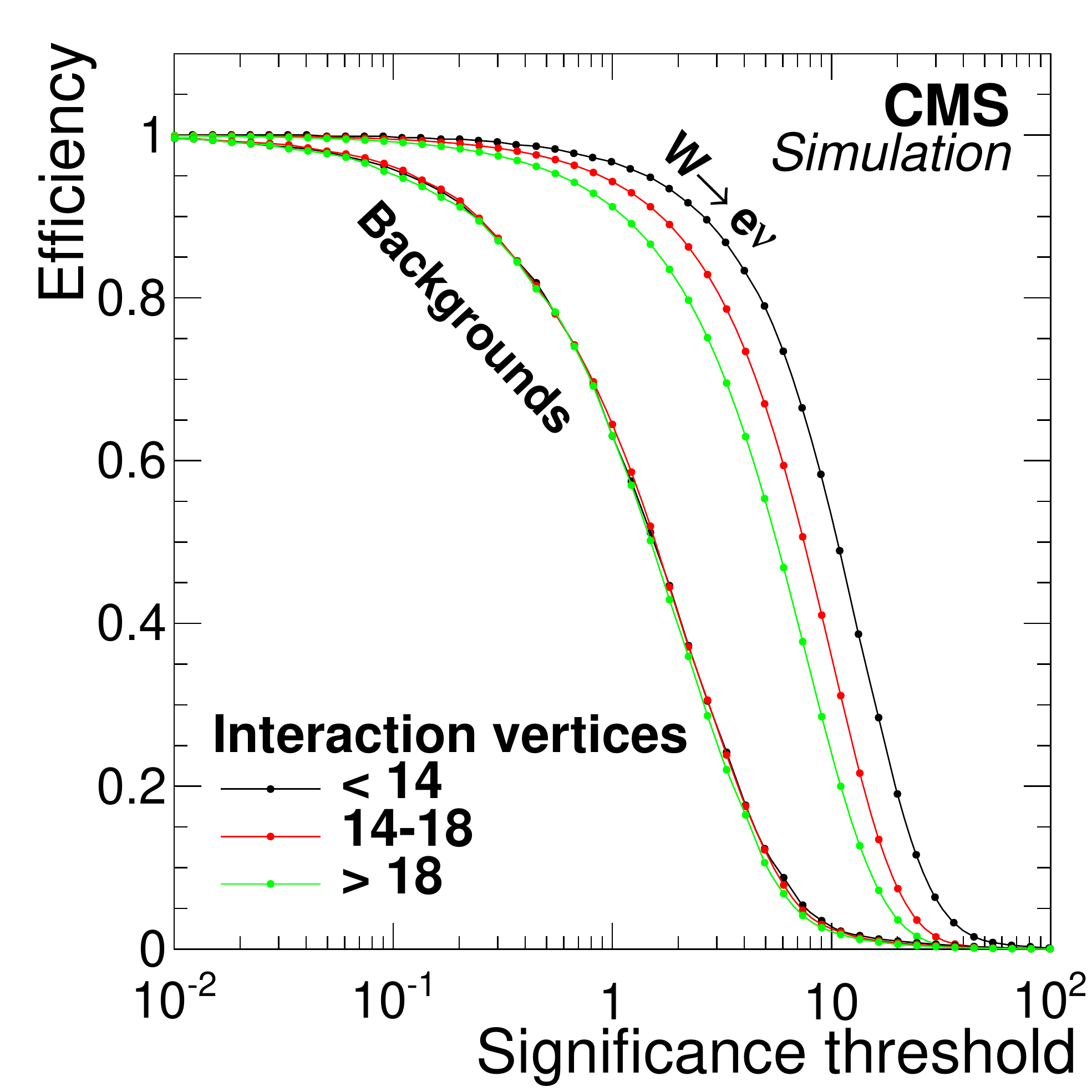}
\caption{Efficiency curves for \vecmet significance in \Wenu\ channel in
three regions defined by the number of reconstructed vertices.
The signal versus background efficiencies are shown in the left pane.
In the right pane, the signal (right) and background (left) efficiencies are
shown separately as a function of the threshold on \sig.}
\label{fig:wenu_roc_pu}
\end{figure}

As a result of the pileup dependence observed for genuine \met events, the
background rejection performance of the \vecmet significance can also exhibit a
dependence on pileup.  This is demonstrated for the \Wenu\ channel in
Fig.~\ref{fig:wenu_roc_pu}.  Here we see a decreasing signal efficiency
as the pileup increases.  It is also apparent that while the efficiencies of
non-zero-\met signal events depend on pileup, the efficiencies for the zero-\met background events are
relatively stable. It should be mentioned that the use of a significance algorithm based on No--PU input
objects would reduce the dependency of \sig with the number of additional pileup interactions.

\subsection{Treatment of non-Gaussian resolutions}\label{s:nongaussian}

As noted earlier, the jet \pt resolution
functions exhibit non-Gaussian tails. The challenge presented by such tails
lies in the convolution integrals needed to compute the \met likelihood
function. This can be done analytically for Gaussian resolutions, but not
when non-Gaussian elements are introduced and direct, numerical
convolution is prohibitively slow. The convolution process, however, can be
reduced under Fourier transformation to a simple multiplication of the
transformed functions. With this approach, each jet resolution function
$R_i(p_x,p_y)$ is transformed to $\widetilde R_i(k_x,k_y)$, and then the
product $\prod_{i=1}^{n}\widetilde R_i(k_x,k_y)$ is computed and
back-transformed to yield the fully convolved result. When computed with
fast Fourier transform (FFT) techniques, this method enables the required
convolutions to be done at a speed that, while slower than the evaluation of
analytic functions, is still well within reason for late stages of
analysis. Both $R$ and $\widetilde R$ are discretized on 2-dimensional
grids in their respective spaces, and the resulting discretized likelihood
function is smoothed by cubic spline interpolation before computing the
significance. Care is taken in defining the grids to avoid artifacts that
can result from aliasing. To verify the validity of this FFT method and its
implementation, we have compared the results of the FFT and analytic
methods for cases where only Gaussian resolutions are used and find the two
methods yield identical results. When introduced into the selection
criteria for \Wenu\ events, the two methods give comparable results, as
seen in Fig.~\ref{fig:realmet_roc}.

\begin{figure}[htb]
\centering
\includegraphics[width=0.45\textwidth]{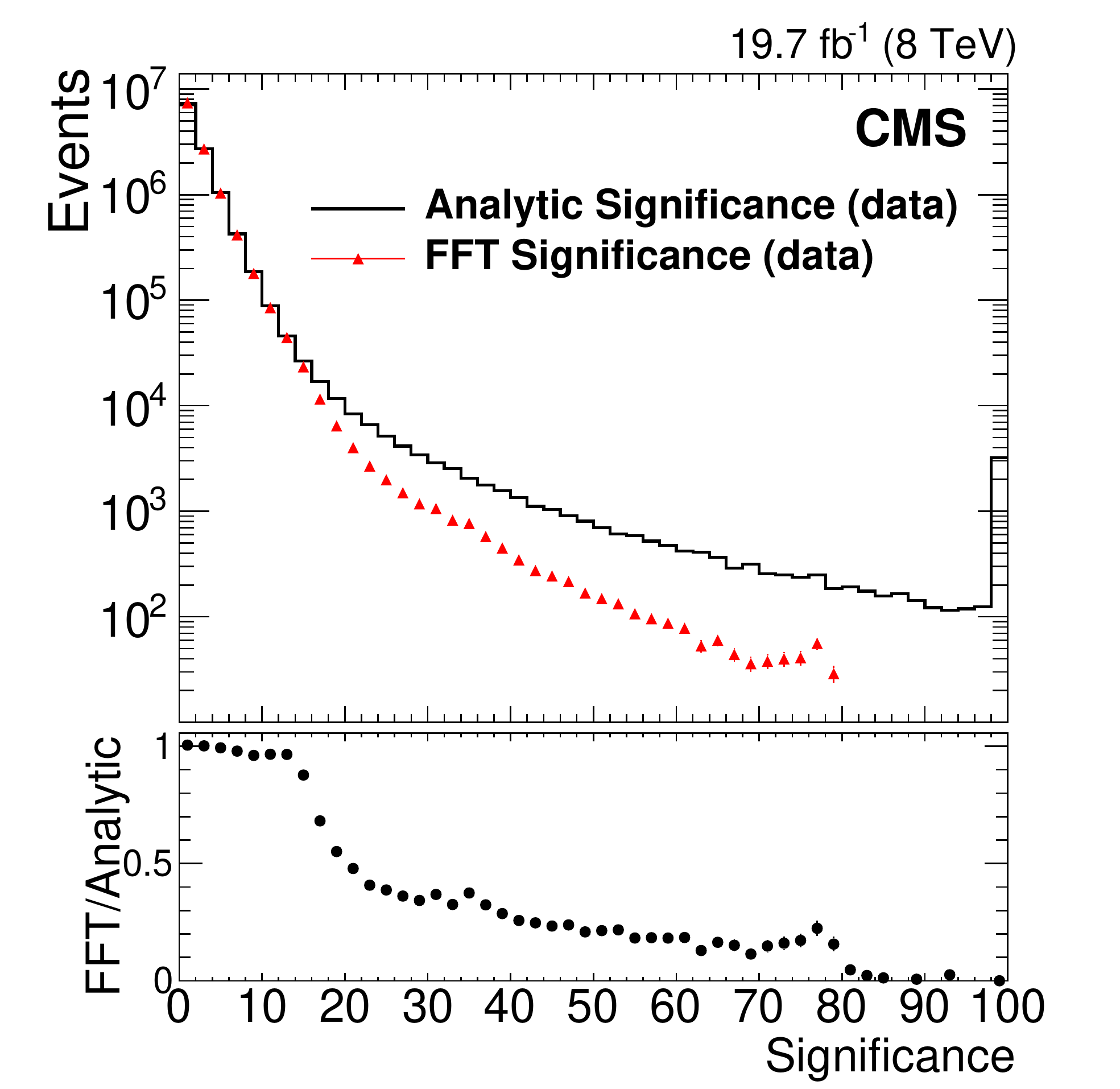}
\includegraphics[width=0.45\textwidth]{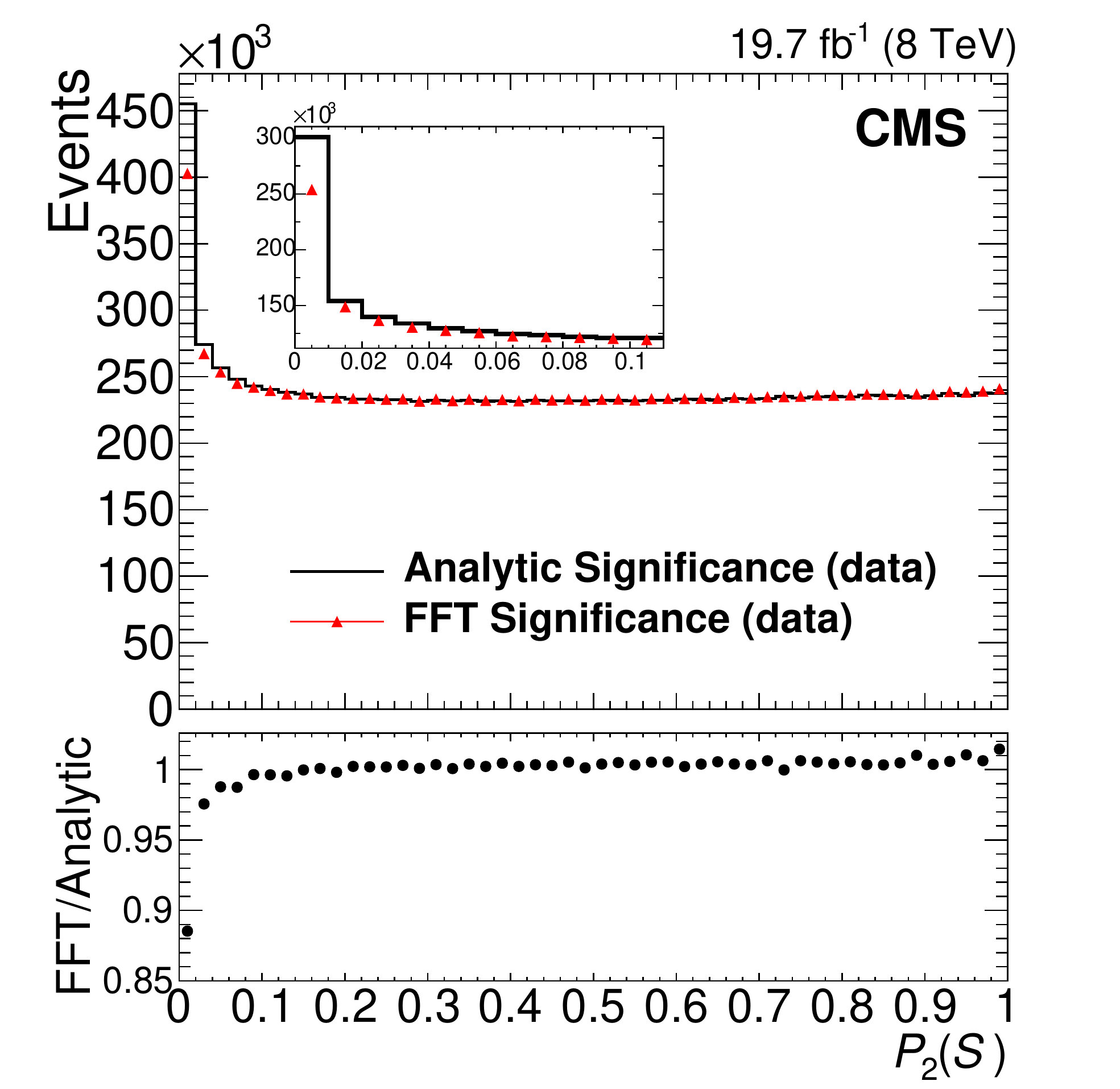}
\caption{
	Comparisons in dijet events of the FFT (non-Gaussian) and analytic (Gaussian) methods
	for calculating \vecmet significance.
	Left: \vecmet significance distribution. Right: \pchisq\ distribution.
	For this figure, both the FFT (red triangles) and analytic (black histogram)
	algorithms are applied only to data.
	The analogous MC distributions for the analytic method are shown in
	Figs.~\ref{fig:zeromet_sig}~and~\ref{fig:zeromet_pchi2}.
	Non-Gaussian significance values of $\mathcal{S}\gtrsim 80$ are suppressed
	due to the finite number of significant figures available to double
	precision variables used in the FFT algorithm. The last bin contains the overflow content.
	}
\label{f:ng}
\end{figure}

To demonstrate the potential utility of the non-Gaussian treatment, we
compare \vecmet significance computed with the FFT and with the analytic method. For the comparison,
we use the dijet event sample, as there is sufficient high-\pt
hadronic activity to exhibit clearly the effects of non-Gaussian contributions to the resolution.
Figure~\ref{f:ng} shows the results of the comparison. The significance
distribution is plotted in the left panel, with the black histogram
computed by the analytic method (i.e. assuming only Gaussian resolutions),
and red data points computed with the FFT
algorithm (using full resolution functions).
The steeper fall of the red points demonstrates that the FFT
algorithm helps to reduce the excess of high-significance values
that arise in the analytic method where the jet measurement uncertainty
is underestimated by the Gaussian approximation.
Events showing non-Gaussian significance values of $\mathcal{S}\gtrsim 80$ are suppressed
due to the finite number of significant digits available to double
precision variables used in the FFT algorithm.

 The right-hand panel
 shows the corresponding reduction of events in the lowest bin
of the \pchisq\ distribution. The remaining excess in that bin is partly
due to events with genuine \met that arise from semileptonic decays of
hadrons. After taking into account these genuine \met components and other
extraneous backgrounds from \ttbar\ and vector boson production,
the net impact of the FFT algorithm is to reduce
the \textit{excess} of zero-\met events in the high-significance, low-\pchisq\ bin
($\pchisq<0.02$) by a factor of two.  Removal of the remaining zero-\met
events in this bin will require deeper understanding of the jet-by-jet
resolution variations that are not captured by the average parametrizations
currently available.

\section{Summary}
\label{sec:conclusions}

\newcommand{\wenu}{\ensuremath{W\rightarrow e\nu}}

The performance of \vecmet reconstruction algorithms has been studied
using data collected in 8\TeV pp collisions with the CMS detector at the LHC.
The data used in this paper were collected from February through December 2012 and correspond to an integrated luminosity up to $\fulllumi\pm\fulllumierr$\fbinv.
The \vecmet reconstruction algorithms and corrections are described
with an emphasis on changes compared to those used with the 7\TeV pp
data collected in 2010~\cite{METJINST}.
Events with artificially high \met in a dijet event sample are examined, and
we find that a majority of such events can be identified and either modified or removed.

We have measured the scale and resolution of \pfvecmet,
as well as the degradation of the \pfvecmet performance due to pileup interactions
in \Zmm, \Zee, and direct-photon events.
The measured \pfvecmet\ scale and resolution in data agree with
the expectations from the simulation after correcting for the jet energy
scale and resolution differences between data and simulation.
We find that pileup interactions
contribute to the degradation of the \pfvecmet\ resolution by 3.3--3.6\GeV
(in quadrature) per additional pileup interaction, similar to the results obtained with the 7\TeV pp data.

We have studied the performance of two novel \vecmet\ reconstruction algorithms
specifically developed to cope with large numbers of pileup interactions.
They show significantly reduced dependence of the \vecmet\ resolution on pileup interactions,
consistently in both data and simulation, although the \vecmet\ response is slightly deteriorated.
With a dedicated configuration of the algorithms, however, the \vecmet\ response can be preserved.

We have also studied the performance of  the \vecmet significance algorithm,
developed to distinguish between events with spurious \vecmet and events with
genuine \vecmet. As an example of its utility, the \vecmet significance shows better discrimination
between \Wenu\ events and QCD or Drell--Yan events compared to a
standard \met reconstruction algorithm.

The studies presented in this paper provide a solid foundation for all the
CMS measurements with \vecmet in the final state,
including measurements involving W bosons and top quarks, searches for
new weakly interacting neutral particles,
and studies of the properties of the Higgs boson.

\begin{acknowledgments}
\label{sec:acknowledgments}
\hyphenation{Bundes-ministerium Forschungs-gemeinschaft Forschungs-zentren} We congratulate our colleagues in the CERN accelerator departments for the excellent performance of the LHC and thank the technical and administrative staffs at CERN and at other CMS institutes for their contributions to the success of the CMS effort. In addition, we gratefully acknowledge the computing centres and personnel of the Worldwide LHC Computing Grid for delivering so effectively the computing infrastructure essential to our analyses. Finally, we acknowledge the enduring support for the construction and operation of the LHC and the CMS detector provided by the following funding agencies: the Austrian Federal Ministry of Science, Research and Economy and the Austrian Science Fund; the Belgian Fonds de la Recherche Scientifique, and Fonds voor Wetenschappelijk Onderzoek; the Brazilian Funding Agencies (CNPq, CAPES, FAPERJ, and FAPESP); the Bulgarian Ministry of Education and Science; CERN; the Chinese Academy of Sciences, Ministry of Science and Technology, and National Natural Science Foundation of China; the Colombian Funding Agency (COLCIENCIAS); the Croatian Ministry of Science, Education and Sport, and the Croatian Science Foundation; the Research Promotion Foundation, Cyprus; the Ministry of Education and Research, Estonian Research Council via IUT23-4 and IUT23-6 and European Regional Development Fund, Estonia; the Academy of Finland, Finnish Ministry of Education and Culture, and Helsinki Institute of Physics; the Institut National de Physique Nucl\'eaire et de Physique des Particules~/~CNRS, and Commissariat \`a l'\'Energie Atomique et aux \'Energies Alternatives~/~CEA, France; the Bundesministerium f\"ur Bildung und Forschung, Deutsche Forschungsgemeinschaft, and Helmholtz-Gemeinschaft Deutscher Forschungszentren, Germany; the General Secretariat for Research and Technology, Greece; the National Scientific Research Foundation, and National Innovation Office, Hungary; the Department of Atomic Energy and the Department of Science and Technology, India; the Institute for Studies in Theoretical Physics and Mathematics, Iran; the Science Foundation, Ireland; the Istituto Nazionale di Fisica Nucleare, Italy; the Korean Ministry of Education, Science and Technology and the World Class University program of NRF, Republic of Korea; the Lithuanian Academy of Sciences; the Ministry of Education, and University of Malaya (Malaysia); the Mexican Funding Agencies (CINVESTAV, CONACYT, SEP, and UASLP-FAI); the Ministry of Business, Innovation and Employment, New Zealand; the Pakistan Atomic Energy Commission; the Ministry of Science and Higher Education and the National Science Centre, Poland; the Funda\c{c}\~ao para a Ci\^encia e a Tecnologia, Portugal; JINR, Dubna; the Ministry of Education and Science of the Russian Federation, the Federal Agency of Atomic Energy of the Russian Federation, Russian Academy of Sciences, and the Russian Foundation for Basic Research; the Ministry of Education, Science and Technological Development of Serbia; the Secretar\'{\i}a de Estado de Investigaci\'on, Desarrollo e Innovaci\'on and Programa Consolider-Ingenio 2010, Spain; the Swiss Funding Agencies (ETH Board, ETH Zurich, PSI, SNF, UniZH, Canton Zurich, and SER); the Ministry of Science and Technology, Taipei; the Thailand Center of Excellence in Physics, the Institute for the Promotion of Teaching Science and Technology of Thailand, Special Task Force for Activating Research and the National Science and Technology Development Agency of Thailand; the Scientific and Technical Research Council of Turkey, and Turkish Atomic Energy Authority; the National Academy of Sciences of Ukraine, and State Fund for Fundamental Researches, Ukraine; the Science and Technology Facilities Council, UK; the US Department of Energy, and the US National Science Foundation.

Individuals have received support from the Marie-Curie programme and the European Research Council and EPLANET (European Union); the Leventis Foundation; the A. P. Sloan Foundation; the Alexander von Humboldt Foundation; the Belgian Federal Science Policy Office; the Fonds pour la Formation \`a la Recherche dans l'Industrie et dans l'Agriculture (FRIA-Belgium); the Agentschap voor Innovatie door Wetenschap en Technologie (IWT-Belgium); the Ministry of Education, Youth and Sports (MEYS) of the Czech Republic; the Council of Science and Industrial Research, India; the HOMING PLUS programme of Foundation for Polish Science, cofinanced from European Union, Regional Development Fund; the Compagnia di San Paolo (Torino); the Consorzio per la Fisica (Trieste); MIUR project 20108T4XTM (Italy); the Thalis and Aristeia programmes cofinanced by EU-ESF and the Greek NSRF; and the National Priorities Research Program by Qatar National Research Fund.
\end{acknowledgments}

\bibliography{auto_generated}  

\cleardoublepage \appendix\section{The CMS Collaboration \label{app:collab}}\begin{sloppypar}\hyphenpenalty=5000\widowpenalty=500\clubpenalty=5000\textbf{Yerevan Physics Institute,  Yerevan,  Armenia}\\*[0pt]
V.~Khachatryan, A.M.~Sirunyan, A.~Tumasyan
\vskip\cmsinstskip
\textbf{Institut f\"{u}r Hochenergiephysik der OeAW,  Wien,  Austria}\\*[0pt]
W.~Adam, T.~Bergauer, M.~Dragicevic, J.~Er\"{o}, C.~Fabjan\cmsAuthorMark{1}, M.~Friedl, R.~Fr\"{u}hwirth\cmsAuthorMark{1}, V.M.~Ghete, C.~Hartl, N.~H\"{o}rmann, J.~Hrubec, M.~Jeitler\cmsAuthorMark{1}, W.~Kiesenhofer, V.~Kn\"{u}nz, M.~Krammer\cmsAuthorMark{1}, I.~Kr\"{a}tschmer, D.~Liko, I.~Mikulec, D.~Rabady\cmsAuthorMark{2}, B.~Rahbaran, H.~Rohringer, R.~Sch\"{o}fbeck, J.~Strauss, A.~Taurok, W.~Treberer-Treberspurg, W.~Waltenberger, C.-E.~Wulz\cmsAuthorMark{1}
% ...
\vskip\cmsinstskip
% ...
1:~~Also at Vienna University of Technology, Vienna, Austria\\
2:~~Also at CERN, European Organization for Nuclear Research, Geneva, Switzerland\\
% ...

\end{sloppypar}
\end{document}